\documentclass[onecolumn,10pt,showpacs,preprintnumbers,nofootinbib,prd,superscriptaddress,english,aps]{revtex4-1}

\usepackage{graphicx,amssymb,amsmath,amsthm,amsfonts,epsfig,bm,upgreek,times}

\usepackage[linktocpage]{hyperref}
\usepackage[usenames]{color}

\def\LL{\left[\left[}
\def\RR{\right]\right]}

\def\nn{\nonumber}

\newcommand{\ben}{\begin{enumerate}}
\newcommand{\een}{\end{enumerate}}

\def\be{\begin{equation}}
\def\ee{\end{equation}}
\def\bea{\begin{eqnarray}}
\def\eea{\end{eqnarray}}
\newcommand{\beq}{\begin{eqnarray}}
\newcommand{\eeq}{\end{eqnarray}}

\begin{document}
\title{Can environmental effects spoil precision gravitational-wave astrophysics?}

\author{Enrico Barausse}\email{barausse@iap.fr}
\affiliation{CNRS, UMR 7095, Institut d'Astrophysique de Paris, 98bis Bd Arago, 75014 Paris, France}
\affiliation{Sorbonne Universit\'es, UPMC Univ Paris 06, UMR 7095, 98bis Bd Arago, 75014 Paris, France}
\author{Vitor Cardoso}\email{vitor.cardoso@tecnico.ulisboa.pt}
\affiliation{CENTRA, Departamento de F\'{\i}sica, Instituto Superior T\'ecnico, Universidade de Lisboa,
Avenida Rovisco Pais 1, 1049 Lisboa, Portugal.}
\affiliation{Perimeter Institute for Theoretical Physics, Waterloo, Ontario N2L 2Y5, Canada.}
%

\author{Paolo Pani}\email{paolo.pani@tecnico.ulisboa.pt}
\affiliation{CENTRA, Departamento de F\'{\i}sica, Instituto Superior T\'ecnico, Universidade de Lisboa,
Avenida Rovisco Pais 1, 1049 Lisboa, Portugal.}
\affiliation{Institute for Theory and Computation, Harvard-Smithsonian
CfA, 60 Garden Street, Cambridge MA 02138, USA}

\date{\today} 

\begin{abstract} 
No, within a broad class of scenarios. 
Gravitational-wave (GW) astronomy will open a new window on compact objects such as neutron stars and black holes (BHs).
It is often stated that large signal-to-noise detections of ringdown or inspiral waveforms can provide estimates of the masses and spins 
of compact objects to within fractions of a percent, as well as tests of General Relativity. These expectations usually neglect the realistic astrophysical environments in which compact objects live.
With the advent of GW astronomy, environmental effects on the GW signal will eventually have to be \emph{quantified}.
Here we present a wide survey of the corrections due to these effects in two situations of great interest for GW astronomy: the BH ringdown emission and the inspiral of two compact objects
(especially BH binaries).
We mainly focus on future space-based detectors such as eLISA, but many of our results are also valid for ground-based detectors such as aLIGO, aVirgo and KAGRA. We take into account various effects such as: electric charges, magnetic fields, cosmological evolution, possible deviations from General Relativity, firewalls, and the effects related to various forms of matter such as 
accretion disks and dark matter halos. 

Our analysis predicts the existence of resonances dictated by the external mass distribution, which dominate the very late-time behavior of merger/ringdown waveforms. 
The mode structure can drastically differ from the vacuum case, yet the BH response to external perturbations is unchanged at the time scales relevant for detectors.
This is because although the vacuum Schwarzschild resonances are no longer quasinormal modes of the system, they still dominate the response at intermediate times. Our results strongly suggest 
that both parametrized and ringdown searches should use at least two-mode templates.

Our analysis of compact binaries shows 
that environmental effects are typically negligible for most eLISA
sources, with the exception of very few special extreme mass ratio inspirals. We show in
particular that accretion and hydrodynamic drag generically dominate over self-force effects for geometrically thin disks, whereas they can be safely neglected for
geometrically thick disk environments, which are the most relevant for eLISA.
Finally, we discuss how our ignorance of the matter surrounding compact objects implies intrinsic limits on the ability to constrain strong-field deviations from General Relativity.
\end{abstract}

\maketitle

\small
\tableofcontents
\normalsize

\section*{Acronyms}
\begin{table}[h]
\begin{tabular}{lllll}
BH     & Black Hole              &\hspace{1cm}     &ISCO   & Innermost Stable Circular Orbit                        \\
DM     & Dark Matter             &\hspace{1cm}      &LIGO   & Laser Interferometric Gravitational Wave Observatory \\
dS     & de Sitter               &\hspace{1cm}      &PN     & Post-Newtonian                                       \\
EoS    & Equation of state       &\hspace{1cm}      &PPN    & parametrized Post-Newtonian                          \\
GR     & General Relativity      &\hspace{1cm}      &QNM    & Quasinormal mode                                 \\
GW     & Gravitational Wave

\end{tabular}
\end{table}

\clearpage
\newpage

\section*{Notation and conventions}

Unless otherwise and explicitly stated, we use geometrized units where
$G=c=1$, so that energy and time have units of length. We also adopt the
$(-+++)$ convention for the metric.  For reference, the following is a list of
symbols that are used often throughout the text.

\begin{table}[h]
\begin{tabular}{ll}
  $g_{\alpha \beta}$ & Spacetime metric; Greek indices run from 0 to $3$. \\
  $M$ & BH mass, defined via the horizon area $A$ as $M^2=A/(16\pi)$. \\
  $M_{\odot}$& Mass of our Sun.\\
  $\delta M$ & Mass of the exterior distribution.\\
  $M_0=M+\delta M$ & Total Arnowitt-Deser-Misner mass of the spacetime. \\
  $D_L$			& Luminosity distance.\\
  $L$ & Typical lengthscale for the extension of the exterior mass $\delta M$ distribution.\\
  $r_0$ & Minimum radial distance at which the exterior mass distribution is non-negligible. \\
        &(Infinitely thin shells of matter are located at $r=r_0$.)\\
  $r_+$ & Radius of the BH event horizon in the chosen coordinates.\\
  $m_{\rm sat}$ & Mass of the small satellite in a binary system.\\
  $M_T=M+m_{\rm sat}$ & Total mass of a binary system. In the extreme mass ratio limit, $M_T\to M$.\\
  $\nu$		& Symmetric mass ratio of a binary system. In the extreme mass ratio limit, $\nu\to m_{\rm sat}/M$ is the mass ratio.\\
  $m_p$		& Proton mass.\\
  $\dot{M}$& Accretion rate onto the BH.\\
  $L_{\rm Edd}$& Eddington luminosity.\\
  $f_{\rm Edd}=\dot{M}/\dot{M}_{\rm Edd}$ & Eddington ratio for mass accretion.\\
  $v_s$& Speed of sound in matter.\\
  $H$	& Height of the disk.\\
  $\eta$	& Radiative efficiency of the accretion process.\\
  $\alpha$	& Viscosity parameter.\\
  $F_{\rm DF}$ & Dynamical friction force. \\
  $\tilde{r}$	& Orbital separation normalized by the gravitational radius ($G M/c^2$ for isolated BHs, $G M_T/c^2$ for binaries).\\
  $v_K$	       & Keplerian velocity [$(GM/r)^{1/2}$ for test particles around BHs, $(GM_T/r)^{1/2}$ for binaries]. \\
  $\Omega_\phi$		& Angular velocity of the binary.\\
  $v_r$			& Radial velocity of the binary.\\
  $\Sigma$ & Surface density of (geometrically thin) accretion disks.\\
  $\sigma$ & Typical lengthscale for the extension of initial data profiles.\\
  $\sigma_T$ & Thomson scattering cross-section.\\
  $\sigma_A$ & Dark matter annihilation cross section.\\
  $\sigma_{\rm DM}$ & Dark matter scattering cross-section.\\
  $\sigma_v=\sqrt{\langle v^2\rangle}$& Velocity dispersion of DM particles.\\
  $\rho_{\rm DM}$ & DM density. Reference value is taken to be $10^3M_{\odot}/{\rm pc}^3$.\\
  $m_{\rm DM}$  & Mass of DM particle.\\
  $\hat\alpha$	& Exponent of a DM density distribution toy model, $\rho\sim \rho_0(R/r)^{\hat \alpha}$.\\
  $a$ & Kerr rotation parameter: $a = J/M \in [0,M]$.\\
  $j$ & Dimensionless Kerr rotation parameter: $j \equiv a/M$.\\
  $Q$		& Black-hole charge\\
  $q=Q/M$		& Dimensionless black-hole charge\\
  $\omega$ & Fourier transform variable. The time dependence of any field is $\sim e^{-i\omega t}$.  \\
  $\Psi(\omega,r)=\int_0^{+\infty} e^{i\omega t}\psi(t,r)$& Laplace-transform of field $\psi(t,r)$.\\
  $\omega_R,\,\omega_I$ & Real and imaginary part of the QNM frequencies.\\
  $(\omega_R^{(0)},\omega_I^{(0)})$   & QNM frequencies of isolated BH spacetime with $\delta M=0$.\\
  $\delta_R=1-\omega_R/\omega_R^{(0)}$& Fractional variation in the (real part of) QNM frequency for finite $\delta M$.\\
  $\delta_I=1-\omega_I/\omega_I^{(0)}$& Fractional variation in the (imaginary part of) QNM frequency for finite $\delta M$.\\
  $\delta_{\rm iso}$& Fractional difference between axial and polar modes (isospectrality holds for $\delta M=0$).\\
  ${\cal Q}=\omega_R/(2|\omega_I|)$	& Quality factor\\
  $s$ & Spin of the field.\\
  $l$ & Integer angular number, related to the eigenvalue $l(l+1)$ of scalar spherical harmonics.\\ 
  $n$ & Overtone number. We conventionally start counting from a ``fundamental mode'' with $n=0$.\\
  $\varphi$ 	& GW phase.\\
  $\Psi$ 	& GW phase in frequency domain.\\
\end{tabular}
\end{table}
%


%
\bigskip
\clearpage
\newpage

\section{Introduction}
The past few decades have seen tremendous activity in gravitational-wave (GW) physics, at both the theoretical and the experimental level. 
Ground-based GW detectors have reached design sensitivity and are currently being upgraded
to sensitivities one order of magnitude higher~\cite{LIGO}, and the first exploratory missions for space-based detectors are upcoming~\cite{AmaroSeoane:2012je,Seoane:2013qna,Kawamura:2011zz}, whereas the prospects for GW detection with pulsar timing arrays have increased tremendously~\cite{Sesana:2013cta}. Altogether, these facilities will provide access to the GW Universe in the full range from nHz to tens of kHz.
Because the main sources of GWs are compact binaries of black holes (BHs) or neutron stars,
a huge theoretical effort has gone into the modeling of these systems. Slow-motion, post-Newtonian expansions and relativistic, extreme-mass ratio
perturbative approaches have been developed to exquisite levels, and matched against state-of-the-art numerical waveforms.

The gravitational waveform from the coalescence of compact objects carries a signature of the dynamics of those objects; accordingly, it can be roughly divided into an inspiral phase, where the two objects follow quasi-Newtonian trajectories, and a ringdown phase describing relaxation of the final merged body to a stationary state. The two phases are connected by a (short) nonlinear regime when spacetime is highly dynamical and relativistic.

Two corollaries of all the intense labor on compact binary coalescence have a direct implication on tests of the nature of the compact objects, and also of the underlying theory of gravity:

\noindent (i) When the final merged object is a BH, measurements of mass and angular momentum as well as tests of the no-hair theorem and of General Relativity (GR) in the strong-field regime can be made by studying the ringdown waveform~\cite{Berti:2005ys,Berti:2007zu,Berti:2009kk,Kamaretsos:2011um}. 
Because BHs in GR are characterized by mass and angular momentum only, their oscillation spectrum is fully described by these two parameters. Measurements of one single characteristic mode (i.e., vibration frequency and relaxation time scale) can be inverted to yield the mass and angular momentum of the BH; measurements of more than one mode can test the theory of gravity and/or the nature of the object~\cite{Berti:2005ys,Berti:2007zu,Berti:2009kk,Kamaretsos:2011um}.
Ref.~\cite{Berti:2005ys} considered the original design of the space-based detector LISA and GWs from supermassive BHs
at redshift $z$ and luminosity distance $D_L$. 
The accuracy with which the redshifted mass ${\cal M}_z=(1+z) M$ of the final BH ($M$ being its the physical mass) can be measured is given by [see Eq.~(4.15a) in Ref.~\cite{Berti:2005ys}]
\begin{equation} 
\label{accuracy_ringdown}
\frac{\Delta {\cal M}_z}{{\cal M}_z}
\sim 4.5\times 10^{-5}\frac{(1+\nu)^2}{\nu}\frac{D_L}{1 {\rm Gpc}}\left(\frac{10^6 M_{\odot}}{{\cal M}_z}\right)^{3/2}\,.
\end{equation}
We have used the realistic estimate that the fraction of total mass going into ringdown is of the order of $\epsilon_{\rm rd}\sim 0.44\nu^2/(1+\nu)^4$,
with $\nu=m_1 m_2/(m_1+m_2)^2$  the symmetric mass ratio of the progenitors [see Eq.~(4.17) in Ref.~\cite{Berti:2007fi}].

\noindent (ii) The inspiral waveform carries information about the two objects, the background geometry and the underlying theory of gravity, allowing in principle a measurement of all spacetime multipoles~\cite{Gair:2012nm,Yunes:2013dva}. 
In particular, again for the same space-based detector LISA, Ref.~\cite{Berti:2004bd} gives
\beq 
\label{accuracy_inspiral}
\frac{\Delta {\cal M}}{{\cal M}}
\sim(-1.1476+ 7.2356z + 5.7376z^2)\times 10^{-6}\,, \qquad z\in[1,10]
\eeq
for the accuracy with which the chirp mass ${\cal M}$  of a $10^6 M_{\odot} -10^6 M_{\odot}$ mass binary at redshift $z$ can be measured.

Both estimates \eqref{accuracy_ringdown} and \eqref{accuracy_inspiral} were derived for space-based detectors,
and should be revised in current designs. Nevertheless, they offer a glimpse of what GW astronomy can become: 
a {\it precision} discipline, able to map nearly the entire Universe in its compact object content to a precision better than $1\%$. 

However, most of the studies to date have considered isolated compact objects, neglecting the realistic astrophysical environment where these objects live.
We are thus faced with the following questions: how much of an impact does our ignorance of realistic astrophysical environments
(accretion disks, magnetic fields, electromagnetic charges, cosmological expansion, etc)
have on GW detection, compact binary parameter estimation, and how does it affect the ability of future GW detectors to test GR? Such questions are usually dismissed on the grounds that these effects are negligibly small for compact binaries.

The sole purpose of this work is to quantify such statements, with the view that \emph{quantitative} estimates are mandatory if one aspires to perform precision GW physics. This is a specially sensitive question as far as tests of GR are concerned, since these require a careful control on environmental effects, as simple explanations are often preferable. 
An interesting example concerns the planet Neptune, whose existence was inferred through unexpected changes in the orbit of Uranus (i.e., changes in Newton's theory were not required). A second historical event, specially amusing for relativists, relates to the first attempts at explaining Mercury's excessive periastron precession, which precisely invoked the existence of unseen disks of matter~\cite{Verrier:1859} (in that case, changes in the theory {\it were} indeed required). 

\subsection*{Attenuation, redshift, wave propagation and generation}
The environment in which compact objects live can influence GW emission by these sources in different ways:
It can affect the generation of the waves themselves (either by altering the motion of the sources via extra forces, accretion etc, or by modifying the local structure of spacetime
for instance due to the self-gravity of dirtiness), or it can affect GWs after they are generated as they
travel to the observer.  An exhaustive review of possible effects on wave propagation is given by Thorne~\cite{Thorne_notes}, the overall conclusion being that there is no known mechanism that can affect
the amplitude or phase of the wave to appreciable amplitude, with the possible exception of weak lensing. Because the latter has been discussed at length by several authors (see e.g. Refs.~\cite{Holz:2005df,Nissanke:2013fka} and references therein), we do not consider it here.

For the remainder of this work we focus on the first, almost unexplored effect, i.e, how astrophysical environments affect GW generation.

\subsection*{Purpose of this work and assumptions: how small is small?}

Any large signal-to-noise detection of a GW can in principle be used as a precision tool, containing information about the merging objects
but also about their environment: accretion disks, galactic halos, strong magnetic fields and all those effects that indirectly leave an imprint in the GW signal. These extrinsic factors will degrade the pure-GR waveform, and hamper our ability to test GR or to measure the merging objects' masses and spins with phenomenal accuracy. On the other hand, these effects can {\it in principle} give us important information about the surrounding medium and galaxy.

We initiate here a broad survey of the corrections due to environmental effects in two situations of great interest for GW astronomy: the ringdown emission of a single BH and the inspiral of two compact objects. We take into account various effects such as: electric charges, magnetic fields, cosmological evolution, possible deviations from GR, and various forms of matter such as accretion disks and dark matter (DM) halos. In many cases, a detailed analysis of these effects is necessarily model-dependent. When possible we have instead opted for a broader (yet less precise) discussion capturing the correct order of magnitude and the essentials of the phenomena under consideration. Our study can be extended in various ways, but we believe it might serve as a basis and guidance for more sophisticated analysis.

Our results are based on the following assumptions:
\begin{itemize}
\item[1] The central BH is spherically symmetric. 

\item[2] The spacetime geometry outside the horizon is modeled as a small perturbation away from the Schwarzschild geometry.

\item[3] For the ringdown studies in Part~\ref{part:ringdown}, the matter around the BH is also spherically symmetric. Therefore, 
the spacetime geometry is spherically symmetric.

\item[4] The inspiral studies in Part~\ref{part:inspiral} assume that the orbits are circular and evolve adiabatically. 

\item[5] When modeling the effect of accretion onto the central BH -- which causes the BH mass to change -- dynamical effects are neglected, i.e, the
central BH grows adiabatically.

\item[6] When we parametrize corrections from GR in Part~\ref{part:testsGR}, we assume a perturbative expansion in the coupling parameters of the modified theory, thus neglecting possible strong-field effects (the latter have to be investigated on a case-by-case basis).
\end{itemize}

This paper is organized as follows. In Sec.~\ref{sec:introdirty} we present the various environmental effects that we wish to quantify.
We found it useful to divide the rest of the paper in three parts, each one starting with an executive summary of our main findings. 
A reading of Sec.~\ref{sec:introdirty} should be sufficient for the busy reader to understand 
the three executive summaries.

In Part~\ref{part:ringdown} we discuss the ringdown emission of ``dirty BHs''. We define the latter to be BH geometries surrounded by matter of possible different nature, so that they are not exactly described by the 
Schwarzschild geometry~\cite{Visser:1992qh}. Unlike several other works on the topic (see e.g. Refs.~\cite{Visser:1992qh,Visser:1993qa,Medved:2003rga,Boonserm:2013dua}), our purpose is to characterize their GW signature in astrophysical contexts.
In Part~\ref{part:inspiral} we discuss environmental effects for a two-body inspiral. We estimate the effects of matter interactions for the satellite and for the central object, as well as the corrections to the gravitational waveforms due to various types of environmental effects.
In Part~\ref{part:testsGR} we discuss the consequences of environmental effects on possibles tests of General Relativity (and theories of gravity alternative to it) with GW astronomy.
Finally, Part~\ref{part:conclusions} contains our conclusions and a discussion of possible future work.

In this paper we will typically use geometrized  units $G=c=1$, but in some sections (especially when estimating the orders of magnitude of the various effects) 
we will restore $G$ and $c$ and use physical units for clarity.

\section{Compact objects and astrophysical environments}\label{sec:introdirty}
Compact objects are not isolated but rather surrounded by a multitude of matter fields in the form of gas, electromagnetic fields, fluids, etc. We refer to all of these configurations as ``dirtiness''. Dirtiness can potentially affect mass and spin measurements and other estimates of the observables associated with the source. In addition, dirtiness might hamper our ability to perform tests of Einstein's gravity, since deviations from GR (e.g. BH hair or modified GW emission) can be degenerate with -- or subdominant to -- other environmental effects.

\subsection{Composite-matter configurations: accretion-disk geometries}
The most common and best understood environmental effect in astrophysical BHs
is the gas in accretion disks. It is customary to normalize the bolometric luminosity of active galactic nuclei (AGNs), $L_{\rm bol}$, to the Eddington luminosity 
\be
L_{\rm Edd}=\frac{4\pi G M m_p c}{\sigma_T}\,,
\ee
where $m_p$ is the proton's mass and $\sigma_T$ is the Thomson cross section. The Eddington luminosity can
be derived by requiring equilibrium between the radiation pressure exerted on the matter surrounding the BH and the gravitational
pull of the BH itself; i.e. the Eddington limit sets an upper limit to the luminosity achievable through accretion onto a BH
and therefore on the mass accretion rate\footnote{Note however that the standard derivation of the Eddington luminosity assumes spherical symmetry, cf. e.g.~Ref.~\cite{Heinzeller:2006tg} for a discussion of the effects and limitations of this assumption.}.
One can then define the Eddington ratio $f^{\rm L}_{\rm Edd} =L_{\rm bol}/L_{\rm Edd}$ for the luminosity, which
ranges from $10^{-2}$ to 1 (with typical values of $f^{\rm L}_{\rm Edd}\sim 0.1$) for quasars and AGNs, and is typically much lower (down to $f^{\rm L}_{\rm Edd}\sim 10^{-9}$
for SgrA$^{*}$) for quiescent galactic nuclei (see e.g. Refs.~\cite{2009ApJ...699..626H,2009ApJ...696..891H,2006ApJ...648..128K,2009MNRAS.397..135K}).

The luminosity is then linked to the mass accretion rate by $\dot{M}=L_{\rm bol}/(\eta c^2)$, where
$\eta$ is the radiative efficiency of the accretion process. While $\eta$ depends on the nature of the accretion
process and on the spin of the BH, Soltan-type arguments~\cite{LyndenBell:1969yx,Soltan:1982vf} (i.e. a comparison between the luminosity function
of AGNs and the mass function of BHs) require an average value $\eta\approx 0.1$ (see e.g. Refs.~\cite{Shankar:2007zg,Li:2012gc,2012MNRAS.423.2533B}).
Taking therefore $\eta= 0.1$ as the fiducial radiative efficiency, one can define the Eddington mass
accretion rate as
\be
\dot{M}_{\rm Edd}
\approx
2.2\times 10^{-2} \left(\frac{M}{10^6 M_\odot}\right) M_\odot {\rm yr}^{-1}\,. \label{Medd}
\ee
One may then try to infer the mass accretion rate by assuming that the Eddington ratio for mass accretion, 
$f_{\rm Edd} =\dot{M}/\dot{M}_{\rm Edd}$, matches the Eddington ratio for luminosity, $f^{\rm L}_{\rm Edd}$.
Clearly, this would be the case only if $\eta$ was roughly constant and $\sim 0.1$ for all galactic nuclei. While that  
is probably a decent assumption for quasars and AGNs, as shown by the Soltan-type arguments mentioned above,
quiescent nuclei probably have considerably smaller radiative efficiencies. This follows mainly from observations 
of SgrA$^*$, for which direct mapping of the gas in its vicinity is possible and gives values of 
$f_{\rm Edd} \sim 10^{-4}$~\cite{1999ApJ...517L.101Q}, which yields $\eta\sim 10^{-5}$. Similar efficiencies
are also measured for virtually all nearby galactic nuclei~\cite{1988Natur.333..829F,1997ApJ...477..585M,2008ARA&A..46..475H}. 
(Clearly, quiescent nuclei are not constrained by Soltan-type argument because
they are too faint.)

Models that allow for such low radiative efficiencies are given by Advection Dominated Accretion Flows (ADAFs)~\cite{adafs},
where the (optically thin) ion-electron plasma accreting onto the BH has
too low a density and too large a temperature to efficiently transfer the accretion energy (which is transferred
from the gravitational field to the ions by viscosity) to the electrons (which are mainly responsible for the electromagnetic emission). 
As a result, the ion component is heated to the virial temperature, 
thus making the disk geometrically thick, and the accretion energy is ``advected'' 
into the BH.
ADAFs are believed to well describe galactic nuclei with $f^{\rm L}_{\rm Edd} \lesssim 0.01$.
In some variations of the model (such as Advection Dominated Inflow-Outflow Solutions~\cite{adios} -- ADIOS's  -- 
or Convection Dominated Accretion Flows~\cite{CDAF,2000ApJ...539..798N} -- CDAFs), 
however, only a small fraction of the plasma that gets close to the BH accretes onto it, generating
enough energy to drive an outflow or a convective process that returns most the plasma at large distances.
This is important to note because it means that in ADAFs the mass accretion rate is given by the observed ``large-scale'' 
rate $f_{\rm Edd} \dot{M}_{\rm Edd}$, but in ADIOS's and CDAFs it will be considerably lower than that, because most of the
plasma will not fall down the horizon. In what follows, however,
we will only need the accretion rate through a surface enclosing the BH but still far away from the horizon,
so we can assume that the actual accretion rate is
the same as the large-scale accretion rate without loss of generality. 

For larger Eddington ratios such as those encountered in AGNs and quasars,
accretion disks are optically thick. Geometrically, they can be described 
by thin disks ~\cite{shakura_sunyaev} for $10^{-2}\lesssim f^{\rm L}_{\rm Edd}\lesssim 0.2$,
while for $f_{\rm Edd}\gtrsim 0.2$ at least a fraction of the accretion energy is advected 
into the BH, as a result of which the disk ``puffs up'' and can be
described by a geometrically slim disk~\cite{slim_disks}, i.e. one with height comparable to the radius.

The gas present in the vicinity of binary systems (and in particular binary BHs) 
affects the orbital evolution and therefore the emission of GWs. The main effects 
can be classified as follows
\begin{itemize}
\item Accretion: the masses and spins of the compact objects change due to accretion onto them~\cite{Barausse:2007dy,Macedo:2013qea}
\item Purely gravitational effects (self-gravity): the matter in accretion disks or in other structures exerts a gravitational pull on the compact objects~\cite{Barausse:2006vt,Macedo:2013qea}
\item Dynamical friction and planetary migration: the gravitational interaction of the compact objects with their own wake in the matter medium~\cite{Barausse:2007dy,Barausse:2007ph,Yunes:2011ws,Kocsis:2011dr,Macedo:2013qea}
\end{itemize}
These effects impact the evolution of the orbital phase of the binary and thus the phase of the gravitational waveforms.
In Section~\ref{sec:matterinspiral} we will estimate the magnitude of these effects by using a simple Newtonian model and insights from astrophysical observations. 

\subsection{Electromagnetic fields and cosmological constant}
In addition to ``ordinary matter'' configurations around compact objects, three standard effects that might contribute to dirtiness are: the presence of an electromagnetic charge and of background magnetic fields and the effects due to the cosmic acceleration.

The observed accelerated expansion of our Universe is most simply described by
Einstein's equations augmented by a cosmological constant $\Lambda$~\cite{1989RvMP...61....1W,Polchinski:2006gy,Bousso:2012dk}.
Cosmological BH solutions are described by the Schwarzschild-de Sitter geometry,
\be
ds^2=-\left(-\Lambda\,r^2/3+1-2M/r\right)dt^2+\left(-\Lambda\,r^2/3+1-2M/r\right)^{-1}dr^2+r^2d\Omega^2\,,\label{eq:dS}
\ee
which satisfies the field equations $G_{\mu\nu}+\Lambda\,g_{\mu\nu}=0$. Observations place the magnitude of $\Lambda$ to be~\cite{1989RvMP...61....1W,Polchinski:2006gy,Bousso:2012dk}
\be
\Lambda \sim 10^{-52}{\rm m}^{-2}\,,
\ee
which we use as a reference value.

Magnetic fields pervade our Universe on cosmic, galactic and stellar scales, with different amplitudes.
At the surface of the Earth for instance, $B\sim 0.5 {\rm Gauss}$.
The magnetic field at the center of our galaxy has been estimated to be of order $B\sim 10^{-4} {\rm Gauss}$
at scales of order 100 pc, and $B\sim 10^{-2} {\rm Gauss}$ closer to the event horizon of the supermassive BH~\cite{2010Natur.463...65C,2006ApJ...636..798L} (and possibly as large as $10^2 {\rm Gauss}$ very near the horizon~\cite{Eatough:2013nva,Moscibrodzka:2009gw,Dexter:2010pe}).
The largest magnetic fields can be found in magnetars
and reach $B\sim 10^{15}{\rm Gauss}$~\cite{2013Natur.500..312T}.
The magnetic field of the supermassive BHs that power active galactic nuclei is believed to be $\lesssim 10^3-10^5$ 
Gauss~\cite{2005ChJAS...5..347Z,2009AA...507..171S,2013AstBu..68...14S}.
We will take $B\sim 10^{8}$ Gauss as reference value for a very large magnetic field.

Not much is known about the spacetime metric describing BHs immersed in magnetic fields. One of the few known exact solutions was discovered by Ernst and describes a nonrotating BH in a poloidal magnetic field~\cite{1976JMP....17...54E},
\beq
ds^2&=&K^2\left(-\left(1-\frac{2M}{r}\right)dt^2+\left(1-\frac{2M}{r}\right)^{-1}dr^2+r^2d\theta^2\right)+\frac{r^2\sin^2\theta}{K^2}d\phi^2\,,\\
K&\equiv& 1+\frac{B^2r^2\sin^2\theta}{4}\,.\label{eq:ernst}
\eeq
Note that this solution is {\it not} asymptotically flat and it is cylindrically symmetric. Nonetheless,
it allows for planar (equatorial) geodesic motion [see Eq.~\eqref{radialpotentialB} below] similarly to the spherically symmetric case and is simple enough to be studied analytically in various cases. 

Finally, concerning the electromagnetic charge, the standard lore is that astrophysical BHs are neutral to a very good approximation because
\begin{enumerate}

\item Quantum Schwinger pair-production effects discharge BHs rather quickly~\cite{Gibbons:1975kk,Hanni:1982}.
        In a back-of-the-envelope calculation, the Schwinger process is efficient whenever the (electric) potential energy across a Compton wavelength is larger than the particle rest mass, yielding rapid quantum discharge unless $q=Q/M\lesssim 10^{-5}M/M_{\odot}$ ($Q$ being the electric charge).
        
\item Even in situations where the Schwinger process is feeble, a vacuum breakdown mechanism can be efficient, whereby a spark gap triggers a cascade of electron-positron pairs~\cite{1969ApJ...157..869G,1975ApJ...196...51R,Blandford:1977ds}.
Blandford and Znajek estimate that this limits the charge of astrophysical BHs to  
$q\lesssim 10^{-13}\sqrt{M/M_{\odot}}$.

\item BHs are never in complete isolation. Intergalactic or accretion disk plasma is sufficient, via selective accretion, to neutralize quickly any charged BH. The reason is that electrons have a huge charge-to-mass ratio, $q_e\sim 10^{21}$. Thus, an accretion of a mass $\sim 10^{-21}M$ is sufficient to neutralize even an extremely charged BH.
     
\end{enumerate}
Nevertheless, there are various attempts at explaining BH-driven high-energy phenomena using charged BHs~\cite{1998ApJ...498..640P}.
These BHs may acquire charge through different mechanisms, all related, in one way or another, to the fact that BHs immersed in magnetic field build up a nonzero net charge. Classical charge induction by external magnetic fields was studied by Wald
\cite{Wald:1974np}, who showed that BHs acquire a charge which satisfies the bound
\be
q\lesssim 1.7\times 10^{-6} \frac{M}{10^6 M_{\odot}}\frac{B}{\rm 10^{8} Gauss}\,.
\ee
We recall that the most extreme magnetic fields in the Universe are found in magnetars and are of the order of $10^{13}$ Gauss, and up to $10^{15}$ Gauss~\cite{2013Natur.500..312T}.
Thus, the overall result of these studies is that astrophysical BHs are likely to be neutral to a high-degree.
We will take $q=10^{-3}$ as a fiducial, overly conservative reference charge-to-mass ratio of BHs.

\subsection{Dark Matter}
\label{sec:dm_intro}
DM is ubiquitous in
the Universe and, in particular, in the vicinity of massive BHs but, unlike baryonic matter, it is not expected to form accretion disks. 
This is because DM has a small interaction cross-section (i.e, behaves almost as collisionless) and it is therefore hard to conceive a way in which it could form a disk-like configuration around a BH. Nevertheless, the dynamical influence of the BH does affect the distribution of DM around it. 

Gondolo and Silk~\cite{GS} showed that the massive BH's adiabatic growth gives rise to a ``spike'' in the DM density, i.e. it 
causes the initial DM density profile (i.e. the one in the absence of the BH) to develop a steeper slope near the BH (assuming that the BH
is at the center of the DM halo). For example, in the case of the 
Navarro-Frenk-White profile~\cite{NFW}, which gives an initial DM density formally diverging as $1/r$ near the center,
the BH's presence would make the DM density diverge as $r^{-7/3}$~\cite{GS}. These DM spikes received much attention
as they would enhance the possible annihilation signal from DM near the Galactic center, but they were later shown
to be quite efficiently destroyed by BH mergers~\cite{DMcoresBin} and by scattering of DM by stars~\cite{CoreScatter1,CoreScatter2}
(although this latter mechanism is less efficient for more massive galaxies such as M87~\citep{silk2}, and 
the cusp may in any case still re-grow in a relaxation time~\cite{cusp_regrowth}). Also, 
even in the absence of stars and BH mergers, it was shown~\cite{ullio} that these spikes
do not form unless \textit{(i)} the original seed from which the massive BH grows has a mass of $\lesssim 1$ \% 
the BH's mass at $z=0$, \textit{and (ii)} this seed forms very close to the DM halo's center (to within 50 pc of it).
This second condition is necessary because seeds are brought to the center of the halo by dynamical friction,
but the time scale of this process exceeds the Hubble time unless the seeds are very close to the center.

Taking into account the interaction of massive BH binaries with the DM cusp, Ref.~\cite{DMcoresBin} finds that the DM density in the halo's center has a shallow slope, or even a core, rather than a spike. This is 
also plausible because stellar luminosity profiles in large ellipticals (which are gas-poor and where therefore stars behave as collisionless 
``particles'') are found to exhibit a shallow slope or even a core, which could be the result of the interaction of the stars with a massive BH binary~\cite{SpikesStars}. More precisely,  Ref.~\cite{DMcoresBin} finds that the DM density in the core created by the massive BH binary
is $\rho_{\rm DM} \sim 10^2 M_\odot/{\rm pc}^3$ for comparable-mass binaries, and $\rho_{\rm DM} \sim 10^3 M_\odot/{\rm pc}^3$ for binaries with mass ratio $1/10$.

The situation for
extreme mass-ratio inspirals (EMRIs, i.e. 
binaries consisting of a stellar-mass BH or a neutron star around
a massive BH) is slightly more complicated. As mentioned above, if the galaxy hosting the EMRIs has recently experienced a major (i.e. comparable-mass) galactic merger and thus a
comparable-mass BH merger, the DM spike is destroyed and a core is formed. At least 90 \% of Milky-Way type galaxies are expected to have 
undergone a major merger after $z=2$,\footnote{Actually, a stellar cusp and later a DM cusp may re-grow after a major merger~\cite{Merritt:2005yt}, but if the merger happens at $z\lesssim 2$ there is no time for such a regrowth to be completed, at least in the case of Milky Way type halos~\cite{Merritt:2005yt}.} based on semianalytical galaxy formation models~\cite{DMcoresBin}. 
However, even though most Milky-Way type galaxies are expected to have undergone a recent major merger, that
may not be the case for a large fraction of \textit{satellite} galaxies. In fact, Ref.~\cite{SatelliteHalos} showed that a Milky-Way type halo may
have a considerable number of BHs that have never undergone mergers and which are hosted in satellite galaxies. These satellite galaxies
may therefore have retained the original DM spike produced by the adiabatic BH growth. The precise number and mass of unmerged BHs in satellite galaxies depend on the mass of the BH seeds at high redshifts. Ref.~\cite{SatelliteHalos} finds that $\sim 1000$ such BHs may be present in each  Milky-Way type halo in a scenario in which the BHs grow from remnants of Population III stars~\cite{light_seeds} (i.e. from ``light'' seeds'' with mass $M_{\rm seed}\sim 10^2 M_{\odot}$ at $z\sim20$), while $\sim 100$ may be present if the BHs grow from ``heavy'' seeds with mass $M_{\rm seed}\sim 10^5 M_{\odot}$, originating at $z\sim 15$ from the collapse of primordial gas in early-forming halos~\cite{heavy_seeds1,heavy_seeds2,heavy_seeds3}. In either case, however, the mass of these BHs at $z\sim0$ is comparable
to $M_{\rm seed}$, so this population will give rise to EMRIs detectable with
future space-based detectors (such as eLISA) only if BHs evolve from heavy seeds.

Under the hypothesis that massive BHs do indeed evolve from heavy seeds, and thus that EMRIs detectable with eLISA may be present in these satellite galaxies, we will examine whether DM may have an effect on the gravitational waveforms. First of all, as mentioned above, the DM spikes
might not form at all unless the BH seed forms within 50 pc of the peak of the DM distribution~\cite{ullio}, which is in itself a very demanding assumption. Second, even if that is the case, the BH must grow its mass by at least a factor of $\sim 100$ relative to the seed's mass for the DM density to be enhanced relative to the initial DM profile~\cite{ullio}, 
e.g. if the BH grows its mass by a factor of $\sim 100$ ($1000$), the DM density around it may be 
$\sim 10^9 M_{\odot}/{\rm pc}^3$ ($10^{12} M_{\odot}/{\rm pc}^3$), while the formally divergent spike 
found in Ref.~\cite{GS} actually appears only in the limit in which \textit{all} of the BH mass is adiabatically added after the BH's formation.
Finally, if the BH accretes a large fraction of its mass after the seed is formed, it is also plausible 
that star formation may have happened in the satellite galaxies. The scattering of stars by the BHs will then reduce the slope of the DM density profile, and the DM density near the BH will be at most $\sim 10^{10}-10^{12} M_{\odot}/{\rm pc}^3$~\cite{CoreScatter1,CoreScatter2}.
These would still be very high densities that could give rise to DM annihilation. In fact, above a certain density threshold $\rho_{\rm DM, max}$
annihilation would happen on time scales shorter than the Hubble time, thus causing the local DM density to plateau at $\rho_{\rm DM, max}$.
This threshold value depends on the annihilation cross section $\sigma_{A}$ and on the mass of the DM particles $m_{\rm DM}$, i.e. $\rho_{\rm DM, max} \sim m_{\rm DM}/(\langle \sigma_A v\rangle t) $,
where $t$ is the Hubble time and $v$ is the velocity of DM particles. Choosing plausible values for these parameters, one finds that 
$\rho_{\rm DM, max}$ may be as large as $10^{11} M_{\odot}/{\rm pc}^3$.

Because of the considerations above, we will take as the reference value for the DM density
\begin{equation}
\rho_{\rm DM}=10^3M_{\odot}{\rm pc}^{-3}\sim 4\times 10^{4}{\rm GeV}/{\rm cm^3}\,,\label{rhoDM}
\end{equation}
but we will also entertain the possibility of higher DM densities $\sim10^{11} M_{\odot}/{\rm pc}^3$ in satellite galaxies.
%

\subsection{Beyond-GR effects}\label{introBeyondGR}
Unlike large-distance corrections --~where cosmological observations are in clear conflict with theoretical expectations~--
the motivation to modify GR in the strong-curvature regime is inspired by conceptual arguments. These arguments rely in one way or another on the conflict of the classical equations of GR with quantum mechanics and lead to the obvious conclusion that the ultimate theory of quantum gravity will differ from GR where curvature effects are important. In this paradigm, curvature effects become important close to singularities -- but these seem to be hidden from us by horizons, as all evidence indicates. The very existence of horizons also sets an upper limit on the energy scale involved, so that putative quantum effects at the Planck scale are negligible for astrophysical BHs.
Therefore, it appears that any problem that might potentially affect the strong-gravity dynamics of astrophysical objects can be understood \emph{within} GR rather than by extending it. 

Nonetheless, paradigms such as this have been challenged time and again in physics. 
Current experiments can probe gravitational potentials which are roughly six orders of magnitude weaker than those experienced near a massive BH and they can probe only curvatures which are thirteen orders of magnitude smaller than those experienced near and within compact objects~\cite{Psaltis:2008gka}. Extrapolating Einstein's theory to those unexplored regions introduces dangerous bias in our understanding of the strong gravitational interaction~\footnote{As a comparison, the gravitational potential on Earth's surface, where Newtonian gravity proved to be extremely successful, is only 4 orders of magnitude smaller than its corresponding value on the Sun surface where relativistic effects are relevant, as shown by the classical tests of GR. Likewise, even a very successful theory as quantum electrodynamics cannot be extrapolated from atomic to nuclear energy scales, the latter being 6 orders of magnitude larger and well described by strong interactions.}.

GW astronomy promises to test GR in the strong-curvature, highly-dynamical regimes which are completely unexplored to date (cf. e.g. Ref.~\cite{Yunes:2013dva}). While waiting for experiments to guide the theoretical efforts, the study of beyond-GR effects on the GW signal faces the problem that each theory is associated with different corrections and that a case-by-case analysis seems to be required. Deviations from GR (as well as matter effects) in spinning geometries were considered in the so-called ``bumpy-BH'' formalism~\cite{0264-9381-9-11-013,Collins:2004ex,Vigeland:2009pr,Vigeland:2011ji} and in other approaches~\cite{Glampedakis:2005cf,Johannsen:2011dh}, although none of them is free from limitations. A different promising approach is the so-called parametrized post-Einstein (ppE) expansion, which attempts to model modified gravitational waveforms directly, in a way that can potentially accommodate most of the corrections to the GR signal~\cite{Yunes:2009ke,Yunes:2013dva}. 

Here we use and extend some of the salient features of these available approaches. In order to parametrize generic corrections to the background metric, we adopt an approach similar to the bumpy-BH case: taking advantage of the assumption of spherical symmetry, we parametrize possible deviations from GR directly at the level of the metric by considering weak-field deformations around the most general static and spherically symmetric geometry. With respect to the bumpy-BH formalism, this approach has the merit to be generic (we do not assume Einstein's equations nor any particular symmetry other than the spherical one) and sufficiently simple to provide direct order-of-magnitude estimates. On the other hand, to parametrize corrections to the GW fluxes, in some sections of Part~\ref{part:testsGR} we use the ppE framework. 

An alternative theory of gravity would modify the GR signal essentially in three ways: \textit{i)} by altering the background solutions, i.e. by deforming the geometry of compact objects; \textit{ii)} by modifying the GW emission, for example introducing monopolar and dipolar radiation, changing the coupling with sources and possibly suppressing some radiation (as in the case of massive fields~\cite{Cardoso:2011xi,Alsing:2011er}); \textit{iii)} by modifying the physical properties of the GWs once they are emitted, e.g. the dispersion relation, the polarization and the way they interact with matter and with the detector. In this section we mostly focus on the ``conservative'' effects of \textit{i)}, which in many situations are the dominant correction. ``Dissipative'' effects related to \textit{ii)} and \textit{iii)} are discussed in Part~\ref{part:testsGR}.

We consider a general static, spherically symmetric spacetime:
\begin{equation}
ds^2=-A(r)dt^2+\frac{1}{B(r)}dr^2+r^2\left(d\theta^2+\sin^2\theta\,d\phi^2\right)\,, \label{ansatz}
\end{equation}
which we treat as a small deformation of the Schwarzschild geometry. Accordingly, we expand the metric coefficients as
\begin{eqnarray}
 A(r)&=&\left(1-\frac{r_+}{r}\right)\left[1+\sum_{i=1}^{N_\alpha} \alpha_i \frac{M^i}{r^i}\right]\,,\label{PRA1}\\
 B(r)&=&\left(1-\frac{r_+}{r}\right)\left[1+\sum_{i=1}^{N_\beta} \beta_i   \frac{M^i}{r^i}\right]^{-1}\,,\label{PRA2}
\end{eqnarray}
where $r_{+}$ is the horizon's radius, and we assume that $\alpha_i$ and $\beta_i$ are small dimensionless parameters. In writing the expansions above, 
we have required regularity of the metric (which implies $A(r_+)=B(r_+)=0$ at the horizon) and asymptotic flatness. 

The asymptotic behavior of the metric reads
\begin{eqnarray}
 A(r)&=&1-\frac{r_+-M\alpha_1}{r}+\frac{M^2\alpha_2-M r_+ \alpha_1}{r^2}+{\cal O}(1/r^3)\,, \label{Asympt}\\
 B(r)^{-1}&=&1+\frac{r_++M\beta_1}{r}+{\cal O}(1/r^2)\,.\label{Bsympt}
\end{eqnarray}
By comparing this with the parametrized post-Newtonian (PPN) expansion of the metric~\cite{Will:2005va},
\begin{eqnarray}
 A(r)&=&1-\frac{2M}{r}+2(\beta-\gamma)\frac{M^2}{r^2}+{\cal O}(1/r^3)\,,\\
 B(r)^{-1}&=&1+2\gamma\frac{M}{r}+{\cal O}(1/r^2)\,,
\end{eqnarray}
we can identify
\begin{equation}
 r_+=2M+M\alpha_1\,,\qquad \beta_1=2(\gamma-1)-\alpha_1\,,\qquad \alpha_2=2(\beta-\gamma)+\alpha_1(2+\alpha_1)\,.\label{PPNconditions0}
\end{equation}
The PPN parameters are very well constrained by observations~\cite{Will:2005va} and their measured value is close to unity, $\delta\gamma\equiv\gamma-1\sim10^{-5}$ and $\delta\beta\equiv\beta-1\sim 10^{-3}$.
In the following we consider $\delta\gamma,\delta\beta\ll1$ and work to first order in all these perturbative quantities.\footnote{Note however that the PPN constraints are derived assuming the central object is a star. Since we are mainly interested in BHs, we make the extra assumption that the asymptotic behaviors~\eqref{Asympt}--\eqref{Bsympt} are the same for a BH geometry and a star. This might not be the case in some modified gravity, for example in theories which allow for some Vainshtein-like mechanism~\cite{Babichev:2013usa}, but is the case, for instance, in scalar-tensor theories, \AE-theory and Ho\v rava gravity. Also, note that the assumption $\delta\gamma,\delta\beta\ll1$ is \textit{weaker} than requiring that Birkhoff theorem be satisfied in modified gravity.}
Equation~\eqref{PPNconditions0} can be written as
\begin{equation}
 r_+=2M+M\alpha_1\,,\qquad \beta_1=2\delta\gamma-\alpha_1\,,\qquad \alpha_2=2(\delta\beta-\delta\gamma)+\alpha_1(2+\alpha_1)\,.\label{PPNconditions}
\end{equation}

This model has $N_\alpha+N_\beta$ parameters, in addition to the mass $M$. The parameters $\delta\gamma$ and $\delta\beta$ are already strongly constrained, but we keep them undetermined to explore the potential of GW measurements to provide stronger constraints than those currently in place. The remaining parameters are currently unconstrained in a PN sense. It is worth stressing that $\alpha_1$ is also unconstrained. This parameter is related to a shift of the event horizon for a given mass (i.e. to a deformation of the BH area formula) and it was neglected in previous analyses (e.g. in Ref.~\cite{Johannsen:2011dh} and in subsequent studies, cf. Ref.~\cite{Joao} for a discussion) although it is the dominant term in a weak-field expansion.

Note that the parametrization~\eqref{PRA1}--\eqref{PRA2} is substantially different from a PN expansion, because it includes strong-gravity effects such as the presence of an event horizon.
The metric above is effectively a weak-field expansion around the Schwarzschild geometry and, as such, it is not unique. Furthermore,  a priori there is no guarantee that the expansion~\eqref{PRA1}--\eqref{PRA2} converges for small values of $N_\alpha$ and $N_\beta$ in the strong-field regime. In other words, there is no reason why any strong-field observable derived from such a deformed metric should depend more strongly on the lowest order coefficients.
We will show, however, that only the first coefficients of the $\alpha_i$ and $\beta_i$ series do indeed give relevant contributions
to GW observables, even at the strongest curvatures that can be probed with GW astronomy. Naively, this is due to the fact that geodesic motion has an upper cutoff in frequency given by the innermost stable circular orbit (ISCO) $r\sim 6M$, whereas the ringdown emission is governed by the light ring at $r\sim 3M$. Both the ISCO and the light ring radii are larger than the central mass $M$, so higher powers of $M/r$ are suppressed. As a result, Eqs.~\eqref{PRA1}--\eqref{PRA2} provide a very efficient parametrization, at least in the static case, where $M/r\lesssim 1/2$ outside the horizon.

\clearpage
\newpage

\part{Ringdown}\label{part:ringdown}
\section{Executive Summary}
%
\begin{table}
\caption{Upper limits on the corrections to the fundamental ringdown frequencies of a Schwarzschild BH in a variety of environments,
with respect to isolated BHs. We compute $\delta_{R,I}=1-\omega_{R,I}/\omega_{R,I}^{(0)}$, where $\omega_{R,I}$ is the real (imaginary)
part of the ringdown frequency of ``dirty'' BHs, whereas $\omega_{R,I}^{(0)}$ is the corresponding quantity for isolated BHs with the same total mass.
We include the effects of a cosmological constant $\Lambda$, electric charge $Mq$, background magnetic field $B$, accretion of gas,
as well as the impact of the DM galactic halo and that of other possible stationary matter configurations with mass $\delta M$ at a distance $r_0$, $6<r_0/M<40$ from the BH 
(see text for more details.) 
All of these results refer to the dominant quadrupole mode, but the order of magnitude is the same for other modes.
}
\begin{tabular}{c|ccc}
 \hline\hline
Correction           	 & $|\delta_R|/P [\%]$ &$|\delta_I|/P[\%]$ &$P$\\
\hline
cosmological constant  &$10^{-32}$           &$10^{-32}$         &$\frac{\Lambda}{10^{-52}{\rm m}^{-2}}\left(\frac{M}{10^6M_{\odot}}\right)^2$  \\
galactic DM halos	 	   &$10^{-21}$       &$10^{-21}$  &$\left(\frac{M}{10^6M_{\odot}}\right)^2\frac{\rho_{\rm DM}}{10^{3}{M_\odot}/{\rm pc}^3}$ \\
accretion              &$10^{-11}$           &$10^{-11}$        &$f_{\rm Edd}\frac{M}{10^{6}M_{\odot}}$ \\
charge 		             &$10^{-5} $           &$10^{-6}$        &$\left(q/10^{-3}\right)^2$  \\
magnetic field         &$10^{-8}$            &$10^{-7}$      &$\left(\frac{B}{10^{8}{\rm Gauss}}\right)^2\left(\frac{M}{10^6M_{\odot}}\right)^2$   \\
rings	 	                &$0.01$	             &$0.01$	         &$\frac{\delta M}{10^{-3}M}$\\	
matter bumpy	 	        &$0.02$	             &$0.04$	        &$\frac{\delta M}{10^{-3}M}$	\\
short hair	  	         &$0.05$	             &$0.03$          &$\frac{\delta M}{10^{-3}M}$ 	\\
thin shell		 	         &$20$	               &$200$	        &$\frac{\delta M}{10^{-3}M}$ \\
\hline\hline
\end{tabular}
\label{bstable}
\end{table}
We summarize here the most important results and conclusions of our study on the ringdown of BHs in realistic astrophysical environments.
For the readers wishing to skip the technicalities in the subsequent pages, we stress that these results were (mostly) obtained from particular models of environments surrounding the BH, but we believe they are conservative reference values. For the extraction of quasinormal frequencies we follow the numerical procedure documented elsewhere~\cite{Berti:2009kk,Pani:2013pma} (cf. also the notebooks and data files freely available at~\cite{rdweb,rdweb2}). We find that:
\begin{enumerate}

\item The system consisting of a BH $+$ surrounding matter has QNMs which can differ substantially from those of vacuum, isolated BHs and are typically localized farther away \footnote{The QNMs of isolated BHs are localized and excited at the light ring, and correspond to poles of the appropriate Green's function~\cite{Leaver:1986gd,Cardoso:2008bp,Berti:2009kk}.}.
The QNMs of isolated BHs seem not to show up in the spectrum of BHs surrounded by matter (i.e, they do not correspond to poles of the associated Green function).
Despite not showing up in the spectrum, the modes of the corresponding isolated BH play an important -- dominant most of the times -- role in time evolutions of the system \footnote{This is an intriguing result which would merit an independent study on its own.}. 

\item Accordingly, the lowest QNMs of {\it isolated} BHs {\it can} be used to estimate the redshifted mass of supermassive BHs to levels of $0.1\%$ or better, even when including our ignorance on the astrophysical environment. This statement is 
supported by Table~\ref{bstable}, which shows an upper limit to the corrections to the ringdown frequencies in a variety of astrophysical scenarios: BHs with nonvanishing electric charge $qM$, immersed in a magnetic field $B$, in a Universe with a cosmological constant $\Lambda$, or surrounded by ``matter'' of different distributions. These numbers are conservative, in the sense that we always take extreme situations as a reference value; actual astrophysical conditions are likely to be less extreme. 
Also note that the estimates shown in Table~\ref{bstable} consider localized matter
situated at distances $6<r_0/M<40$ from the BH. For matter localized farther away (or much closer to the horizon), the QNMs differ very strongly from those of isolated BHs, but the dominant energy emission in an inspiral and merger is carried by ringdown modes of {\it isolated BHs} (see Figs.~\ref{fig:scattering},~\ref{fig:infall},~\ref{fig:scattering_bumpyII}).

\item The QNM spectrum is extremely rich. For each mode of an isolated Schwarzschild BH we find an \emph{infinite} family of matter-driven modes plus
one other mode which is a parametric correction to the isolated BH QNM.
When the matter distribution is localized at large distance $\sim r_0$, we observe a constant drift of the QNMs as $r_0$ increases. In other words, the farther away the matter is from the BH, the more the QNM spectrum of the matter-BH system differs from those of isolated BHs. 
This surprising effect is due to the QNM exponential sensitivity $e^{i\omega r_*}$ to ``small corrections''.
We prove this effect using three different models: a one-dimensional quantum-mechanical toy model, a thin-shell distribution around a Schwarzschild BH, and a deformed Schwarzschild metric due to the presence of unspecified matter fields.

\item The QNMs of ``dirty'' BHs typically play a subdominant role in time evolutions, as they are localized farther away from the BH. 
During the merger of two BHs, these modes are excited to low amplitudes and at very late times. Accordingly, they play little role in the merger waveform of BHs, but they will likely dominate over Price's power-law tails~\cite{Price:1971fb,Ching:1995tj}.
On the other hand, these modes might in principle be excited to large amplitudes by inspiralling matter (e.g. by an EMRI), and give us important information on matter surrounding the BH. A recent example of excitation of similar modes by orbiting matter is given by floating orbits, i.e. the resonant excitation 
of modes of putative massive fields in the vicinities of supermassive BHs~\cite{Cardoso:2011xi}. Similar studies for accretion disks, magnetic fields or DM halos are lacking.

\item Our results are in agreement with some sparse results in the literature and also seem to solve one puzzling finding, as we describe below in Section~\ref{sec:previous}.

\end{enumerate}
In summary, the presence of matter changes the QNM spectrum drastically, but it does not affect the ability of GW observatories to detect BH ringdowns in realistic environments using templates of isolated BHs. The tabulated values of isolated BH QNMs in the literature~\cite{rdweb,rdweb2,Berti:2009kk} can also be used efficiently to estimate the mass and spin of realistic BHs. In optimistic scenarios, detections could even be used to investigate accretion disks and DM halos. As we noted, for realistic configurations the dominant ringdown phase will be identical to that of isolated BHs, but with an additional stage, which could dominate over Price's power-law tails (see for instance Figs.~\ref{fig:scattering},~\ref{fig:infall},~\ref{fig:scattering_bumpyII}). For large SNR events, such imprints could be detectable.
On the other hand, the effects listed above and discussed in the what follows are interesting \emph{per se}. Although beyond the scope of this work, a more rigorous understanding of these phenomena is highly desirable.

\subsection{Detectability \label{sec:detection}}
Before venturing in a detailed analysis, let us consider the accuracy in a ringdown detection required to observe deviations from the vacuum Schwarzschild case. The changes in the ringdown frequencies induced by environmental effects can be observationally scrutinized, but require very sensitive GW telescopes.
With a detection with signal-to-noise ratio SNR, each ringdown mode is determined with an accuracy in frequency of roughly
\be
\sigma_{\omega_R}/\omega_R\sim 1/{\rm SNR}\,,
\ee
where we used the fact that the quality factors of BHs is of order $3-20$ \cite{Berti:2005ys,Berti:2007zu,Berti:2009kk}.
A crude estimate of the minimum SNR {\it necessary} to perform ringdown null tests can be obtained
by requiring that the measurement error is at least smaller than any putative environmental effect (see Table~\ref{bstable}),
\be
1/{\rm SNR}<\delta_R\,.
\ee
where $\delta_{R,I}=1-\omega_{R,I}/\omega_{R,I}^{(0)}$, and $\omega_{R,I}$ is the real (imaginary)
part of the ringdown frequency of matter-surrounded BHs, whereas $\omega_{R,I}^{(0)}$ is the corresponding quantity for isolated BHs with the same total mass.
Thus, using the results of Table~\ref{bstable}, signal-to-noise ratios of at least a few $10^3$ are necessary to see the impact of environmental effects on ringdown modes.

\section{Ringdown of ``dirty'' black holes: fundamental fields and classical hairs}
Our definition classifies BHs as ``dirty'' if they are surrounded by {\it any} kind of matter.
Some types of matter, such as electromagnetic fields or a cosmological constant are more ``fundamental''
than others and can sometimes even be called classical hair instead of dirtiness. We decided therefore to start by analyzing BHs
surrounded by this type of matter, and by computing the corrections to the ringdown frequencies due to a putative small charge, background magnetic field and the cosmological constant. We will compare our results against those of an equal mass isolated BH.

In order to make our approach and motivation clear, consider the following mock-problem: assume that BHs all have a very small rotation. What error (in the ringdown frequency) do we incur in by assuming that they are Schwarzschild BHs instead? This can be answered very quickly with the fitting formulae in Refs.~\cite{Berti:2005ys,Berti:2009kk}. These authors provided fits for the dimensionless vibration frequency $M\omega_R$ and for the quality factor ${\cal Q}\equiv \omega_R/(2|\omega_I|)$,
as functions of the dimensionless spin parameter, $j\equiv J/M^2$ ($J$ being the BH's angular momentum):
\beq
M\omega_R&=&f_1+f_2(1-j)^{f_3}\,,\\
{\cal Q}&=&q_1+q_2(1-j)^{q_3}\,.
\eeq
For a Schwarzschild BH, we therefore get an uncertainty
\be
\delta_R=f_2f_3/(f_1+f_2)\, j\,,\qquad \delta {\cal Q}=q_2q_3/(q_1+q_2)\, j\,,
\ee
at small $j$.
For the dominant $l=m=2$ mode, $(f_1,f_2,f_3)=(1.5251,-1.1568,0.1292)$ and $(q_1,q_2,q_3)=(0.7000,1.4187,-0.4990)$.
We obtain
\be
\delta_R=-0.41 j\,,\qquad \delta {\cal Q}=-0.33 j\,,\qquad \delta_I=-0.07 j\,.
\ee
Thus, finally we would know that a small spin of $j\sim 10^{-3}$ would impact on mass measurements at a level of 1 part in $10^4$.
Notice also that highly spinning BHs are in general more sensitive to changes in parameters.
From here on, we consider only nonspinning BHs. 
\subsection{Cosmological constant\label{sec:cosmologicalconstant}}
The quasinormal modes of the Schwarzschild-dS BH given in Eq.~\eqref{eq:dS} can be computed via accurate continued fraction representations~\cite{Yoshida:2003zz}. It can be shown that both axial and polar gravitational perturbations give rise to the same spectrum. 
Using the approach of Ref.~\cite{Yoshida:2003zz}, we find 
\beq
&&\left(\delta_R,\delta_I\right)=M^2\Lambda\left(4.5,3.8\right)\,,\quad l=2\,,\label{deltaLambda}\\
&&\left(\delta_R,\delta_I\right)=M^2\Lambda\left(3.7,3.4\right)\,,\quad l=3\,.
\eeq
and we therefore get
\beq
&&\left(\delta_R,\delta_I\right)=\left(4.5,3.8\right)\times 10^{-34} \left(\frac{M}{10^6M_{\odot}}\right)^2\left(\frac{\Lambda}{10^{-52}{\rm m}^{-2}}\right)\,,\quad l=2\,,\\
&&\left(\delta_R,\delta_I\right)=\left(3.7,3.4\right)\times 10^{-34}\left(\frac{M}{10^6M_{\odot}}\right)^2\left(\frac{\Lambda}{10^{-52}{\rm m}^{-2}}\right)\,,\quad l=3\,,
\eeq
so that the corrections are completely negligible even for extremely massive BHs.
\subsection{Magnetic fields}
To study the effects of magnetic fields we consider the spacetime~\eqref{eq:ernst}, which is cylindrically symmetric and is not asymptotically flat. Separating the relevant gravitational perturbations in this spacetime seems like a formidable task. We will estimate the changes in ringdown frequencies by using the Klein-Gordon equation as a proxy. Scalar fields in this metric were studied in Refs.~\cite{Aliev:1988wy,Konoplya:2007yy}, which concluded
that the ringdown is equivalent to that of a scalar field with mass $2Bm$ in the background of a Schwarzschild BH, where $m$ is an azimuthal number (perturbations have a dependence $\sim e^{im\phi}$)\footnote{This equivalence is somewhat hand-waving, but we will take it for granted. It is clearly desirable to have a more robust analysis of perturbations in this background.}.
We find the following behavior 
\be
\left(\delta_R,\delta_I\right)=\left(3.9,-7.0\right)\times 10^{-10}\left(\frac{B}{10^{8}{\rm Gauss}}\right)^2\left(\frac{M}{10^6M_{\odot}}\right)^2\,,\quad l=m=2\,.\label{eq:magnetic}
\ee
While this prediction is most certainly correct up to factors of order $2$ or so for the gravitational case, it would be interesting to devise ways to solve the full set of the perturbed Einstein equations in the background \eqref{eq:ernst}.

Eq.~\eqref{eq:magnetic} was derived using a very specific background, but we now show that the estimate is robust.
Consider a BH immersed in an approximately constant magnetic field spacetime, with energy density (in geometrized units) $\rho_B=B^2/(8\pi)$. As we show below in Section~\ref{sec:halos} this gives rise to an approximately Schwarzschild de Sitter solution with cosmological constant $\Lambda_{\rm eff}=24\pi\rho_B$. We can thus use the result of the previous Section~\ref{sec:cosmologicalconstant} to get 
\be
\left(\delta_R,\delta_I\right)=\left(2.4,2.0\right)\times 10^{-10}\left(\frac{B}{10^{8}{\rm Gauss}}\right)^2\left(\frac{M}{10^6M_{\odot}}\right)^2\,,\quad l=2\,,\label{eq:magnetic2}
\ee
in agreement, to within a factor two, with the estimate from an exact spacetime solution. We will take the more conservative estimate \eqref{eq:magnetic} as reference.

\subsection{Charge}
Perturbations of charged black holes couple the gravitational and electromagnetic fields together,
and excite electromagnetic degrees of freedom, such as vectorial-type QNMs~\cite{MTB,Pani:2013ija}.
While this is already an interesting result {\it per se}, i.e. that arbitrarily small charges will give rise
to an entirely new set of modes in the spectrum, we focus here on the purely  gravitational modes, i.e. those which
join smoothly to a neutral BH in the zero charge limit.
Working in the small-charge limit, we can use the fitting formulae provided in Ref.~\cite{Pani:2013ija} (where the charged, slowly-rotating case was considered; here we simply set the rotation to zero):
\begin{equation}
M\omega_{R,I}= f_0 + f_1 y + f_2 y^2 +f_3 y^3+f_4 y^4\,,
\end{equation}
where $y=1-\sqrt{1-q^2}$. Using the values listed in Table~I of Ref.~\cite{Pani:2013ija} for the $l=2$ fundamental mode, we obtain
\be
\delta_R=-7\times 10^{-8} \left(\frac{q}{10^{-3}}\right)^2\,,\qquad \delta_I=-3\times 10^{-8} \left(\frac{q}{10^{-3}}\right)^2\,.
\ee
Note that, because of the symmetries of the Einstein-Maxwell equations when $q\to-q$, the corrections to the uncharged case are proportional to $q^2$. Therefore, even in the overly optimistic scenario, $q\sim 10^{-3}$, the correction to the ringdown frequencies are less than 1 part in $10^7$. Note also that both gravitational sectors have the same spectra~\cite{Berti:2009kk,Pani:2013ija}. As we mentioned, a different set of electromagnetic-type modes are also excited when charged BHs are perturbed~\cite{Zilhao:2012gp}, and these may give rise to a new observable set of ringdown frequencies.

\section{Ringdown of nonisolated black holes: composite matter configurations and dark matter\label{sec:dirty}}
In this section we initiate the study of the ringdown emission by matter-BH systems, i.e. BHs surrounded by some sort of composite matter. The structure of the QNMs of BHs surrounded by matter is extremely rich. We found it convenient to start by analyzing a simple toy model which presents most of the features we shall encounter in more realistic configurations.

\subsection{A toy model for matter bumps: two rectangular barriers\label{sec:toy}}
%
\begin{figure}[thb]
\begin{center}
\begin{tabular}{cc}
\epsfig{file=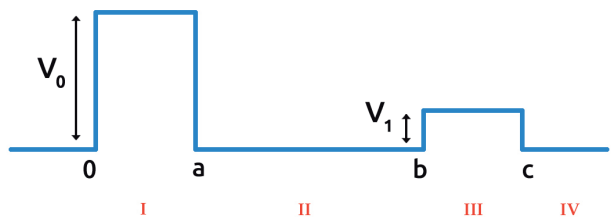,width=8.4cm,angle=0,clip=true}
\end{tabular}
\caption{Toy model of two rectangular barriers of height $V_0$ and $V_1\ll V_0$. For $V_1=0$ one recovers the original toy model
considered by Chandrasekhar and Detweiler~\cite{Chandrasekhar:1975zza}.
\label{fig:double_barrier}}
\end{center}
\end{figure}
We consider an extension of Chandrasekhar and Detweiler toy model~\cite{Chandrasekhar:1975zza}, which has the advantage that no differential equations need to be numerically solved, therefore reducing the likelihood
of numerical error influencing the results.

In this problem, one considers a one-dimensional potential consisting of two rectangular barriers of height $V_0$
and $V_1\ll V_0$, as in Fig.\ref{fig:double_barrier}. The role of $V_0$ is to mimic the potential barrier at $r\sim 3M$ in the Schwarzschild geometry, whereas $V_1$ would describe a small amount of matter some distance away.

If we define the Fourier transform of the field $\psi(t,x)$ as $\Psi(\omega,x)=\int_0^{+\infty}dt\,\psi(t,x)e^{i\omega t}$, then the
wave equation $\Box\psi=0$ is brought into the form
\be
\Psi''+k^2\Psi=0\,,
\ee
where $k=\sqrt{\omega^2-V(r)}$ and primes stand for derivatives with respect to the $x-$coordinate. In this form, it is clear that $x$ plays the role of the tortoise coordinate
in curved backgrounds. 

\subsubsection{Mode structure}
Following Chandrasekhar and Detweiler~\cite{Chandrasekhar:1975zza}, it is natural to define the QNMs of this system as the solutions which are purely left-moving at $x=-\infty$ and purely right-moving at $x=+\infty$. 
Since the potential is piecewise constant, the left-moving solution reads 
\be
\Psi_L=\left\{\begin{array}{l}
	       e^{-i\omega x} \qquad x\leq0 \\
               A_{\rm in}^{I}e^{-i k_I x}+A_{\rm out}^{I}e^{ik_I x} \qquad 0\leq x\leq\, a \\
               A_{\rm in}^{II}e^{-i \omega x}+A_{\rm out}^{II}e^{i\omega x} \qquad a\leq x\leq\, b \\
               A_{\rm in}^{III}e^{-i k_{III} x}+A_{\rm out}^{III}e^{ik_{III} x} \qquad b\leq x\leq\, c \\
               A_{\rm in}^{IV}e^{-i \omega x}+A_{\rm out}^{IV}e^{i\omega x} \qquad c\leq x\,, 
              \end{array}\right.\,,\label{PsiL}
\ee
where $k_{I}=\sqrt{\omega^2-V_0}$ and $k_{III}=\sqrt{\omega^2-V_1}$.
Notice that the coefficients $(A_{\rm in}^{X},A_{\rm out}^{X})$ are obtained via algebraic relations, by requiring continuity of 
$\Psi$ and $\Psi'$ at $x=(0,a,b,c)$. The requirement that the solution be purely right-moving for $c\leq x$, is equivalent to
\be
A_{\rm in}^{IV}(\omega)=0\,,
\ee
and is an (algebraic) eigenvalue equation for the frequencies $\omega$.

\begin{table}[hbt]
\scriptsize
\centering \caption{QNM spectrum of the double barrier potential with $V_0=16/a^2$, $V_1=10^{-3}/a^2$, $b=10a$, $c=11a$. The fundamental mode in vacuum ($V_1\equiv0$) reads $\omega a=4.66-0.710i$.} 
\vskip 12pt
\begin{tabular}{@{}cccccccccccccccccccccc@{}}
\hline \hline
&\multicolumn{21}{c}{$n$}\\ \hline
               &0    &1    &2    &3    & 4   &5    &6    &7    &8    &9    &10   &11   &12   &13   &14   &15   &16   &17   &18   &19   &20\\
\hline \hline
$a\,\omega_R$ &0.186&0.529&0.858&1.182&1.503&1.821&2.136&2.447&2.751&3.045&3.334&3.629&3.928&4.224&4.502&4.748&5.001&5.272&5.553&5.836&6.115\\
$-a\,\omega_I$  &0.340&0.368&0.392&0.411&0.430&0.449&0.469&0.491&0.512&0.529&0.533&0.527&0.518&0.508&0.495&0.503&0.542&0.573&0.593&0.610&0.625 \\
\hline \hline
\end{tabular}
\label{tab:spectrum}
\end{table}
\begin{figure}[thb]
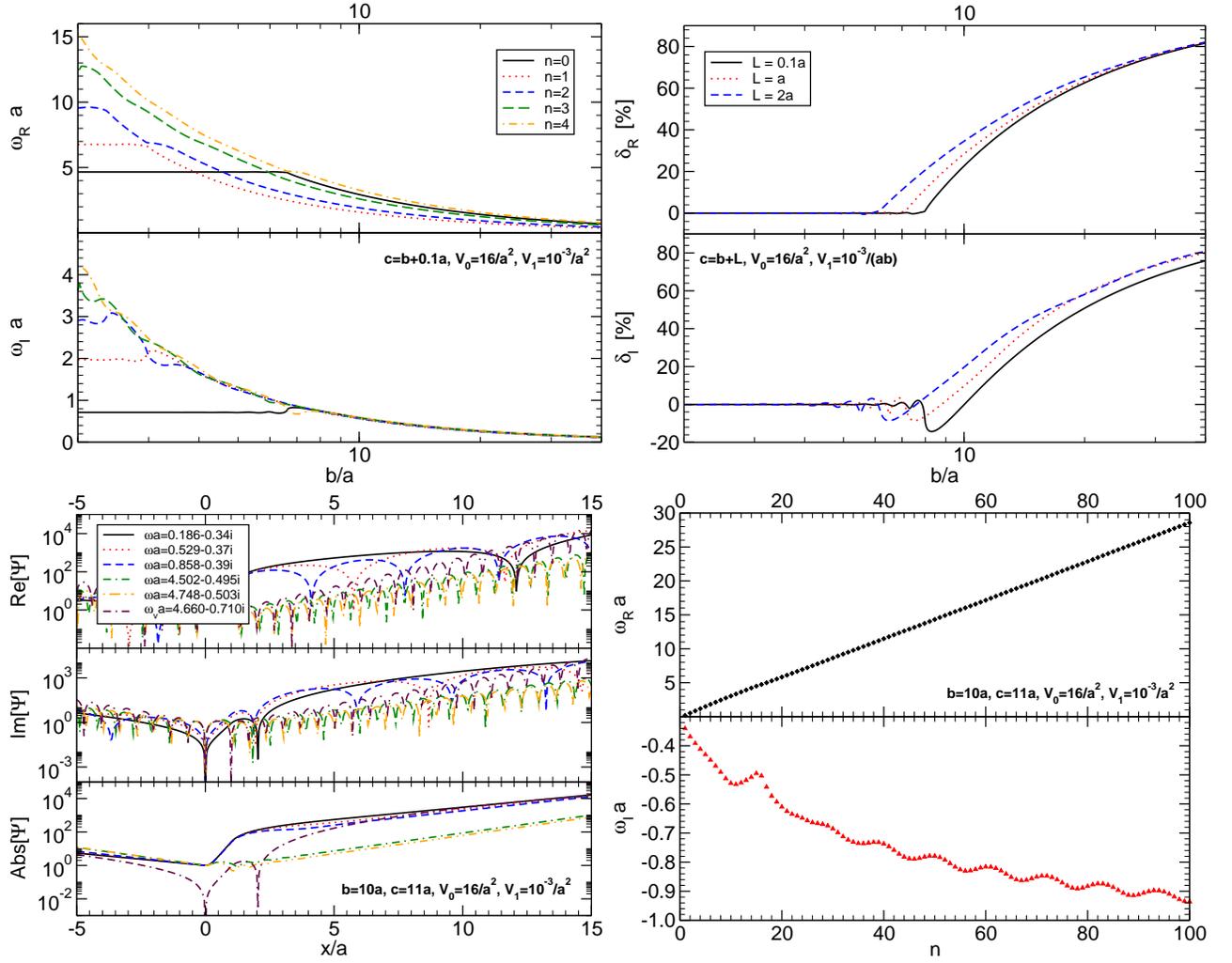

\begin{center}
\begin{tabular}{cc}
\epsfig{file=toy_model_modes.eps,width=8.4cm,angle=0,clip=true}&
\epsfig{file=toy_model_modes_L.eps,width=8.4cm,angle=0,clip=true}\\
\epsfig{file=eingenfunctions_double_barr.eps,width=8.4cm,angle=0,clip=true}&
\epsfig{file=toy_model_spectrum.eps,width=8.4cm,angle=0,clip=true}
\end{tabular}
\caption{
QNMs of the double-barrier system.
Top left: Some modes of the toy model of two rectangular barriers of height $V_0$ and $V_1\ll V_0$ as functions of the second barrier position, $b$. For definiteness, we focus on $V_0=16/a^2$, $V_1=10^{-3}/a^2$ and $c-b=0.1a$. Top right: the same but for the fundamental mode only as a function of $b$ with $V_1=10^{-3}/(ab)$ and for different values of $c-b\equiv L$. Qualitatively similar results hold for different choices of the parameters. 
Bottom left: Eigenfunctions of the double barrier potential. We show the first three QNMs, the fundamental mode in the vacuum case ($\omega_v a=0.466-0.710i$) and two intermediate modes whose real part is close to that of $\omega_v$. Bottom right: QNM spectrum (cf. also Table~\ref{tab:spectrum}). In both panels we set $V_0=16/a^2$, $V_1=10^{-3}/a^2$, $b=10a$, $c=11a$, but different choices of the parameters would give similar results.
\label{fig:toy_model_modes}}
\end{center}
\end{figure}
Even in this simple toy model, the QNM spectrum is extremely rich, as shown in Fig.~\ref{fig:toy_model_modes}
and Table~\ref{tab:spectrum}.
We fix $a=1,V_0=16$ for ease of comparison with Chandrasekhar and Detweiler's results. The single barrier case (i.e., $V_1=0$)
yields a fundamental frequency $\omega a=4.660-0.710i$.

When the second barrier is turned on, $V_1\neq0$, the structure of the spectrum changes drastically. 
To understand the spectrum we first take the five lowest-lying modes (i.e, those with smaller imaginary component) 
of the single-barrier and track them when $V_1$ grows from $0$ to $10^{-3}/a^2$. We show in the top left panel of Fig.~\ref{fig:toy_model_modes} 
how these modes change when we now let $b$ increase.  The top right panel shows the variation of the single-barrier fundamental mode as the width of the barrier increases.
We see that all these modes become longer-lived when $b$ increases.
In particular, for any $V_1\neq0$, there exist modes with $\omega_R<\sqrt{V_0}$, which do not exist in the single-barrier case. 

In the large $b$ limit, we observe some surprising features:
\begin{itemize}
 \item As shown in the top right panel of Fig.~\ref{fig:toy_model_modes}, the deviations from the vacuum case \emph{increase} as $b$ increases. That is, the more distant the small bump of matter is from the BH, the more it affects the QNM spectrum.
 \item Those modes that are the lowest-lying ones in the vacuum ($V_1=0$) case, are no longer such when $V_1\neq0$. Indeed, new families of modes appear, some of which become the new fundamental and lowest-lying modes. The 20 lowest-lying modes are given in Table~\ref{tab:spectrum} for $b=10, c=b+1$.
 \item For any $V_1\neq0$, a family of modes appears whose spacing in the real part is roughly constant.
These modes and respective eigenfunctions are plotted in the bottom panels of Fig.~\ref{fig:toy_model_modes}.
\end{itemize}
These peculiar features of the QNMs can be understood analytically in two limiting cases.

\noindent {\bf Perturbative limit.} In the limit, $V_1/ V_0\equiv \epsilon\ll1$, the problem can be solved analytically. Expanding $\omega=\omega^{(0)}+\epsilon \omega^{(1)}$, we obtain
\be
 \omega^{(1)}=\frac{ \left(e^{2 i c \omega^{(0)}}-e^{2 i b \omega^{(0)}}\right) k_I^3}{\omega^{(0)} \left(4 k_I-\omega^{(0)}  \log\left[\frac{V_0-2 \omega^{(0)} \left(\omega^{(0)}+k_I\right)}{V_0-2 \omega^{(0)} \left(\omega^{(0)}-k_I\right)}\right]\right)}\left(\frac{V_0-2 \omega^{(0)} \left(\omega^{(0)}+k_I\right)}{V_0-2 \omega^{(0)} \left(\omega^{(0)}-k_I\right)}\right)^{-\frac{\omega^{(0)}}{k_I}}\,,
\ee
where $\omega^{(0)}$ solves the zeroth order problem~\cite{Chandrasekhar:1975zza}. In the large $b$ limit, the deviation scales as $-e^{2i b \omega^{(0)}}$. Because $\text{Im}[\omega^{(0)}]\equiv\omega^{(0)}_I<0$, the linear corrections \emph{diverges} exponentially, $\omega^{(1)}\propto e^{2 \omega^{(0)}_I b}$. However, in this limit the perturbative procedure does not converge and one has to resort to the exact numerical solutions above. The latter displays a milder (power-law) drifting from the vacuum modes in the large $b$ limit.

\noindent {\bf Double Dirac Delta system.} In order to get some further insight on the problem, let us consider an even simpler model, in which $V=V_0\delta(x-a)+V_1\delta(x-b)$. By integrating the Schroedinger equation across the shell, one gets the junction condition $\Psi'_+-\Psi'_-=V_i \Psi_i$, where $i=a,b$ for the two Dirac deltas, respectively. Without loss of generality, we choose the coordinates such that $a=0$. It is straightforward to derive the QNM condition:
\be
e^{2 i b \omega } V_0 V_1+(i V_0+2 \omega ) (i V_1+2 \omega )=0\,.\label{eqFdelta}
\ee
When $V_1=0$, the only solution of the equation above is $\omega=-i V_0/2$. However, for any $V_1\neq0$, there are \emph{infinite} QNMs whose structure is remarkably rich. When $V_1/V_0=\epsilon$, a perturbative solution is
\be
\omega=-\frac{V_0}{2}i(1+\epsilon e^{b V_0})\,.
\ee
Again, this solution is consistent only when $\epsilon  e^{b V_0}\ll1$. An analytical solution exists also in the opposite (large $b$) limit. First, we separate real and imaginary parts and write Eq.~\eqref{eqFdelta} in the form
\beq
 -(V_0+2 \omega_I) (V_1+2 \omega_I)+4 \omega_R^2+e^{-2 b \omega_I} V_0 V_1 \cos(2 b \omega_R)&=&0\,,\\
 2 (V_0+V_1+4 \omega_I) \omega_R+e^{-2 b \omega_I} V_0 V_1 \sin(2 b \omega_R)&=&0\,.
\eeq
In the limit where $\omega_R\gg V_i$, $|\omega_I|\ll V_i$ and $b|\omega_I|\gg1$, the equations above admit the following solution:
\be
 \omega_R=\frac{n\pi}{2b}\,,\qquad e^{-2 b \omega_I}=(-1)^{n+1}\frac{n^2 \pi ^2}{b^2 V_0 V_1}\,,
\ee
which is consistent only when $n\gg1$, $n$ being an odd integer. Note that there exist infinite QNMs even if the corresponding vacuum problem admits one single mode only. Furthermore, the real part is separated by a roughly constant spacing, whereas the imaginary part decreases as $b$ increases. This is consistent with our previous findings and, indeed, this simple double Dirac delta model displays most of the salient features of the more realistic configurations that we discuss in the following. These generic features are also in agreement with what was found by Leung et al.~\cite{Leung:1999rh,Leung:1999iq} in the case of scalar perturbations of a Schwarzschild BH enclosed by a thin-shell.

Perhaps the single most important aspect of this example is that at sufficiently large distances $b$, the fundamental, isolated-barrier mode
at $\omega=4.660-0.710i$ is no longer the fundamental mode nor does it play any obviously outstanding role in the spectrum.
Nevertheless, physical intuition would guarantee that at sufficiently large separations $b$ this isolated-barrier mode must be excited
for at least a finite duration. In fact, we will now see that this is the case.
\subsubsection{Mode excitation: scattering}
To discuss mode excitation, we set up initial conditions for the scalar field in this one-dimensional problem such that
\be
\frac{\partial \psi}{\partial t}(t=0,x)=e^{-(x-x_0)^2/\sigma^2}\,,\qquad \psi(t=0,x)=0\,.
\ee
Laplace-transforming the Klein-Gordon equation we get
\be
\Psi''+k^2\Psi=I(x)\equiv -\dot{\psi}(t=0,x)\,.
\ee
We require the solution to be right- and left-directed at $x=\pm \infty$ respectively, i.e. $\Psi\sim e^{\pm i kx}$.
The general solution can be found with the help of Green's functions and reads
\be
\Psi(\omega,x)=\Psi_R\int_{-\infty}^{x}\frac{I(x') \Psi_L(x')}{2i\omega A_{\rm in}^{IV}}dx'+\Psi_L\int_{x}^{\infty}\frac{I(x') \Psi_R(x')}{2i\omega A_{\rm in}^{IV}}dx'
\ee
where $\Psi_L$ was defined in Eq.~\eqref{PsiL} and $\Psi_R$ is a homogeneous solution which behaves as
$\Psi_R= e^{i\omega x}$ and $x\geq c$.      
Thus, if the observer is located close to $+\infty$, he sees a waveform
\be
\Psi(\omega,x)=e^{i\omega x}\int_{-\infty}^{+\infty}\frac{I(x') \Psi_L(x')}{2i\omega A_{\rm in}^{IV}}dx'\,.
\ee
For the initial data we consider, this is simply
\be
\Psi(\omega,x)=\frac{e^{i\omega x}}{2i\omega A_{\rm in}^{IV}}\sqrt{\pi}\,\sigma\,e^{-ik x_0-k^2\sigma^2/4}\left(A^{X}_{\rm in}+A^{X}_{\rm out}e^{2ik_X x_0}\right)\,,
\ee
with $X=I,II,III,IV$ depending on where the initial data are localized (we focus on initial data that does not overlap between different regions).
Because all coefficients $(A_{\rm in}^{X},A_{\rm out}^{X})$ are obtained via algebraic relations, no integrations are necessary to find $\Psi(\omega,x)$. The time-domain solution at $x>c$ is then given by 
\be
\psi(t,x)=\int_{-\infty}^{+\infty} \frac{d\omega}{2\pi} \frac{e^{i\omega (x-t)}}{2i\omega A_{\rm in}^{IV}}\sqrt{\pi}\,\sigma\,e^{-ik x_0-k^2\sigma^2/4}\left(A^{X}_{\rm in}+A^{X}_{\rm out}e^{2ik_X x_0}\right)\,,
\ee
\begin{figure}[thb]
\begin{center}
\begin{tabular}{cc}
\epsfig{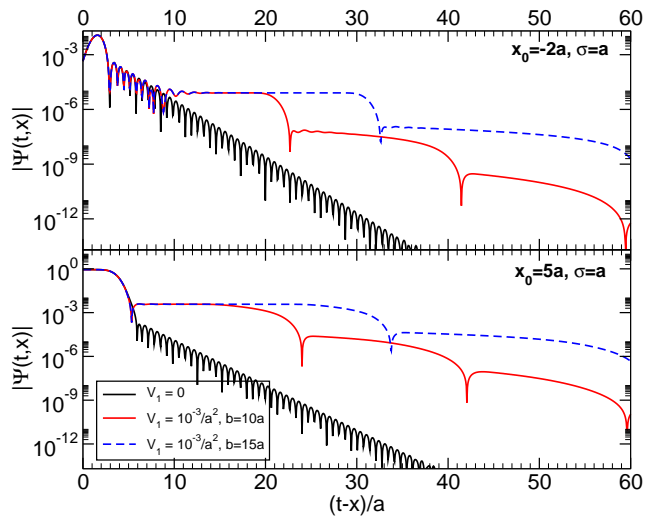}
\end{tabular}
\caption{Left: Waveforms for a gaussian packet in the double barrier potential. Top panel: the initial packet is located on the left of the first barrier, $x_0=-2a$ and $\sigma=a$. Bottom panel: the initial packet is located between the two barriers, $x_0=5a$ and $\sigma=a$. For both panels: $V_0=16/a^2$ and $c=b+a$. Other choices give qualitatively similar results.
\label{fig:scattering}}
\end{center}
\end{figure}
We have considered various configurations which all give qualitatively similar results. For concreteness, we present results for barriers with $V_0=16,V_1=10^{-3},a=1,c=b+1$ and two values of $b$, $b=10,15$.
The time-evolution of the scalar field is depicted in Fig.~\ref{fig:scattering}, for a Gaussian of unit width centered at $x_0=-2$
(top panel) and at $x_0=5$ (bottom panel). Thus, the initial data are localized to the left of both barriers in the top panel and in between the barriers in the bottom panel.
The black solid line refers to $V_1=0$, and a very clear ringdown waveform is seen after an initial transient. We find that the ringdown is very well described by the fundamental mode of the isolated barrier, $\omega a=4.660-0.710i$.

The red (solid) and blue (dashed) lines show the response in the double-barrier case. Let us focus on the top panel. The initial transient is the same as the isolated barrier, and for a finite time it also displays the same ringdown, with the same parameters as the \emph{isolated} barrier. Eventually, the system ``thermalizes'' to the double-barrier vibrations and we see ringdown which corresponds to the fundamental mode of the double-barrier case, in particular the fundamental mode of Table~\ref{tab:spectrum}. Thus, even though the isolated-barrier mode does not play any seemingly special role
in the QNM spectrum of the composite system, it does play a major role in the transient evolution.

The bottom panel of Fig.~\ref{fig:scattering} shows that, by tuning the initial perturbations and the model parameters, one can circumvent the isolated-barrier ringdown in the double-barrier system, by exciting to higher levels
the double-barrier modes. This will happen if the initial data ring directly $V_1$ to a large amplitude, as is the case when we let the gaussian
evolve from in-between both barriers. We believe that such an analogous situation is hard to come by in the astrophysical setups.
\subsubsection{Lessons from the double-barrier toy model}
The double-barrier model is instructive in many respects. First, it shows that the QNM spectrum of matter-BH systems can be drastically different from that of their vacuum counterparts. New families of modes exist and they might contain lowest-lying modes. The fundamental mode of the vacuum case does not seem to play any special role in the spectrum. However, a simple scattering experiment shows an interesting ``memory effect'', in which the fundamental mode of the isolated BH appears at intermediate time. This opens up the interesting prospect of detecting \emph{both} the modes of vacuum BH and those associated with the surrounding matter from the ringdown waveform of BHs in astrophysical environments.

Perhaps one of the most relevant lessons from this toy model is the existence of two very different regimes: one in which the matter fields are localized near the BH and the other where matter is many Schwarzschild radii away. In the latter case, two very separated scales exist so that, to some extent, one would still expect the modes of the isolated BH to play a role in the waveform (even though they no longer belong to the QNM spectrum). On the other hand, in the former case there is no separation of scales and one would expect a genuine deviation from the quasinormal ringdown modes of the vacuum case. In the next sections we will indeed discuss various models by dividing the matter distributions into two classes: those that are localized near the BH (i.e. at some $r\lesssim 10M$) and those localized farther away ($r\gtrsim 10M$).

\subsection{Spherical thin shells}
As far as we are aware the first, and only, attempt to understand the impact of surrounding matter on the QNMs of
astrophysical BHs to date was that of Leung et al~\cite{Leung:1999rh,Leung:1999iq}. These authors focused on spherical, infinitely-thin shells
of matter around nonrotating BHs and studied in some detail the Klein-Gordon equation in this spacetime.
We now substantially generalize their results, by considering gravitational perturbations as well as scalar fields.
We thus consider a thin-shell distribution with metric
\be
ds^2=\left\{\begin{array}{l}
             -\bar\alpha\left(1-\frac{2M}{r}\right)dt^2+\left(1-\frac{2M}{r}\right)^{-1}dr^2+r^2\left(d\theta^2+\sin^2\theta\,d\phi^2\right)\,,\qquad r<r_0  \\
             -\left(1-\frac{2M_0}{r}\right)dt^2+\left(1-\frac{2M_0}{r}\right)^{-1}dr^2+r^2\left(d\theta^2+\sin^2\theta\,d\phi^2\right)\,,\qquad r>r_0
            \end{array}\right.\,, \label{metricshell}
\ee
where $\bar\alpha=\frac{1-2M_0/r_0}{1-2M/r_0}$ and $\delta M=M_0-M$. Here $M$ is the BH horizon mass and $M_0$ the total spacetime ADM mass.
%

\subsubsection{Scalar fields}
If we write the metric above as Eq.~\eqref{ansatz} \footnote{We use the inverse definition for $B(r)$ as compared to Leung et al~\cite{Leung:1999iq}.}
and use the decomposition $\Phi(t,r,\theta,\phi)=\sum_{lm}\frac{\Psi(r)}{r}e^{-i\omega\,t}Y_{lm}(\theta,\phi)$, the Klein-Gordon equation $\Box\Phi=0$ is reduced to
\be
\sqrt{AB}\left(\sqrt{AB}\Psi'\right)'+\left(\omega^2-A\left(\frac{l(l+1)}{r^2}+\frac{(AB)'}{2Ar}\right)\right)\Psi=0\,.
\ee
Integrating across the shell location $r=r_0$ we find the jump condition
\be
\sqrt{AB}_+\Psi'(r_0+\epsilon)-\sqrt{AB}_-\Psi'(r_0-\epsilon)=\frac{\Psi(r_0)}{r_0}\left(\sqrt{AB}_+-\sqrt{AB}_-\right)\,.
\ee
Here primes stand for derivatives with respect to the radial coordinate $r$ and the subscripts stand for quantities evaluated to the right (+)
and left (-) of the shell. Using Eq.~\eqref{metricshell}, we finally find
\be
\left(1-\frac{2M_0}{r}\right)\Psi'(r_0+\epsilon)-\sqrt{{\bar \alpha}}\left(1-\frac{2M}{r}\right)\Psi'(r_0-\epsilon)=-\frac{2(M_0-M)}{r_0^2(1+1/\sqrt{{\bar \alpha}})}\Psi(r_0)\equiv \kappa \Psi(r_0)\,.
\ee
This junction condition is equivalent to expression (4.24) in Leung et al.~\cite{Leung:1999iq} and can be used to match the scalar perturbation inside the shell to those in the exterior. The eigenvalue problem is then entirely specified. One can compute the QNMs of the system by, e.g., robust continued-fraction representations or also by direct integration, as discussed below. Since the results are qualitatively similar to those obtained in the more realistic case of gravitational perturbations, we focus only on a detailed discussion of the latter.
\subsubsection{Gravitational perturbations}
Gravitational perturbations of a Schwarzschild BH surrounded by a thin shell can be studied by using the thin-shell formalism developed in Ref.~\cite{Pani:2009ss}. The details of the procedure are given in Appendix~\ref{app:shell}. In brief, the strategy is to solve the Regge-Wheeler-Zerilli equations in vacuum and connect the interior and the exterior solutions through the Israel conditions. For axial variables, these matching conditions, in the Regge-Wheeler gauge, read~\cite{Pani:2009ss}
\begin{equation}
\LL h_0\RR=0\,,\qquad \LL\sqrt{B} h_1\RR=0\,,
\label{junctionaxial}
\end{equation}
where $\LL\dots\RR$ denotes the ``jump'' of a given quantity across the shell, i.e. $[[A]] \equiv {A(r_0^+)-A(r_0^-)}$. The treatment of polar perturbations is more involved and it yields the
following relations for the jump of the polar metric functions across the
shell~\cite{Pani:2009ss}:
\begin{equation}
[[K]]=0\,,\qquad [[K']]=-\frac{1}{2A(1+2v_s^2)}\left(\frac{2M}{r_0^2}\LL
H\RR-[[H\,A']]-2A(r_0)[[H']]+4i\omega[[H_1]]\right)\,,
\label{junctionpolar}
\end{equation}
The parameter $v_s$ is related to the EoS of the thin shell, $\Theta=\Theta(\Sigma)$, through
\be 
v_s^2\equiv-\left(
\frac{\partial\Theta}{\partial\Sigma}
\right)_{\Sigma=0}\,,\label{eq:v}
\ee
and it has the dimensions of a velocity. Here, $(\Sigma,\,\Theta)$ are the shell's
surface energy density and surface tension respectively. A microscopic model of matter on the
thin shell is needed for a microphysical interpretation of $v_s$, but (roughly
speaking) this parameter is related to the sound speed in the shell.  

As explained in Appendix~\ref{app:shell}, the matching conditions above, together with the linearized Einstein equations, are sufficient to solve the dynamical equations for the axial and polar variables. After physically motivated boundary conditions at the horizon and at infinity are imposed, the problem is reduced to a simple one-dimensional eigenvalue problem for the complex frequency $\omega=\omega_R+i\omega_I$. We have computed the eigenfrequencies with a direct integration. We have carried out an extensive analysis of the QNMs of the system as a function of $\delta M$, $r_0$, $v_s$ and $l$. The parameter space is very rich and a summary of our main results is presented below.

\begin{figure*}[thb]
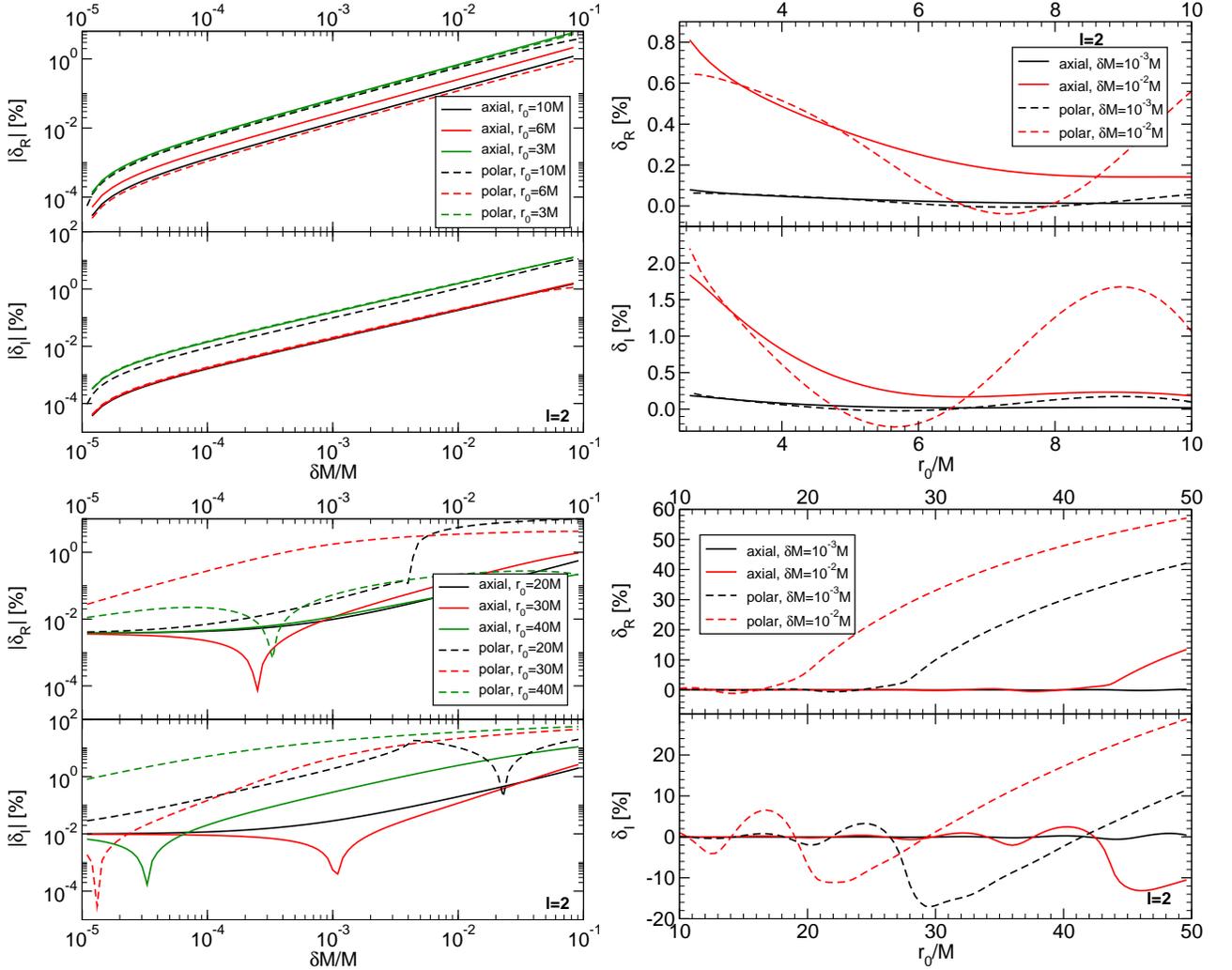

\begin{center}
\begin{tabular}{cc}
\epsfig{file=modes_shell_l2_near.eps,width=8.4cm,angle=0,clip=true}&
\epsfig{file=modes_shell_radius_l2_near.eps,width=8.4cm,angle=0,clip=true}
\\
\epsfig{file=modes_shell_l2_far.eps,width=8.4cm,angle=0,clip=true}&
\epsfig{file=modes_shell_radius_l2_far.eps,width=8.4cm,angle=0,clip=true}
\end{tabular}
\caption{Left panels: percentage deviations of the real and imaginary parts of the QNMs of a Schwarzschild BH surrounded by a thin-shell with respect to the case of an isolated BH (with the same horizon mass) as a function of the shell mass $\delta M$ and for different values of the shell radius, $r_0$ and $l=2$. Note that isospectrality is mildly broken and that polar modes refer to $v_s=0$ (the dependence on $v_s$ is shown in Fig.~\ref{fig:modes_EOS}). Right: The same as a function of $r_0$ for different values of $\delta M$. Top panels refer to to the case $r_0\lesssim10 M$, whereas bottom panels refers to $r_0\gtrsim 10M$.
\label{fig:modes}}
\end{center}
\end{figure*}
%
Figure~\ref{fig:modes} shows a summary of the fundamental $l=2$ modes in the case of $r_0<10M$ (top panels) and when $r_0>10M$ (bottom panels). Here and in the following, we present the results by keeping the BH mass $M$ fixed, i.e. the total mass of the spacetime, $M+\delta M$, changes linearly with $\delta M$.
In the left panels we show the percentage deviations of $\omega_R$ and $\omega_I$ as functions of $\delta M$ for fixed values of the shell radius. In the case $r_0<10M$, energy conditions impose $\Sigma>|\Theta|$ and this translates into a lower limit on $r_0$ for given $\delta M$. When $\delta M\ll M$, the limit reads $r_0\gtrsim 2.5M$ and we shall limit our results to this region. In the right panels we present the dependence of the modes on $r_0$ for fixed values of $\delta M$.

It is clear from Fig.~\ref{fig:modes} that the behavior at small $r_0$ is qualitatively different from that at large $r_0$, in agreement with the toy model discussed above. While the behavior for $r_0\lesssim10M$ is linear in $\delta M/M$,  
the dependence when $r_0\gtrsim 10M$ is more involved. We also observe the same qualitative behavior of the modes at large $r_0$ as previously discussed.

\begin{figure}[thb]
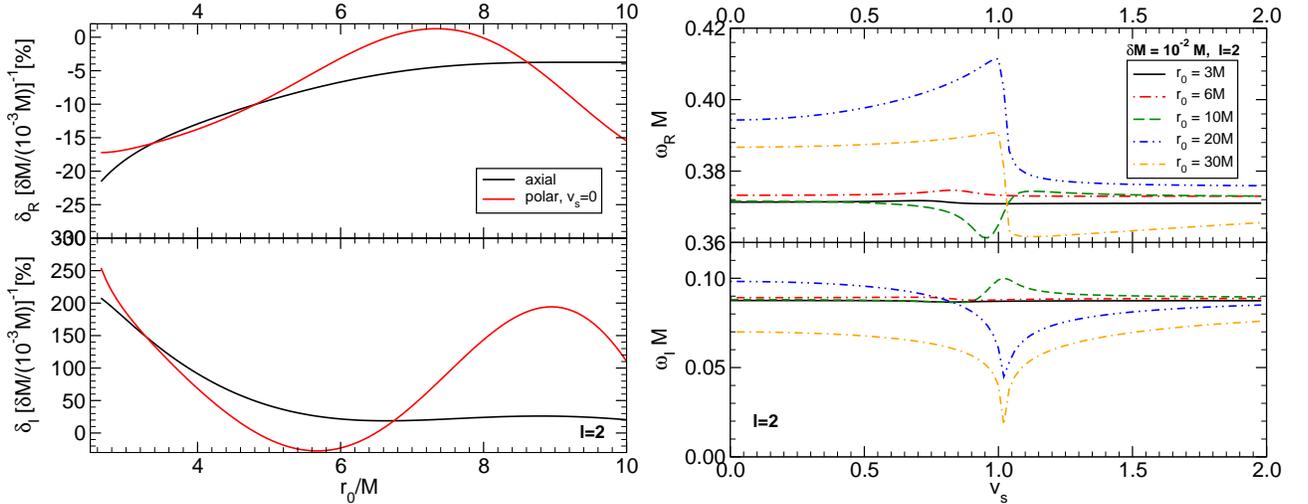

\begin{center}
\begin{tabular}{cc}
\epsfig{file=coeff_VS_rS.eps,width=8.4cm,angle=0,clip=true}&
\epsfig{file=modes_shell_l2_EOS.eps,width=8.4cm,angle=0,clip=true}
\end{tabular}
\caption{Left panels: linear coefficients of $\delta_R$ and $\delta_I$ in the small $\delta M/M$ limit as functions of the shell radius $r_0$ for $l=2$ and for the real and imaginary parts. In the polar case we have considered $v_s=0$. Right panels: real and imaginary parts of the QN fundamental frequency for $l=2$ polar modes as a function of $v_s$ for $\delta M=10^{-2}M$ and different values of $r_0$. 
\label{fig:modes_EOS}}
\end{center}
\end{figure}

Let us focus on the most interesting region $r_0\lesssim10M$ and $\delta M/M\ll1$. In this case, the corrections to the vacuum modes are linear in $\delta M/M$. The linear coefficient is shown in the left panel of Fig.~\eqref{fig:modes_EOS}.   Finally, in the right panel of Fig.~\eqref{fig:modes_EOS} we show the dependence of the fundamental $l=2$ polar mode with the EoS parameter $v_s$ for fixed values of $\delta M$ and $r_0$. Note that the dependence is very mild when $v_s\lesssim0.3$, which is the most interesting region from a phenomenological point of view. For completeness, we also show the region when the speed of sound at the shell is superluminal, as might be the case for exotic forms of matter. Interestingly, the peculiar behavior of the relativistic regime, $v_s\sim1$, emerges naturally from our method and it is more pronounced when $r_0\gtrsim 10 M$.

%

\subsubsection{Isospectrality breaking}
\begin{figure}[thb]
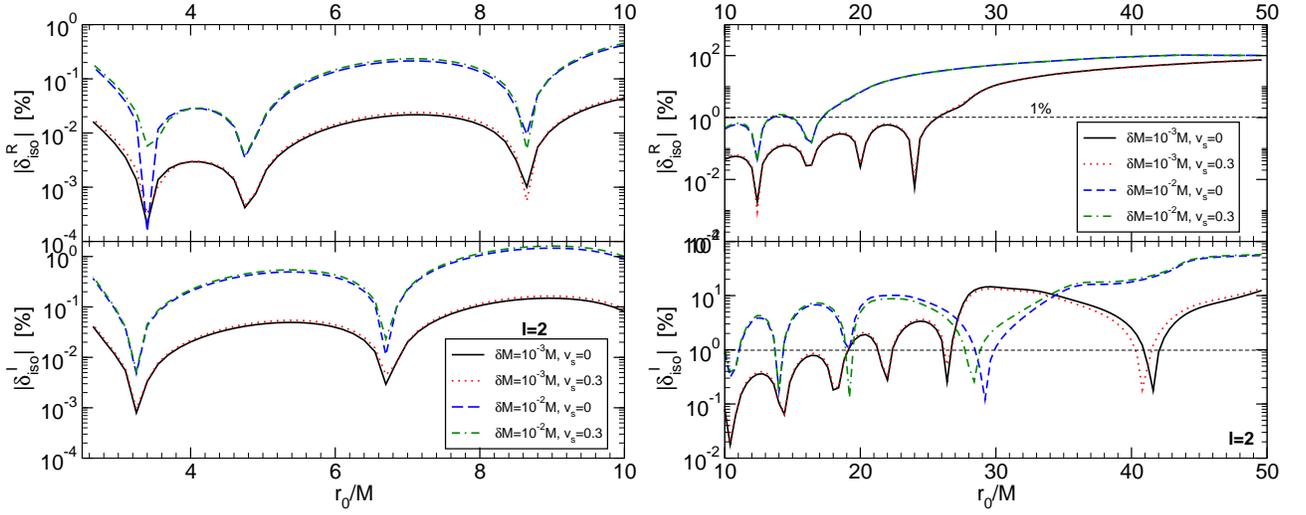

\begin{center}
\begin{tabular}{cc}
\epsfig{file=l2_iso_near.eps,width=8.4cm,angle=0,clip=true}&
\epsfig{file=l2_iso_far.eps,width=8.4cm,angle=0,clip=true}
\end{tabular}
\caption{Fractional difference between axial and polar modes (normalized by the polar modes) of a thin-shell model as a function of the shell location, $r_0$, for different values of $\delta M$ and $v_s$.
\label{fig:modes_iso}}
\end{center}
\end{figure}

A remarkable property of a Schwarzschild BH in vacuum is the isospectrality of the axial and polar sector. Although this classification does not have a particular meaning for spinning BHs, these overlapping modes stay isospectral also in the case of Kerr BHs. The presence of matter surrounding the BH breaks this degeneracy. Furthermore, polar modes also depend on the matter EoS through the parameter $v_s$. In Fig.~\ref{fig:modes_iso} we show the fractional difference between axial and polar modes for the thin-shell model for two representative values of $v_s$.

This isospectrality breaking is a robust, general prediction of our model. Indeed, because isospectrality is a very fragile property, it is generically broken if the object is not a BH \emph{in isolation}, or also if the underlying theory of gravity is not GR. Thus, a model-independent signature of deviations from isolated BHs in GR is the presence of two lowest-lying modes which are very close to each other. Different types of deviations (matter or modified gravity) predict different amounts of isospectrality breaking, and the possibility of resolving the two modes from the gravitational waveform is an interesting open problem. 

\subsection{Matter-bumpy black holes}
A thin-shell distribution is unrealistic and has the disadvantage of introducing discontinuities in the metric. In this section we focus on two different models where the metric is $C^1$ everywhere. We take the ansatz~\eqref{ansatz} with $A(r)=B(r)=1-2m(r)/r$ and two models (``I'' and ``II'') for $m(r)$:
\beq
m_I(r)&=&\left\{\begin{array}{l}
            M \qquad \hspace{5cm} r<r_0\\
            M+3 \delta M \left(\frac{r-r_0}{L}\right)^2-2 \delta M \left(\frac{r-r_0}{L}\right)^3 \quad r_0<r<r_0+L\\
            M_0 \qquad \hspace{4.8cm} r>r_0+L
           \end{array}\right.\,.   \label{bumpyI} \\
m_{II}(r)&=&M+\frac{\delta M}{2}\left(1 + {\rm erf}[\left(r +2M\log[r-2M]-r_0\right)/L]\right)           \,.\label{bumpyII}
\eeq
Here $\delta M=M_0-M$, and erf is the error function. These models are specified by 3 parameters: $\delta M$, $L$ and $r_0$. When $r_0>2M$, this metric describes a Schwarzschild BH where some matter field localized at $[r_0,r_0+L]$ has been superimposed. Note that this is a consistent solution in the Newtonian limit, i.e. $\delta M\ll M$ and $r_0\gg M$. 

\begin{figure}[thb]
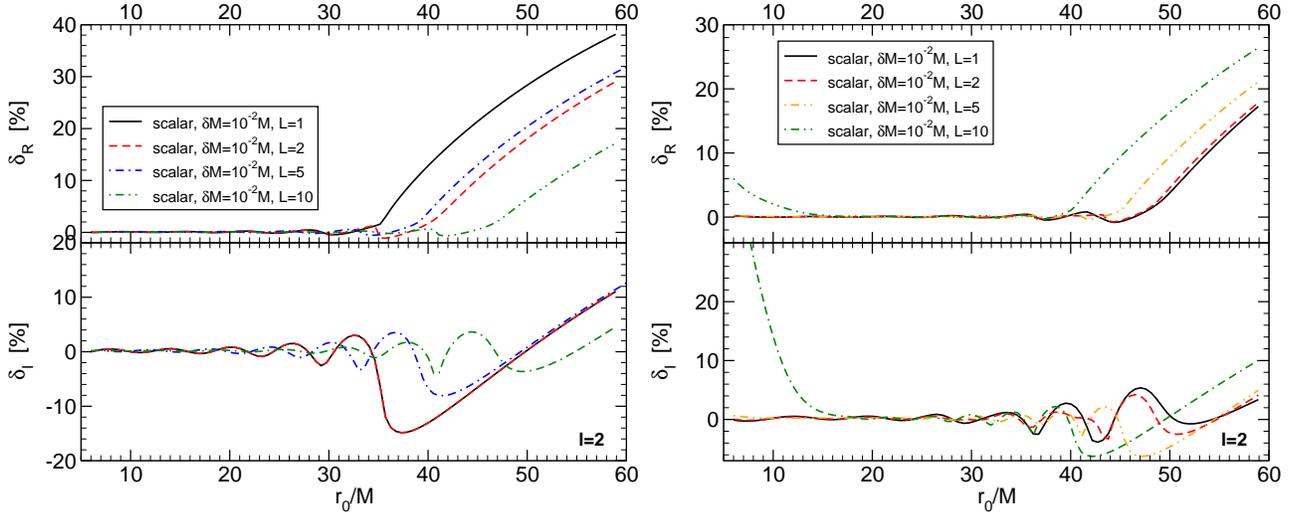

\begin{center}
\begin{tabular}{cc}
\epsfig{file=modes_bumpy.eps,width=8.4cm,angle=0,clip=true}&
\epsfig{file=modes_bumpyII.eps,width=8.4cm,angle=0,clip=true}
\end{tabular}
\caption{Percentage deviations of the real and imaginary parts of the fundamental $l=2$ mode of a matter-bumpy BH as a function of $r_0$ for different values of $\delta M$ and $L$. Left and right panels refer to the models I and II in Eqs.~\eqref{bumpyI} and \eqref{bumpyII}, respectively.
\label{fig:modes_bumpy}}
\end{center}
\end{figure}

Computing the gravitational perturbations of this spacetime would require an explicit stress-energy tensor for this matter distribution. In addition, the metric perturbations and the fluid perturbations would be coupled, rendering the analysis unnecessarily involved. To circumvent these technicalities, we consider a probe scalar field on this background. The Schroedinger-like potential reads
\begin{equation}
V=\left(1-\frac{2m(r)}{r}\right)\left[\frac{l(l+1)}{r^2}+\frac{2m(r)}{r^3}-\frac{2m'(r)}{r^2}\right]\,, \label{potscalar}
\end{equation}
and is continuous everywhere by virtue of the smoothness of the metric. By a direct integration, we have computed the scalar QNMs of the system, which depend on $l$, $\delta M$, $r_0$ and $L$. It is interesting to track the modes as functions of $r_0$ at large distance and for fixed $\delta M$. This is shown in Fig.~\ref{fig:modes_bumpy} for the two profiles $m_I(r)$ and $m_{II}(r)$.

We observe the same qualitative features as the toy model and the thin-shell model. Namely, when tracked as a function of $r_0$, the fundamental mode oscillates around the vacuum mode for $r_0\lesssim30M$ (the precise number depends on the model parameters), whereas large deviations from the vacuum case occur at large distances. 
%

\begin{figure}[thb]
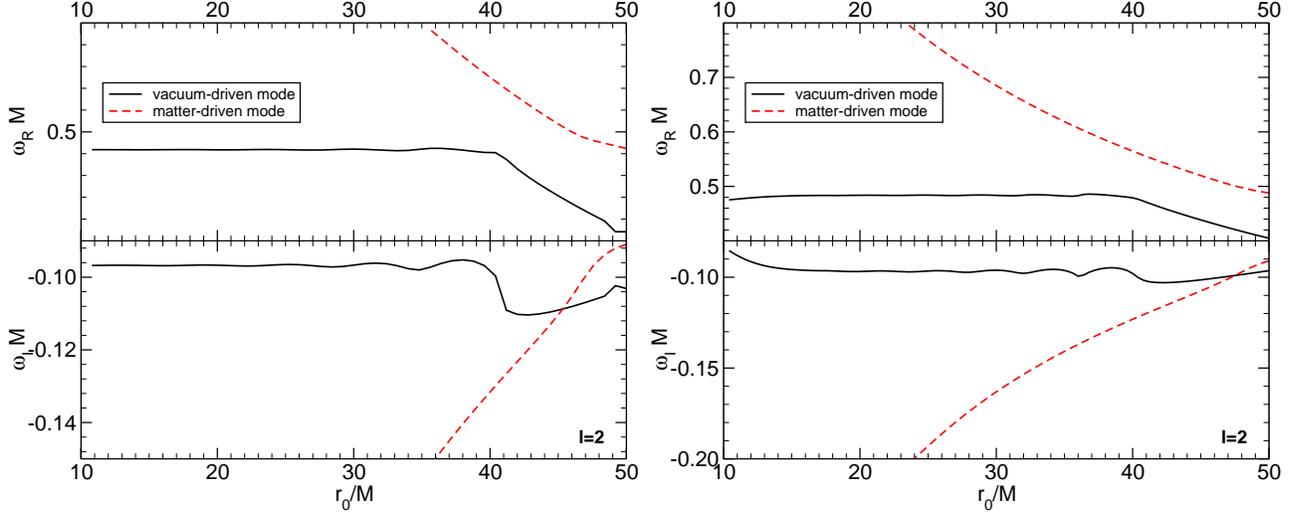

\begin{center}
\begin{tabular}{cc}
\epsfig{file=2modes_bumpyI.eps,width=8.4cm,angle=0,clip=true}&
\epsfig{file=2modes_bumpyII.eps,width=8.4cm,angle=0,clip=true}
\end{tabular}
\caption{Example of two modes of the bumpy BH configuration I (left panels) and II (right panels) for $\ell=2$, $\delta M=0.01M$, $L=10M$ as a function of $r_0$ in the large $r_0$ limit. The black curve represents the ``vacuum-driven'' mode, i.e. the one that resembles the fundamental mode of the vacuum BH at intermediate distance. The red curve represents the ``matter-driven'' mode, i.e. a mode which exists only in the presence of matter and that approaches $(\omega_R,-\omega_I)\to\infty$ as $r_0$ decreases. In this limit, the mode becomes difficult to track further. Note that, at large distance, this mode becomes lower lying with respect to the vacuum-driven mode.
\label{fig:2modes_bumpy}}
\end{center}
\end{figure}

\begin{figure}[thb]
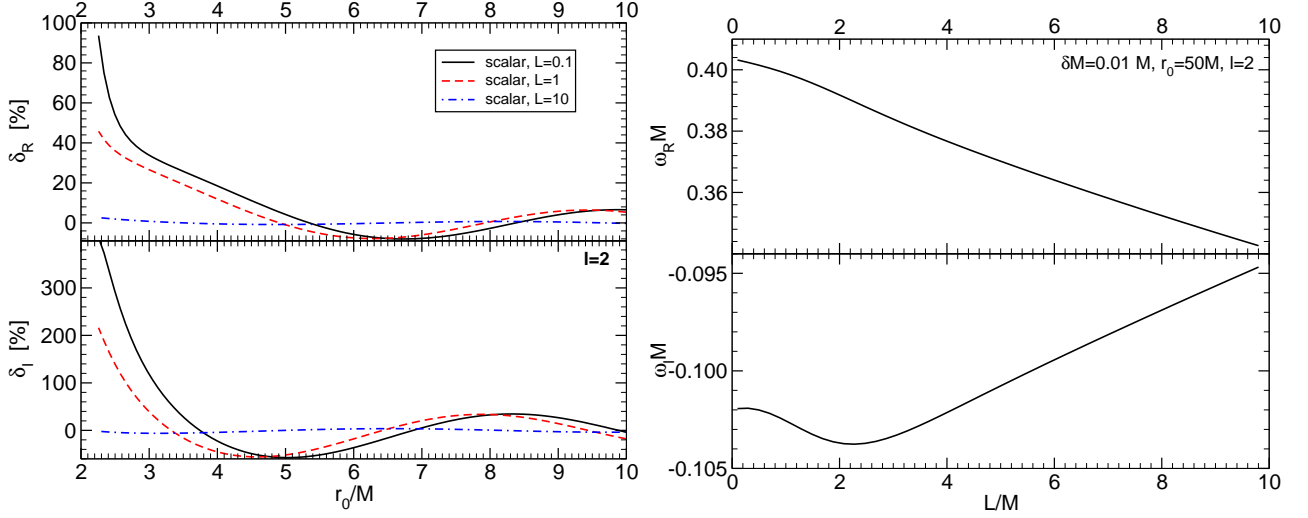

\begin{center}
\begin{tabular}{cc}
\epsfig{file=modes_bumpy_linear.eps,width=8.4cm,angle=0,clip=true}&
\epsfig{file=bumpyII_VS_L.eps,width=8.4cm,angle=0,clip=true}
\end{tabular}
\caption{Left: The linear in $\delta M/M$ correction to $\delta_{R,I}$ as a function of $r_0$ for the bumpy BH model I and for various values of $L$. Right: example of a mode tracked as a function of $L$ for fixed $\delta M$ and $r_0\ll M$.
\label{fig:bumpy_linear}}
\end{center}
\end{figure}

In addition, we also observe the appearance of \emph{new} modes which are not present in the vacuum case. A representative example is shown in Fig.~\ref{fig:2modes_bumpy} and the results are discussed in the caption.
In Fig.~\ref{fig:2modes_bumpy} we show only two of these modes, but there is actually an infinite set, whose structure is qualitatively similar to that shown in Fig.~\ref{fig:toy_model_modes} for the toy model discussed above. Note that some of these modes can have the same real or imaginary part for specific values of the parameters (cf. bottom panels of Fig.~\ref{fig:2modes_bumpy}). Correspondingly to these crossings, and also because of the presence of a multitude of modes, it might be challenging to track a single mode in the entire parameter space. This is the reason for the nonmonotonic behavior of the modes as a function of $L$ that is shown in Fig.~\ref{fig:modes_bumpy}: at large distances, it is likely that the root finder simply starts tracking a different mode. Nonetheless, the structure of the large-$r_0$ corrections is clear and qualitatively similar for the two models. A representative example of a mode tracked as a function of $L$ for fixed $r_0$ and $\delta M$ in the large $r_0$ limit is shown in the right panel of Fig.~\ref{fig:bumpy_linear}.

In Table~\ref{tab:QNMs_bumpyII}, we present a selection of the lowest-lying QNM frequencies for the matter profile $m_{II}(r)$. At this point, it is important to stress that, even if the spectrum contains new lowest-lying modes and modes which deviate substantially from the vacuum case, this does not necessarily imply that the ringdown waveforms would be so drastically affected. We shall prove this statement in the next section, confirming our findings for the previous models.

\begin{table}[hbt]
\centering \caption{Two lowest-lying QNM frequencies for the matter profile $m_{II}(r)$ for $l=2$.} 
\vskip 12pt
\begin{tabular}{@{}cccc@{}}
\hline \hline
$(\delta M,\,L)$     &$r_S$&$(\omega_R,-\omega_I)$&$(\delta_R,\delta_I)(\%)$  \\
\hline \hline
                      &10    &(0.4520,\,0.005018)  &(6.5,\,94.8)\\

                      &15    &(0.4514,\,0.006244)  &(6.7,\,93.5)\\
                      
(2,\,1.5)             &20    &(0.4513,\,0.005693)  &(6.7,\,94.1)\\
                      
                      &50    &(0.5062,\,0.06286)   &(-4.7,\,35.0)\\
                      
                      &100   &(0.4518,\,0.005923)  &(6.6,\,93.9)\\ \hline \hline
(0.1,\,2)             &3     &(0.4640,\,0.08289)   &(4.1,\,14.3)\\

                      &100   &(0.4533,\,0.005903)  &(6.3,\,93.9)\\ \hline \hline
\end{tabular}
\label{tab:QNMs_bumpyII}
\end{table}

Taking aside the large-$r_0$ corrections, we now focus on the region $r_0\lesssim10M$. In this region, the behavior of the modes is much more clear and the corrections are linear in the small $\delta M$ limit. We define:
\begin{equation}
 \delta_{R,I} = \delta_{R,I}^{(1)}(l,r_0,L)\frac{\delta M}{M}+{\cal O}(\delta M^2)\,.
\end{equation}
The linear corrections to $\delta_{R,I}$ are shown in the left panel Fig.~\ref{fig:bumpy_linear}. When $r_0$ is located a few Schwarzschild radii away from the BH, the corrections increase as $L$ decreases, i.e., when the matter density increases. In fact, our results should reduce to those of the thin-shell in the $L\to0$ limit. We estimate a deviation of the order $\delta_{R,I}\sim{\cal O}(10)\delta M/M$ for $L\lesssim M$. This is the value reported in Table~\ref{bstable}.

%

%
%

\subsubsection{Mode excitation: scattering and infall}
Our excursion into the double-barrier toy model suggested that, as interesting and rich as the spectra of bumpy potentials may be,
their excitation is hard to achieve with transient sources. More importantly, on time scales relevant for detectors,
it is possible that the free modes (i.e., the QNMs of the composite system) are completely irrelevant. To test this, we consider the excitation of scalar fields.
For weak enough scalar fields, the Klein-Gordon equation sourced by the radial infall of a unit scalar charge, $\Box\Phi=-4\pi/(\gamma\,r^2)\delta(r-R(t))\delta(\cos\theta)\delta(\phi)$, can be studied as a perturbation in the background spacetime [we use standard Schwarzschild coordinates $(t,r,\theta,\phi)$]. Here,
$\gamma$ is the Lorentz factor of the scalar charge at infinity, as measured by a static observer with respect to the BH, whereas $R(t)$ is the radial position of the infalling scalar charge, whose worldline reads $(T(t),R(t),0,0)$. Decomposing the field in spherical harmonics $Y_{lm}$
\be
\Phi(t,r,\theta,\phi)=\sum_{lm} \frac{\psi_{lm}(t,r)}{r} Y_{lm}(\theta,\phi)
\ee
and performing a Laplace transform,
\be
\psi_{lm}(t,r)=\frac{1}{2\pi}\int_{-\infty}^{+\infty}d\omega e^{-i\omega t}\Psi(\omega,r)\,,
\ee
one gets the following ordinary differential equation
\be
\frac{d^2\Psi}{dr_*^2}+\left(\omega^2-V(r)\right)\Psi=I=-\dot{\psi_0}+i\omega\psi_0+{\cal S}\,.
\ee
For concreteness we focus on two cases: 

\noindent{\bf Infalls.} Here $\psi_0=\psi(t=0,r),\,\dot{\psi_0}=\dot{\psi}(t=0)$ and
\be
{\cal S}=-A\frac{e^{i\omega T(r)}\sqrt{2+4l}}{4\pi r \gamma dR/dT}\,,
\ee
where the geodesic equation yields
\be
\frac{dT}{dR}=-\frac{\gamma}{A\sqrt{\gamma^2-A}}\,.
\ee
We take the scalar particle to materialize at $t=\theta=\phi=0,r_p=6M$ with gamma factor $\gamma=2$.
This is an ad hoc source, intended to mimic the plunge of particles at the last stable circular orbit.

\begin{figure}[thb]
\begin{center}
\begin{tabular}{cc}
\epsfig{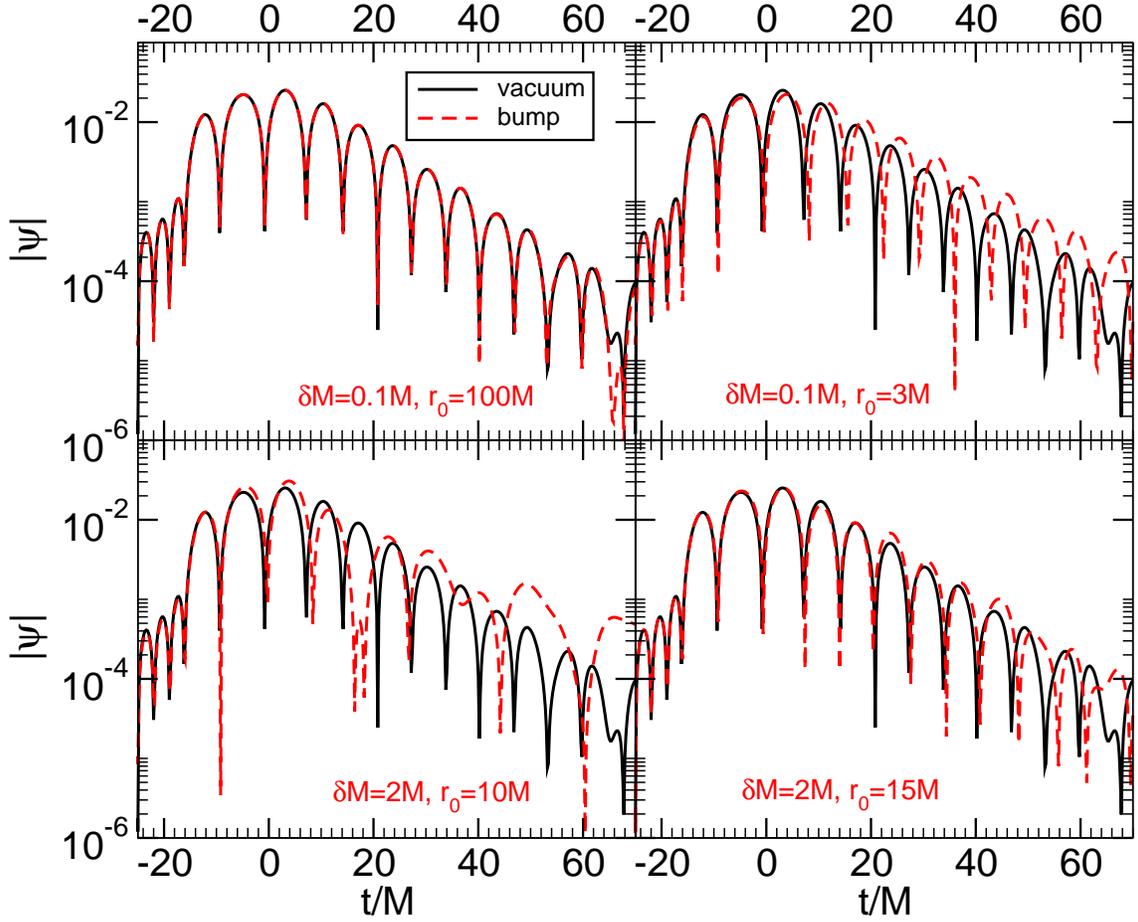}
\end{tabular}
\caption{Scalar $l=2$ waveforms for a point particle falling into a Schwarzschild BH.
The particle ``suddenly'' materializes at $r_p=6M$ with gamma factor $\gamma=2$
and subsequently falls into the BHs. The matter profile has extension $L=2,1.5$
in the top and bottom panels respectively.
\label{fig:infall}}
\end{center}
\end{figure}

The scalar waveforms for $l=2$ are shown in Fig.~\ref{fig:infall} for different configurations. In the top left panel, we compare the waveform obtained for $\delta M=0.1M$ and $r_0=100M$ with the one corresponding to the vacuum (Schwarzschild) case. The waveforms perfectly match and we observe the typical ringdown of the \emph{isolated} BH. This is the same ``memory effect'' we discussed for the toy model. It is worth stressing that, for this choice of the parameters, the QNM spectrum does \emph{not} contain the modes of the isolated BHs. The situation is different in the other panels, where we consider smaller values of $r_0$ or larger values of $\delta M$. In the bottom panels, in order to display large deviations, we have considered the value $\delta M=2M$ although the latter is not consistent with a Newtonian approximation.

\begin{figure*}[thb]
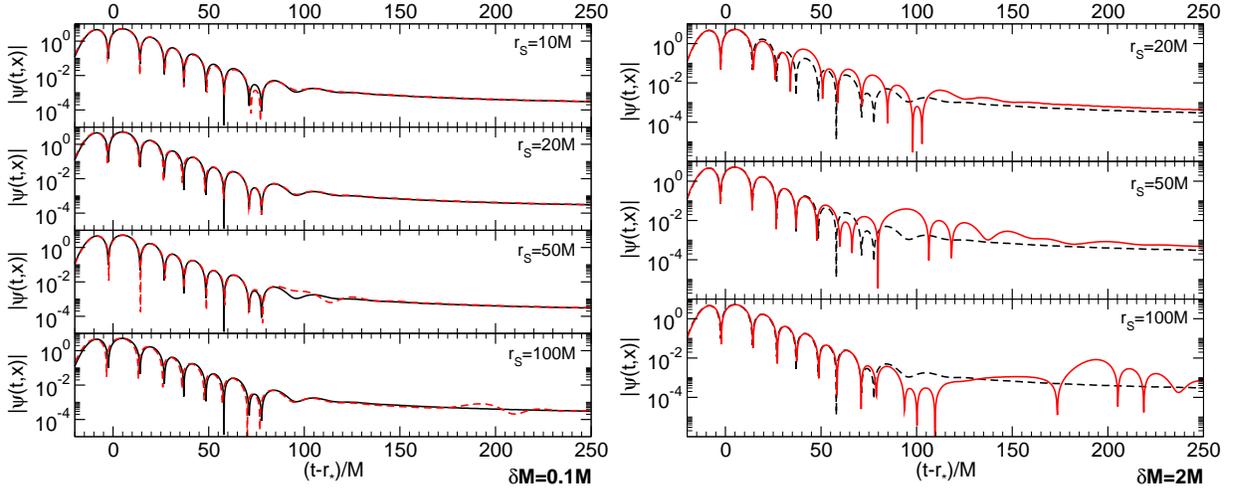

\begin{center}
\begin{tabular}{cc}
\epsfig{file=scattering_bumpyII.eps,width=8cm,angle=0,clip=true}&
\epsfig{file=scattering_bumpyII_largeM.eps,width=8cm,angle=0,clip=true}
\end{tabular}
\caption{Scalar $l=1$ waveforms for a scattering in a bumpy BH background. Initial data are ${\cal S}=\psi_0=0$ and $\dot{\psi_0}=fe^{-(r_*-r_0)^2/\sigma^2}$ with $\sigma=6M$ and $r_p=10M$. We used our bumpy profile II with $L=2M$ and $\delta M=0.1 M$ (left panels) and $\delta M=2M$ (right panels), respectively. Each single panel corresponds to a different value of $r_S$ in the bumpy metric.
\label{fig:scattering_bumpyII}}
\end{center}
\end{figure*}

\noindent{\bf Scattering}. We have considered also the scattering of a sourceless scalar field with ${\cal S}=\psi_0=0$ and $\dot{\psi_0}=A(r)e^{-(r_*-r_p)^2/\sigma^2}$. This is shown in Fig.~\ref{fig:scattering_bumpyII}. Again, the signal at intermediate time is governed by the ringdown of the isolated BH, and the effects of the surrounding matter appear only at late time.

These results confirm that the QNMs of ``dirty'' BHs typically play a subdominant role in time evolutions, particularly when matter is localized farther away from the BH. Although the QNM spectrum is dramatically different from the case of an isolated BH, in dynamical situations (e.g. during the merger of two BHs) the modes of the matter-BH system are excited to low amplitudes and at very late times. Accordingly they play little role in the merger waveform of BHs, but they will likely dominate over Price's power-law tails~\cite{Price:1971fb,Ching:1995tj}.
Furthermore, these modes can be excited to large amplitudes by inspiralling matter. For example, extra modes associated to massive fields may give rise to ``floating orbits'', which correspond to resonances in the gravitational flux~\cite{Cardoso:2011xi}.

\subsection{Effects of Dark-Matter Halos\label{sec:halos}}
In this section, we wish to describe a different source of deformation of an isolated Schwarzschild BH. 
Namely, massive BHs immersed in DM halos.
Because the DM density in the vicinity of the BH is small, we model the geometry by the approximate ansatz given by Eq.~\eqref{ansatz} with $A(r)=B(r)=1-2m(r)/r$, and 
\begin{equation}
 m(r)=M+\frac{4\pi}{3} r^3 \rho_{\rm DM}\,,
\end{equation}
where $\rho_{\rm DM}$ is the DM density. This approximation is valid as long as the correction due to the DM mass is small compared to the BH mass $M$. In this approximation, the geometry takes the form of a Schwarzschild-de Sitter spacetime $A(r)=B(r)=1-2M/r-\Lambda_{\rm eff} r^2/3$ with
\begin{equation}
 \Lambda_{\rm eff}=8\pi \rho_{\rm DM}\,.
\end{equation}
Thus, in this approximation we can directly use the results of Eq.~\eqref{deltaLambda},
\begin{equation}
 \left(\delta_R,\delta_I\right)=\left(6.0,5.2\right)\times 10^{-24} \left(\frac{M}{10^6M_{\odot}}\right)^2\frac{\rho_{\rm DM}}{10^{3} {M_\odot}/{\rm pc}^3}\,,
\end{equation}
for the $l=2$ mode.

%

\subsection{Black holes with short hair}
We have already studied several ad-hoc matter models and corresponding QNMs in the previous sections. 
The general case of extended matter distributions outside BHs
is a rather unexplored subject. Let us explore this subject a bit further by assuming a spherically-symmetric
ansatz which describes a Schwarzschild BH surrounded by a small amount of matter,
\be
ds^2=-\left(1-\frac{2M}{r}+\epsilon H\right)dt^2+\frac{1}{1-\frac{2M}{r}+\epsilon F}dr^2+r^2d\Omega^2\,,\nonumber
\ee
where $H$ and $F$ are radial functions to be determined, and $\epsilon$ is a small, bookkeeping parameter. Einstein equations yield
\beq
\epsilon (rF)'&=&-\frac{8\pi G r^3}{r-2M}T_{tt}\,,\\
\epsilon (r^2H')'&=&-\frac{8\pi G\left(r^3(M-r)T_{tt}+(r-2M)^2\left(r(M-r)T_{rr}-2T_{\theta\theta}\right)\right)}{(r-2M)^2}\,.
\eeq
Thus, given a matter distribution (i.e., given $T_{\mu\nu}$), one can in principle compute the metric coefficients and from these the spacetime QNMs.
Unfortunately, the dynamical perturbations of this spacetime depend on the metric functions $H$ and $F$, so that it does not seem possible to express all equations solely in terms of the matter content. 
In other words, it is impossible to solve for the QNMs, in closed form, in terms of a generic $T_{\mu\nu}$. One has to resort to a case-by-case analysis, which in fact corresponds to our procedure in the 
previous sections.

We end this QNM study with an interesting, analytic solution describing a BH surrounded by 
an anisotropic fluids~\cite{Brown:1997jv}. This ``short-hair'' BH solution is described by
\beq
ds^2&=&-f dt^2+\frac{dr^2}{f}+r^2d\theta^2+r^2\sin^2\theta\,d\phi^2\,,\\
f&=&1-\frac{2M}{r}+\frac{Q_m^{2k}}{r^{2k}}\,, \label{solBrown}\\
\rho&=&\frac{Q_m^{2k}(2k-1)}{8\pi r^{2k+2}},\,\quad P=k\rho\,, \label{rhoBrown}
\eeq
where $\rho$ and $P$ are the density and pressure, respectively, of the anisotropic fluid, and $Q_m$ is a constant, describing the ``matter-hair''. The solution above corresponds to an anisotropic stress-tensor, specified by (3.2) in Brown and Husain's work~\cite{Brown:1997jv}, and it reduces to the Reissner-Nordstrom BH when $k=1$.

\subsubsection{Scalar QNMs of the Brown-Husain solution}
The scalar wave equation in the Brown-Husain solution reduces to
\begin{equation}
 V=f\left(\frac{l(l+1)}{r^2}+\frac{2M}{r^3}-\frac{2k Q_m^{2k}}{r^{2k+2}}\right)\,.
\end{equation}
Let us consider a simple case and set $k=3/2$ in the equation above, so that the behavior of the potential is analytical. This case is qualitatively similar to the Reissner-Nordstrom one: if $Q_m<2^{5/3} M/3$ the solution has two horizons located at $r>0$, whereas in the extremal case these two horizon merge at $r=r_+=4M/3$. The solution for $k=2$ displays a similar behavior.

\begin{figure}[thb]
\begin{center}
\begin{tabular}{cc}
\epsfig{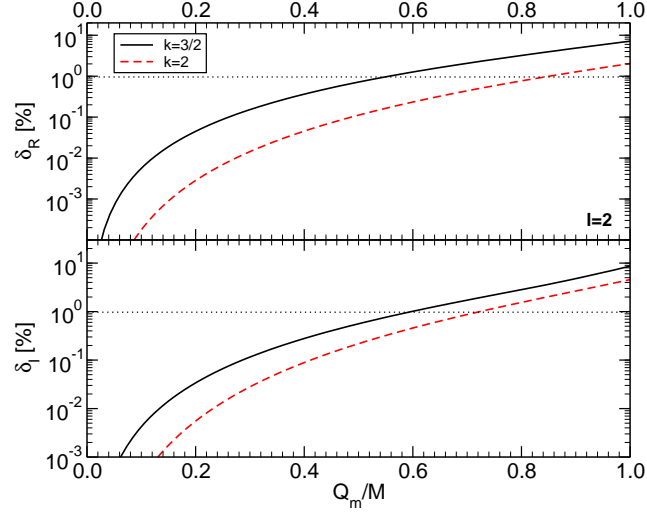}
\end{tabular}
\caption{Percentage difference for the fundamental $l=2$ scalar QNM of the Brown-Husain solution~\eqref{solBrown} as a function of $Q_m$. The extremal limit corresponds to $Q_m/M=2^{5/3}/3\sim1.06$ for $k=3/2$ and to $Q_m/M=3^{3/4}/2\sim1.14$ for $k=2$.
\label{fig:Brown_Husain}}
\end{center}
\end{figure}

We have computed the scalar QNMs using a direct integration. Our results for the fundamental $l=2$ mode as a function of $Q_m$ are shown in Fig.~\ref{fig:Brown_Husain} for $k=3/2$ and $k=2$. The deviation from the Schwarzschild case decreases as $k$ increases, because the hair becomes subdominant in a post-Newtonian expansion. For $Q_m/M\lesssim0.6$, deviations are smaller than $1\%$. 
In the small-$Q_m$ limit, our results are well fitted by:
\begin{eqnarray}
 & k=3/2:&\qquad \delta_R\sim 0.06 \left(\frac{Q_m}{M}\right)^3\sim 0.5\frac{\delta M}{M} \qquad \delta_I\sim 0.04 \left(\frac{Q_m}{M}\right)^3\sim 0.3\frac{\delta M}{M} \,,\\ 
 & k=2:&\qquad \delta_R\sim 0.02 \left(\frac{Q_m}{M}\right)^4\sim 0.3\frac{\delta M}{M} \qquad \delta_I\sim 0.03 \left(\frac{Q_m}{M}\right)^4\sim 0.4\frac{\delta M}{M} \,. 
\end{eqnarray}
where we have used the relation $\delta M/M=2^{-2k} (Q_m/M)^{2k}$, where $\delta M$ is the Newtonian mass associated with $\rho$ in Eq.~\eqref{rhoBrown}.

%

\subsection{Accretion\label{sec:ringdownaccretion}}
Previous results all considered stationary matter outside the horizon. We now briefly turn our attention to accretion, which changes the BH mass as a function of time. A detailed discussion on the effects related to accretion is given in Part~\ref{part:inspiral} below. We can estimate how much accretion will impact the QNMs of astrophysical BHs by computing how much mass a BH accretes during the ringdown phase. The latter lasts typically $\tau\sim20 M$ or less in geometrical units, or
\be
\tau \sim 100 \frac{M}{10^6M_{\odot}}{\rm s}\,.
\ee
By using Eq.~\eqref{Medd}, during this time the BH accretes a mass $\delta M$ of the order
\be
\frac{\delta M}{M}\sim 7\times 10^{-14} f_{\rm Edd}\frac{M}{10^6M_{\odot}}\,.
\ee
Because the instantaneous ringdown frequency $\omega \propto 1/M$, and because accretion is a very slow process compared to the light-crossing time, we estimate
that $\delta_R\propto \delta M/M$. This intuition is confirmed by several distinct linearized calculations of BH ringdown
in (spherically symmetric) Vaidya spacetimes, which model accreting spacetimes and which therefore take dynamical effects into account~\cite{Abdalla:2006vb,Abdalla:2007hg,Chirenti:2011rc,Shao:2004ws}. These studies show that the QNMs correspond to those of standard momentarily stationary BHs, that is, the mass evolves adiabatically on the ringdown time scale.
Thus, we estimate
\be
(\delta_R,\delta_I)\sim (7,7)\times 10^{-14}f_{\rm Edd}\frac{M}{10^6M_{\odot}}\,,
\ee
As mentioned in Sec.~\ref{sec:introdirty}, as well as explained in Part~\ref{part:inspiral} below, $f_{\rm Edd}$ depends on the details of the accretion disk surrounding the BH, but is at most $\sim {\cal O}(1)$. Therefore, the effects of accretion on the ringdown frequencies are completely negligible, as summarized in Table~\ref{bstable}.

\subsection{Nonspherical configurations: self-gravity effects of rings}
Up to now we have investigated spherically symmetric matter distributions only. Many astrophysical BHs
are surrounded by disks and rings of nontrivial angular distribution, which makes the system nontrivial 
to handle from a numerical point of view. However, QNMs have been shown to be associated with radiation
leaking from the unstable light ring, and our previous results are consistent with this picture~\cite{Cardoso:2008bp,Berti:2009kk}.
Thus we now turn briefly our attention to null geodesics of BHs surrounded by an axisymmetric distribution of matter.
Such systems were studied by Will~\cite{Will:1974zz,Will:1975zza} in a setup where the mass of the ring is much smaller
than that of the BH. Will finds an upper bound to the change to the circular photon orbit in the presence of a thin ring of matter localized near the BH. In particular, the frequency of the photon orbit is shifted by
\begin{equation}
 \frac{\delta \Omega_{\rm LR}}{\Omega_{\rm LR}}\lesssim 0.1 \frac{\delta M}{M}\,.
\end{equation}
According to Refs.~\cite{Cardoso:2008bp,Berti:2009kk} the impact on the QNMs is then estimated to be
\be
(\delta_{R},\delta_I)\lesssim 10^{-4} \frac{\delta M}{10^{-3}M}\,,\label{ring}
\ee
which is the value quoted in Table~\ref{bstable}.
These bounds are of the same order as those relating to spherical matter distributions and despite referring to a very specific model,
we have no reason to believe that other distributions will modify Eq.~\eqref{ring} by more than a factor two or so.
\section{Previous results and puzzles in the literature\label{sec:previous}}
As we mentioned earlier, it is somewhat surprising that the work on QNMs of ``dirty'' BHs is very scarce 
\footnote{We are excluding from this list important work on the {\it excitation amplitude} of QNMs from infalling shells or oscillating torii~\cite{Papadopoulos:1998nc,Nagar:2005cj,Nagar:2004ns,Nagar:2006eu}.
These works address the degree to which ringdown is excited in physical processes, concluding that these modes are only excited when matter crosses the light ring, and that they can be strongly suppressed by
interference effects. These studies do {\it not} tackle the main problem we are discussing, which is that of corrections to the QNM frequencies themselves by backreaction.
We are also excluding interesting works on the {\it asymptotic} structure of highly damped QNMs nonisolated BHs, since these are less relevant for GW physics; for more on this topic see for instance Ref.~\cite{Medved:2003rga}.
}.
To the best of our knowledge, Leung and collaborators~\cite{Leung:1999iq,Leung:1999rh} seem to have been the only ones seriously addressing this question, because in their own words, ``inasmuch as the goal
of the GW observatories is to obtain astrophysical
information of our universe...there is no doubt that we will eventually have to face this
problem of the QNM spectra of dirty BHs''.

As we showed, for a given matter profile and at fixed $\delta M$, the corrections $\delta_R,\delta_I$ grow linearly with the distance of the localized source from the BH. This is the most counterintuitive result of our analysis 
and is unexpected, because, at fixed $\delta M$, increasing the radial position
of the matter distribution is equivalent to decreasing its density. In Refs.~\cite{Leung:1999iq,Leung:1999rh}, the authors studied in detail the scalar QNMs of nonrotating BHs surrounded by thin shells of matter, both numerically and perturbatively. Their results and conclusions are in full agreement with our own results reported earlier. Their analysis, in a vein similar to perturbation theory familiar from quantum mechanics, shows that it is the exponential dependence of the QNM wave functions on the radial coordinate that is responsible for the overall behavior at large distances. 
Their analysis fits in very nicely with the analytic results of our toy model, Section~\ref{sec:toy}.

The overall structure of this problem is also in agreement and consistent with a well-known phenomena in BH physics, concerning the behavior of massive fields around BHs~\cite{Detweiler:1980uk,Cardoso:2005vk,Dolan:2007mj,Rosa:2011my,Pani:2012vp,Brito:2013wya}. These studies have shown that fundamental fields of mass $\mu$ give rise to long-lived (or even unstable) states whose lifetime $\tau$ scales as $\tau \sim (M\mu)^{-p}$, with $p$
some integer power. Thus in the zero mass limit, $M\mu\to 0$, these modes do not coalesce to any of the Schwarzschild QNMs. The reason for this behavior is, again, that these modes are localized far away from the BH, at a distance comparable with the Compton length of the massive field, $r/M \sim 1/(M\mu)$.

Another related work is that of Nollert~\cite{Nollert:1996rf}, who considered QNMs of the discretized Regge-Wheeler potential.
Nollert shows that the QNMs of the discretized potential differ substantially from the QNMs of the Regge-Wheeler potential but that, nonetheless, time evolutions
using the discretized potential ringdown according to the Regge-Wheeler potential modes. These findings are in complete consistency with ours, and most likely also explained by the exponential sensitivity at large distances.

These results all highlight a clear need to understand better the relation between time-domain waveforms and the QNMs of perturbed potentials, which has been surprisingly overlooked in the past.

\section{Relation with ``firewalls'' and other exotica around massive BHs\label{sec:firewall}}
Our results for the ringdown modes of matter-deformed Schwarzschild BHs have also a direct application in the context of the recent ``firewall'' proposal~\cite{Almheiri:2012rt} (see also Ref.~\cite{Braunstein:2009my} for a similar earlier proposal). Roughly speaking, under quite generic assumptions on the quantum properties of an event horizon, Ref.~\cite{Almheiri:2012rt} suggests that an infalling observer is expected to encounter Planck-density fluctuations in a region near the horizon, whose thickness is of the order of the Planck length. Here we do not enter the debate about the fate of the observer or about the very existence of such ``firewall'' (cf. Refs.~\cite{Susskind:2012rm,Almheiri:2013hfa} for a detailed discussion). Rather, we are interested in understanding possible gravitational signatures of this proposal at the classical level.

We model such a configuration as a spherical thin shell of mass $\delta M$ located at $r_0\sim 2M_0+\ell_P$, where $\ell_P\sim 1.6\times 10^{-35}{\rm m}$ is the Planck length, $M_0=M+\delta M$ is the total mass of the spacetime and $M$ is the BH mass. Such approximation should be reliable because the thickness of the firewall is negligible with respect to the size of the BH. Although the energy condition $\Sigma>|\Theta|$ is violated when the shell is only some Planck length away the horizon, putative quantum-gravity effects (ignored here) should cure this problem. 

This system resembles other objects in the literature, such as gravastars, aimed at mimicking BH properties (QNMs of gravastars were computed in Refs.~\cite{oai:arXiv.org:0706.1513,Pani:2009ss} and the overall behavior is consistent with the one we discuss below for firewalls).
These other objects have a surface but no horizon and were all but ruled out by observations of the galactic center object, on the basis of the large luminosities that the existence of a surface would give rise to~\cite{Broderick:2005xa}. By energy conservation arguments, infalling accreted matter would release huge amounts of potential energy which could only be converted into radiation at infinity; this radiation is however not observed. These arguments rely heavily on conversion of potential energy into electromagnetic energy, and can be circumvented if the central object is exotic and energy is preferably converted through other channels, like neutrinos or gravitons. For firewalls though, the arguments do not apply due to the existence of the horizon behind the ``firewall,'' which can absorb essentially all the converted energy. To be more precise, consider the infall of accreting matter into the firewall, modeled as a shell, and assume conservatively that all the potential energy is converted into electromagnetic radiation, as in Ref.~\cite{Broderick:2005xa}. Assume also that the produced photons have an isotropic distribution in the frame of an observer at rest with the shell. It is easy to show that of all the photons produced, only a vanishingly small fraction $\sim \sqrt{1-2M_0/r_0}\sim \sqrt{\ell_P/M}$ is able to escape to infinity~\cite{Page:1974he,Bardeen:1973}, while all those remaining are absorbed. This then invalidates the arguments of Ref.~\cite{Broderick:2005xa} when applied to firewalls.

The analysis of the previous section suggests that, in this configuration, the ringdown frequencies of the massive BH would be drastically modified. Indeed, the Schroedinger-like potential of a Schwarzschild BH --~when written in tortoise coordinates $r_*=r+2M\log(r/(2M)-1)$~-- is roughly symmetric around the light-ring at $r\sim 3M$~\cite{Berti:2009kk}. This fact, together with the double-barrier toy model discussed in Sec.~\ref{sec:toy}, suggests that the corrections to the QNMs when $r_0\to r_+$ (i.e. when $r^*_0\to-\infty$ in tortoise coordinates) would be similar to the case $r_0\to \infty$ discussed above and shown in the bottom right panel of Fig.~\ref{fig:modes}. Therefore we expect that, for any $\delta M$, the QNMs would deviate parametrically from the vacuum case as $r_0\to r_+$. Because $r_0$ is expected to be only some Planck length away from the horizon, the firewall proposal provides a natural candidate where very large corrections to the vacuum QNMs are expected.

\begin{figure}[thb]
\begin{center}
\epsfig{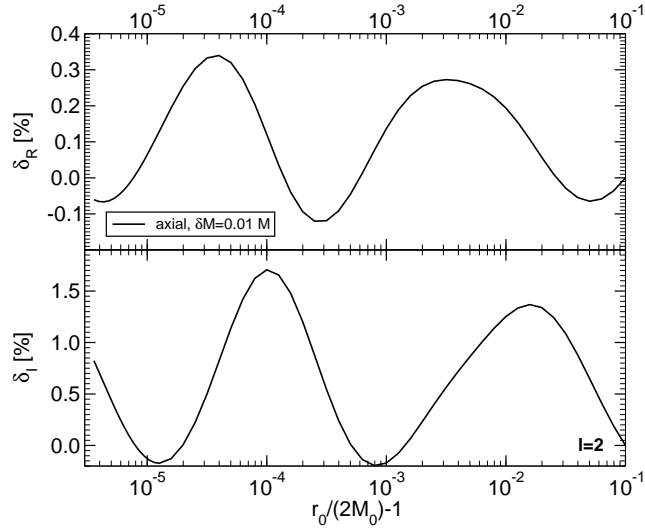}
\caption{
Relative corrections to the fundamental axial QNM for a ``firewall model'' using a thin-shell with mass $\delta M=0.01 M$ located at $r_0$. Top and bottom panels respectively refer to the real and to the imaginary part of the frequency as a function of the shell location in the limit where $r_0$ approaches the horizon (i.e. $r_0\to-\infty$ in tortoise coordinates). The oscillatory behavior is similar to that shown in the bottom right panel of Fig.~\ref{fig:modes}. Finite numerical accuracy prevents us from tracking the modes when $r_0/(2M_0)-1\ll10^{-6}$, where much larger corrections are expected. The modes were computed by using an accurate continued-fraction method adapted from Ref.~\cite{Pani:2009ss}. Polar modes and different values of $\delta M$ would give qualitatively similar results.
\label{fig:firewall}}
\end{center}
\end{figure}

Computing the modified ringdown frequencies explicitly turns out to be extremely challenging, mostly due to the huge difference in scale between the BH mass $M$ for a massive BH and the Planck scale.
In Fig.~\ref{fig:firewall} we plot the modified axial fundamental mode for a thin-shell model in the limit $r_0\to 2M_0$. This figure has to be compared with the bottom right panel of Fig.~\ref{fig:modes} which shows the same corrections in the opposite limit, $r_0\to \infty$. In line with Fig.~\ref{fig:modes}, also in Fig.~\ref{fig:firewall} we observe the same oscillatory behavior as the distance between the shell and the light-ring (in tortoise coordinates) increases. However, finite numerical accuracy prevents us from computing the modes when $r_0/(2M_0)-1\ll10^{-6}$, i.e. in a region where we expect the same power-law drift as in the bottom panel of Fig.~\ref{fig:modes}. Nonetheless, in light of our results in Sec.~\ref{sec:dirty}, there is no reason to expect that such parametric deviation does not occur if the shell is sufficiently close to the horizon, as it is in the firewall proposal.

From the results of Fig.~\ref{fig:modes}, the correction to the real part of the frequency is roughly linear in $r_0$ in the large-$r_0$ limit. Assuming the same behavior occurs in the $r^*_0\to -\infty$ limit (in tortoise coordinates) and converting to Schwarzschild coordinates, we obtain the numerical fit
\begin{equation}
\delta_R\sim 2\times 10^{-2}\log\left(\frac{r_0}{2M_0}-1\right)\frac{\delta M}{10^{-2} M}\,,
\end{equation}
as $r_0\to 2 M_0$. Because the firewall is located at a Planck distance from the horizon, $r_0\sim 2M_0+\ell_P$, we can extrapolate the correction
\begin{equation}
\delta_R\sim -2\left[1-0.01\log\left(\frac{10^6M_\odot}{M_0}\right)\right] \frac{\delta M}{10^{-2} M} \,. \label{deltaRfirewall}
\end{equation}
Because of the logarithmic dependence, this number is only mildly sensitive to the BH mass. For $M\sim 10 M_\odot$, $\delta_R$ would be only $10\%$ smaller than the estimate above for $M\sim 10^6 M_\odot$. Although approximate, our results suggest that a firewall of proper mass as low as $\delta M\sim 10^{-4}M$ would introduce a correction of the order of a few percent in the ringdown frequencies of a massive BH. Such corrections might be observable with advanced GW detectors, provided the modes are excited to appreciable amplitudes. Now, one of the main conclusions of Section~\ref{sec:dirty} is that the modes of the composite system are excited
a light-crossing time $t\sim \log\left(\frac{r_0}{2M_0}-1\right)$ after the main burst of radiation produced at the light ring. Even for a shell a (coordinate) Planck $\ell_P$ distance away from the horizon, this is a very small time interval. Thus, this back-of-the-envelope
calculation predicts that the main splash of radiation, consisting on ringdown of an isolated BH, is quickly followed by ringdown in the modes of the composite BH plus firewall system.
In summary, we predict changes in the GW signal which can be significant if the firewall model is correct, the actual number depends (linearly) on the total mass of the firewall shell, which is still a subject of debate~\cite{Abramowicz:2013dla,Israel:2014eya}.

\section{Parametrized Ringdown Approach}
Finally, we conclude this part by illustrating a powerful method to study small corrections from the Schwarzschild ringdown frequencies in a model-independent fashion. The idea is to use the parametrized metric presented in Sec.~\ref{introBeyondGR} and study probe fields in this background. 

In Ref.~\cite{Pani:2012bp}, a ``master equation'' for scalar and electromagnetic probe perturbations of a generic (slowly-rotating) metric has been derived. In the nonrotating case, the master equation reads
\begin{equation}
\frac{d^2\Psi}{dr_*^2}+\left[\omega^2-A\left(\frac{l(l+1)}{r^2}+\mu^2+(1-s^2)
\left\{\frac{B'}{2r}+\frac{B A'}{2r A}\right\}\right)\right]\Psi=0\,,\label{master}
\end{equation}
where $s$ is the spin of the perturbation ($s=0$ for scalar perturbations and $s=\pm1$ for vector perturbations with axial parity), $\mu$ is a generic mass term and we have introduced a generalized tortoise coordinate $r_*(r)$ such that $dr/dr_*=\sqrt{AB}$.

Inserting the expansion~\eqref{PRA1}-\eqref{PRA2} in Eq.~\eqref{master}, we find an eigenvalue problem that depends parametrically on the infinite set of parameters $\alpha_i$ and $\beta_i$. This problem can be solved with standard methods~\cite{Berti:2009kk,Pani:2013pma} to compute the eigenfrequencies as functions of the parameters. For obvious reasons we dubbed this the ``parametrized ringdown approach''.

\begin{figure}[thb]
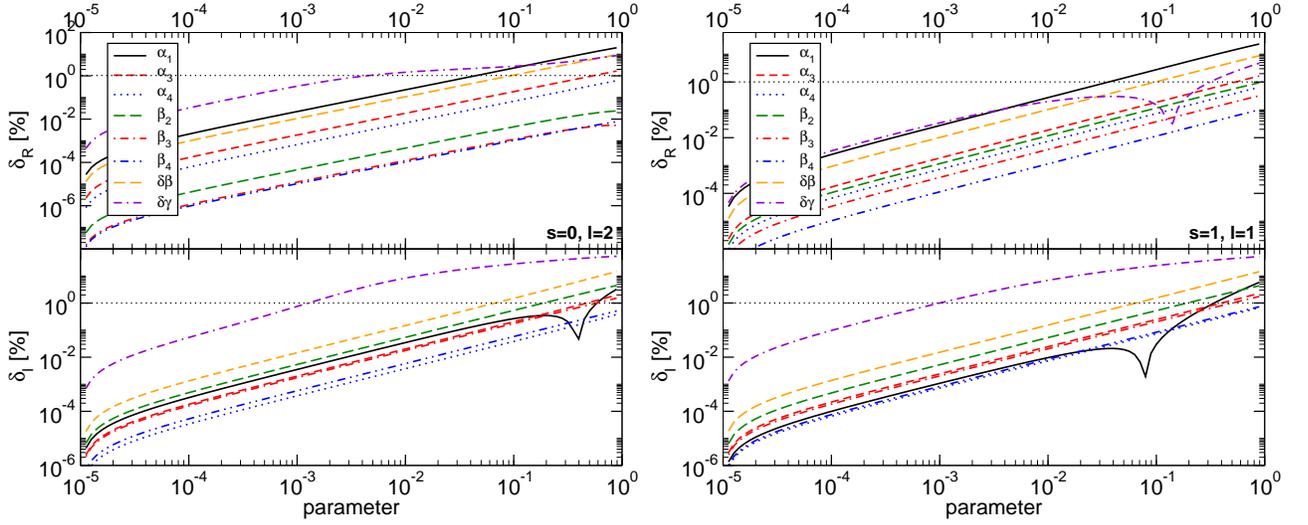

\begin{center}
\begin{tabular}{cc}
\epsfig{file=modes_PRA.eps,width=8.4cm,angle=0,clip=true}&
\epsfig{file=modes_PRA_s1l1.eps,width=8.4cm,angle=0,clip=true}
\end{tabular}
\caption{Percentage difference for the fundamental $l=2$ scalar and $l=1$ electromagnetic QNMs of the parametrized metric~\eqref{PRA1}-\eqref{PRA2} as a function of various parameters. Each curve corresponds to a single parameter turned on, while the others are set to zero.
\label{fig:PRA}}
\end{center}
\end{figure}

This technique will be discussed in detail elsewhere. Here, as a first example of its effectiveness, we have solved the parametrized ringdown equation using direct integration, in order to compute the corrections to the fundamental $s=0,1$ modes. In the small coupling limit, these corrections can be written as a linear combination of the parameters:
\begin{eqnarray}
 \delta_{R}&=&\delta_{R,\beta}^{sl}\delta\beta+\delta_{R,\gamma}^{sl}\delta\gamma+\sum_{i=2}^{N_\alpha}\delta_{R,\alpha_i}^{sl}\alpha_i+\sum_{i=2}^{N_\beta}\delta_{R,\beta_i}^{sl}\beta_i \,,\label{fitR}\\
 \delta_{I}&=&\delta_{I,\beta}^{sl}\delta\beta+\delta_{I,\gamma}^{sl}\delta\gamma\sum_{i=2}^{N_\alpha}\delta_{I,\alpha_i}^{sl}\alpha_i+\sum_{i=2}^{N_\beta}\delta_{I,\beta_i}^{sl}\beta_i  \label{fitI}\,.
\end{eqnarray}
The first corrections for different values of $s$ and $l$ are presented in Table~\ref{tab:PRA} and shown in Fig.~\ref{fig:PRA}.
\begin{table}[htb]
\scriptsize
 \begin{tabular}{cc|cccccccc}
  &	($s$,$\ell$,$n$)	& $(\delta_{R,\beta},\delta_{I,\beta})$& $(\delta_{R,\gamma},\delta_{I,\gamma})$& $(\delta_{R,\alpha_1},\delta_{I,\alpha_1})$& $(\delta_{R,\alpha_3},\delta_{I,\alpha_3})$& $(\delta_{R,\alpha_4},\delta_{I,\alpha_4})$& $(\delta_{R,\beta_2},\delta_{I,\beta_2})$& $(\delta_{R,\beta_3},\delta_{I,\beta_3})$ & $(\delta_{R,\beta_4},\delta_{I,\beta_4})$ \\
\hline
& (0,2,0) &$(0.10,0.15)$  &$(3.7,5.7)$  &  $(0.22,0.02)$ &  $(0.02,4\times10^{-3})$ &  $(7\times10^{-3},4\times10^{-3})$ &  $(5\times10^{-4},0.05)$ &  $(10^{-4},0.02)$ &  $(10^{-4},6\times 10^{-3})$  \\
\hline
& (0,2,0) &$(0.10,0.15)$  &$(3.8,11)$  &  $(0.28,0.01)$ &  $(0.02,0.02)$ &  $(7\times10^{-3},7\times10^{-3})$ &  $(0.01,0.05)$ &  $(4\times10^{-3},0.02)$ &  $(10^{-3},8\times 10^{-3})$  \\
\hline
\hline
\end{tabular}
\caption{Coefficients of the fits~\eqref{fitR} and~\eqref{fitI} for the real and
  imaginary parts of scalar ($s=0$) and electromagnetic ($s=1$) modes of the parametrized metric~\eqref{PRA1}-\eqref{PRA2}.
\label{tab:PRA}}
\end{table}

A detection of a ringdown mode to the level of one percent can constrain the parameter $\delta\gamma$ to the level of roughly one part in $10^4$ and the parameters $\delta\beta$ and $\alpha_1$ to the level of one part in $10^3$. Remarkably, higher-order coefficients in the expansion~\eqref{PRA1}-\eqref{PRA2} are less constrained, showing that a weak-field expansion is still meaningful for ringdown tests. This is due to the fact that the ringdown modes are governed by the light ring located roughly at $r\sim 3M$ \cite{Cardoso:2008bp} so that higher powers of $M/r$ are still mildly suppressed there. While the putative constraints on $\delta\gamma$ and $\delta\beta$ are less stringent than or comparable to those currently in place, ringdown tests have the potential to put very competitive bounds on the coefficients of the higher-order expansion.

A detailed investigation of this parametrized ringdown approach and its generalization to spinning geometry is an interesting extension of our work, which we leave for future work.

\clearpage
\newpage

\part{Inspirals}\label{part:inspiral}

\section{Executive Summary}
The inspiral of compact binaries can be affected by environmental ``dirtiness'' in a variety of ways, including the gravitational pull exerted by the matter configuration, accretion onto the binary's components, dynamical 
friction and planetary migration.
We summarize here the most important results and conclusions of our study on the compact-binary inspiral in realistic astrophysical environments, which are quantified in Tables~\ref{tab:periastron} and \ref{tab:dephasing} and detailed in the rest of this Part~\ref{part:inspiral}. Our main findings are the following:
\begin{enumerate}
\item Environmental effects can be safely neglected for most sources of GWs (especially in the case of space-based detectors such as eLISA).
This result is apparent from Tables~\ref{tab:periastron} and \ref{tab:dephasing}, where we show the corrections to the periastron shift and GW phase, respectively, in a variety of environments.

\item The only exception to the above concerns thin accretion disks, where accretion, dynamical friction, planetary migration and the gravitational pull of the disk are large enough to spoil parameter estimation and possibly detection for EMRIs.
The effects mentioned above are in fact generally more important than second-order self-force effects~\cite{2ndorderSF} in thin-disk environments. Furthermore, accretion, dynamical friction and planetary migration can even be more important than GW emission itself at separations larger than $\sim 30 -40M$. 
On the other hand, all environmental effects in thick disks are typically subdominant relative to second-order self-force corrections. Overall, our results confirm that EMRI detection should not be significantly affected by environmental effects, because the majority of events detectable by eLISA will likely occur in thick-disk environments. Indeed, as we show in Section~\ref{sec:prospects}, the fraction of EMRIs observable by eLISA and which evolve in thin-disk environments is at most a few peryielcent.
\item The effects of dynamical friction from DM halos are comparable or dominant with respect to the gravitational pull of DM. As a by-product, our analysis shows that the results of Ref.~\cite{Eda:2013gg} would be drastically modified, because they were obtained by considering self-gravity only and neglecting dynamical-friction effects. 
\item Detection of the inspiral GW signal can potentially improve the constraints on the PN parameter $\beta$ by two orders of magnitude, as well as provide novel constraints for higher-order coefficients of a weak-field expansion of the metric. This is also discussed in Part~\ref{part:testsGR} in the context of tests of GR.
\end{enumerate}
A discussion of these issues is presented in Sec.~\ref{sec:matterinspiral}, whereas specific observables are computed in the remaining sections. 

\begin{table}
\caption{Corrections to the periastron shift of a particle 
on a circular orbit with dimensionless radius $\tilde{r}_c=r_c/\left(GM/c^2\right)$, where $r_c$ is the (areal) radius of orbit, around a BH.
We consider the corrections due to a cosmological constant, electric charge, magnetic field,
accretion onto the central BH, as well as the 
effects of the gravitational pull (``self gravity'') of a thick/thin disk, of DM halos, and of a DM ``spike'' modeled
by a power-law density profile $\eqref{powerlawdensity}$ with ${\hat \alpha}=7/3$. (Other values of ${\hat \alpha}$ produce comparable corrections.)
Dissipative effects such as GW radiation reaction and hydrodynamic drag are negligible and are not considered here.
}
\begin{tabular}{c|cc}
 \hline\hline
Correction           	 &$\delta_{\rm per}/P$   &$P$              \\
\hline
Cosmological constant  &$10^{-31}$   &$\frac{\Lambda}{10^{-52}{\rm m}^{-2}}\left(\frac{M}{10^6M_{\odot}}\right)^2(\tilde{r}_c/10)^4$ \\
Galactic DM halos 	     &$10^{-21}$ 	 &$\left(\frac{M}{10^6M_{\odot}}\right)^2\frac{\rho_{\rm DM}}{10^{3}{M_\odot}/{\rm pc}^3}(\tilde{r}_c/10)^4$ \\ 
Thick accretion disk   & $10^{-16}$	 &$\frac{f_{\rm Edd}}{10^{-4}}\,\frac{M}{10^6M_{\odot}}\frac{0.1}{\alpha}(\tilde{r}_c/10)^{5/2}$   \\
Accretion              &$10^{-8}$   &$f_{\rm Edd}$           \\
Thin disk (assuming Eq.~\eqref{surfacedensity} and $\tilde{r}_c=10$) &$10^{-8}$   &${f_{\rm Edd}}^{7/10} \left(\frac{M}{10^6 M_\odot}\right)^{6/5}\left(\frac{\alpha}{0.1}\right)^{-4/5}$ \\
Charge 		             &$10^{-7} $   &$\left(q/10^{-3}\right)^2$  	     \\
DM distribution $\rho\sim r^{-{\hat \alpha}}$ &$10^{-21}$ &$\left(\frac{M}{10^6M_{\odot}}\right)^2\frac{\rho_{\rm DM}}{10^{3}{M_\odot}/{\rm pc}^3}(\tilde{r}_c/10)^{4-\hat\alpha}\left(\frac{R}{7\times 10^6 M}\right)^{\hat \alpha}$\\
Magnetic field         &$10^{-8}$    &$\left(\frac{B}{10^{8}{\rm Gauss}}\right)^2\left(\frac{M}{10^6M_{\odot}}\right)^2(\tilde{r}_c/10)^4$ \\
\hline\hline
\end{tabular}
\label{tab:periastron}
\end{table}

\begin{table}
\caption{Correction $\delta_\varphi$ to the GW phase of 
a quasicircular EMRI due to various effects (the same as in Table \ref{tab:periastron}, with the addition of some dissipative effects that modify the GW emission.). 
The coefficients $c_X(\chi)$ represent the terms written between square brackets in Eqs.~\eqref{cLambda}, \eqref{cq}, \eqref{cB} and \eqref{calpha}, respectively, and are functions of the combination~\eqref{chi}. The behavior of the coefficients is shown in Fig.~\ref{fig:coefficients_dephasing}. For the ``self-gravity'' corrections due to a power-law distribution, we considered ${\hat \alpha}=7/3$. (Other values induce corrections of the same order of magnitude.) The quantity $\langle\rho_{\rm DM}\rangle$ is the volume-average of the DM density profile $\rho(r)=\rho_0(R/r)^{\hat \alpha}$ between the initial and final orbital radii of the inspiral. Self-gravity effects of thin-accretion disks are modeled using a combination of Kusmin-Toomre and exponential disks discussed below Eq.\eqref{surfacedensity}.
For these cases, we consider an observation time of $T=$ 1 year and a final radius $r_f=6M$. DF and GP stand for ``dynamical friction'' and ``gravitational pull'', respectively.
}
\begin{tabular}{c|cc}
 \hline\hline
Correction           	&$|\delta_\varphi|/P$   &$P$               	\\
\hline
planetary migration & $<10^4$	& cf. Refs.~\cite{Yunes:2011ws,Kocsis:2011dr} \\
thin accretion disks (DF) & $\lesssim 10^2$ & $f_{\rm Edd}\left(\frac{0.1}{\alpha}\right) \left(\frac{\nu}{10^{-5}}\right)^{1/2} \left(\frac{M}{10^6 M_{\odot}}\right)^{-0.3}$ (cf. Sec.~\ref{dephasing_DF_disks})	\\
thin accretion disks (GP) & $\lesssim 10^{-3}$ & cf. Fig.~\ref{fig:dephasing_disks}	\\
magnetic field         &$10^{-4}$  &$\left(\frac{B}{10^{8}{\rm Gauss}}\right)^2\left(\frac{r_f}{6M}\right)^{9/2}\left(\frac{M}{10^6 M_\odot}\right)^{2}\frac{10^{-5}}{\nu}\frac{c_B(\chi)}{2538}$   \\
charge 		            &$10^{-2} $   &$\left(\frac{q}{10^{-3}}\right)^2\left(\frac{r_f}{6M}\right)^{3/2} \frac{10^{-5}}{\nu} \frac{c_q(\chi)}{-0.08}$  \\
gas accretion onto the central BH             &$10^{-2}$   &$f_{\rm Edd}\left(\frac{M}{10^6 M_\odot}\right)^{-5/8}\left(\frac{\nu}{10^{-5}}\right)^{-3/8}\left(\frac{\tau}{1{\rm yr}}\right)^{5/8}$         \\
thick accretion disks (DF)  & $10^{-9}$		&$\frac{f_{\rm Edd}}{10^{-4}}\left(\frac{0.1}{\alpha}\right)\left(\frac{\nu}{10^{-5}}\right)^{0.48} \left(\frac{M}{10^6 M_{\odot}}\right)^{-0.58}$(cf. Sec.~\ref{dephasing_DF_disks})\\
DM accretion onto central BH		& $10^{-8}$		& $\left( \frac{M}{10^6 M_\odot}  \right)\left( \frac{\langle\rho_{\rm DM}\rangle}{10^3 M_\odot {\rm pc}^{-3}}  \right) 
\left(\frac{T}{1\, {\rm yr}}\right) \left( \frac{{\sigma_v}}{220\, {\rm km}/{\rm s}}  \right)^{-1}$ \\
thick accretion disks (GP)  &$10^{-11}$	&$\frac{f_{\rm Edd}}{10^{-4}} \left(\frac{r_f}{6M}\right)^{4} \left(\frac{M}{10^6 M_\odot}\right)^2\frac{10^{-5}}{\nu}\frac{0.1}{\alpha}\frac{c_{\hat \alpha=3/2}(\chi)}{0.3}$	   \\
DM distribution (DF)	& $10^{-14}$	& $\left(\frac{\langle\rho_{\rm DM}\rangle}{10^3 M_\odot/{\rm pc}^3}\right)\left(\frac{\nu}{10^{-5}}\right)^{0.65} \left(\frac{M}{10^6 M_{\odot}}\right)^{0.17}$ \\
DM distribution $\rho\sim r^{-{\hat \alpha}}$ (GP)   & $10^{-16} $	&$\left(\frac{R}{7\times 10^6 M}\right)^{\hat \alpha} \frac{\langle\rho_{\rm DM}\rangle}{10^3 M_\odot/{\rm pc}^{3}}\left(\frac{r_f}{6M}\right)^{11/2-{\hat \alpha}} \left(\frac{M}{10^6 M_\odot}\right)^{2}\frac{10^{-5}}{\nu}\frac{c_{\hat \alpha}(\chi)}{0.15}$	   \\
galactic DM halos 	 	&	$10^{-16}$ 	 &$\frac{\langle\rho_{\rm DM}\rangle}{10^{3}{M_\odot}/{\rm pc}^3}\left(\frac{r_f}{6M}\right)^{11/2}\left(\frac{M}{10^6 M_\odot}\right)^{2}\frac{10^{-5}}{\nu}\frac{c_\Lambda(\chi)}{68}$	\\
cosmological constant &$ 10^{-26}$   &$\frac{\Lambda}{10^{-52}{\rm m}^{-2}}\left(\frac{r_f}{6M}\right)^{11/2}\left(\frac{M}{10^6 M_\odot}\right)^{2}\frac{10^{-5}}{\nu} \frac{c_\Lambda(\chi)}{68}$ \\
\hline\hline
\end{tabular}
\label{tab:dephasing}
\end{table}
%

\section{The effects of matter distributions in the two-body inspiral}\label{sec:matterinspiral}
In this section we estimate how accretion, self-gravity, dynamical friction and planetary migration affect the GW signal in a binary inspiral. 
More specifically, we will consider both comparable-mass binaries ($\nu\sim 1/4)$ and EMRIs ($\nu\ll1$), and analyze the effects of baryonic matter (i.e.
gas in accretion disks) and DM on these systems, comparing their magnitude with that of GW emission and self-force effects~\cite{Barack:2009ux,Poisson:2011nh}. The gravitational self-force acting on a particle in orbit around a large BH introduces corrections to the observables that, at first order, are $\sim {\cal O}(\nu)$, and which include in particular the backreaction of the GW fluxes on the dynamics. Second-order self-force effects are instead $\sim{\cal O}(\nu^2)$. Thus, if the relative corrections due to environmental effects with respect to the vacuum case are roughly larger than ${\cal O}(\nu)$ [${\cal O}(\nu^2)$], such effects would also be stronger than first (second) order self-force corrections and may have important implications for the dynamics of the inspiral.

The computation of some observables (periastron shift, GW dephasing and changes in the orbital motion) is detailed in the remaining sections of this part. In this section, we explicitly restore the factors $G$ and $c$ for clarity.
\subsection{The effect of accretion onto the binary's components}

It is particularly easy to estimate the effect of accretion on the two BHs constituting a binary system, 
and therefore on their orbital evolution and GW emission.  
As in the previous section, we can parametrize the mass accretion rate with the (mass) Eddington ratio
$f_{\rm Edd}$, and thus write the change in the BH masses $\Delta M$ during a typical space-based mission's lifetime $\Delta t \sim 1$ year
as
\be
\frac{\Delta M}{M}=\frac{\dot{M}\Delta t}{M}=2.2\, f_{\rm Edd} \times 10^{-8}\,,\label{eq:accr}
\ee
independently of the mass of the BH. This calculation therefore
applies to both comparable-mass binaries and to EMRIs (i.e., in the latter case Eq.~\eqref{eq:accr} applies both to the central BH and to the satellite).
A similar order of magnitude calculation applies to the spins of the BHs.
This simple analysis shows that for $f_{\rm Edd}\gtrsim 10^{-4}$ (which marginally includes also SgrA$^{*}$), the relative accreted mass is $\Delta M/M\gtrsim 10^{-12}$. This suggests that the effects of accretion might be comparable to or larger than second-order self-force corrections for a typical EMRI with $M\sim4\times 10^6 M_\odot$ and $\nu\sim 10 M_\odot/M\sim 2\times 10^{-6}$, while clearly accretion effects
will be completely negligible for comparable-mass binaries compared to GW emission, because ${\Delta M}/{M}\ll {\cal O}(\nu)\sim1/4$.

\subsection{Gravitational effect of matter}\label{sec:disks}
To estimate the order of magnitude of the effect of matter on the orbital evolution of binaries,
let us note that because of the conservation of mass, for a steady-state accretion disk
\be\label{continuity_eq}
\dot{M} = 2 \pi r  H \rho v_r\,,
\ee
where $\rho$ is the density, $r$ the radius, $H$ the height of the disk, $v_r$ the radial velocity and $\dot{M}=$ const is the steady-state accretion rate. Note
that this equation must be valid at all radii $r$. Also, the radial velocity is related to the speed of sound $v_s\sim (p/\rho)^{1/2}$ 
(where $p$ is the pressure) by~\cite{2002apa..book.....F,shakura_sunyaev}
\be\label{vr}
v_r\sim \frac{\alpha v_s H}{r}\,,
\ee
where $\alpha\sim 0.01 - 0.1$ is the viscosity parameter, and the height 
of the disk is given by~\cite{2002apa..book.....F,shakura_sunyaev}
\be\label{height}
H\sim \frac{v_s r}{v_K}\,,
\ee
where $v_K\approx (G M/r)^{1/2}$  is the local Keplerian velocity.

For $f_{\rm Edd}\lesssim 10^{-2}$ or $f_{\rm Edd}\gtrsim 0.2$, 
the disk is geometrically thick or slim ($H\sim r$) and close to the virial temperature, 
so the speed of sound $v_s$ is comparable to the Keplerian velocity,
consistently with Eq.~\eqref{height}, and $v_r\sim \alpha v_K$ from Eq.~\eqref{vr}.
Equation~\eqref{continuity_eq}, Eq.~\eqref{Medd} and $\dot M=f_{\rm Edd} \dot M_{\rm Edd}$ then imply 
\be
\rho \sim 3.4\times 10^{-6} 
\left(\frac{0.1}{\alpha}\right) \left(\frac{10^6 M_\odot}{M}\right) \frac{f_{\rm Edd}}{\tilde{r}^{3/2}} \mbox{ kg}/\mbox{m}^3 \,,\label{rhoThick}
\ee
where $\tilde{r}=r/\left(GM/c^2\right)$. 

For geometrically thin disks such as those suitable for describing systems with $10^{-2} \lesssim f_{\rm Edd}\lesssim 0.2$,
one can solve the equations describing the disk's structure exactly in Newtonian theory and in a steady-state regime~\cite{2002apa..book.....F,shakura_sunyaev}: 
\be
\rho\approx 169 \frac{f_{\rm Edd}^{11/20}}{\tilde{r}^{15/8}} \left(1-\sqrt{\frac{\tilde{r}_{\rm in}}{\tilde{r}}}\right)^{11/20}\,  \left(\frac{0.1}{\alpha}\right)^{7/10} \left(\frac{10^6 M_\odot}{M}\right)^{7/10} \mbox{ kg}/\mbox{m}^3\,, \label{eq:rhoF}
\ee
\be\label{eq:H}
\frac{H}{GM/c^2}\approx 3\times 10^{-3} f_{\rm Edd}^{3/20} \left(1-\sqrt{\frac{\tilde{r}_{\rm in}}{\tilde{r}}}\right)^{3/20}\,  \left(\frac{0.1}{\alpha}\right)^{1/10} \left(\frac{10^6 M_\odot}{M}\right)^{1/10} {\tilde{r}^{9/8}} \,,
\ee
where $\tilde{r}_{\rm in}\sim 6$ is the radius of the inner edge of the disk in gravitational radii.

The calculation of the Newtonian potential of a thin disk with arbitrary surface density can be a complicated problem. However, certain
functional forms for the surface density yield an analytical expression for the Newtonian potential on the equatorial plane, e.g. the Kusmin-Toomre model and the exponential disk model
used to describe galactic disks~\cite{2000MNRAS.316..540C}:
\bea
\Sigma(r)_{\rm Kusmin}&=&\frac{M_{\rm disk}}{2\pi R^2\left(1+(r/R)^2\right)^{3/2}}\,,\qquad 
\Phi(r)_{\rm Kusmin}=-\frac{GM}{r} - \frac{G M_{\rm disk}}{\sqrt{r^2 + R^2}}\,, \label{disk1}\\
\Sigma(r)_{\rm exp}&=&\frac{M_{\rm disk}}{2\pi R^2}e^{-r/R}\,,\qquad 
\Phi(r)_{\rm exp}=-\frac{GM}{r} - 
\frac{G M_{\rm disk}}{2R^2} r\left(I_0[y]K_1[y]-I_1[y]K_0[y]\right)\,,\label{disk2}
\eea
where $\Sigma$ and $\Phi$ are the surface density and the Newtonian potential, $r$ is the distance from the BH along the disk plane, $y=r/(2R)$, $I_n,\,K_n$ are modified Bessel functions 
of the first and second kind, and $M_{\rm disk}$ and $R$ are two free parameters (respectively the total mass -- if the disk extends from $r=0$ to $r\to\infty$ -- and its scale radius). While neither of
these models alone can be tuned to approximate the surface density given by Eqs.~\eqref{eq:rhoF} and \eqref{eq:H}, i.e.
\begin{equation}
\Sigma_{\rm disk}\approx \rho H\sim 7\times 10^8 \frac{f_{\rm Edd}^{7/10}}{{\tilde r}^{3/4}}  \left(1-\sqrt{\frac{\tilde{r}_{\rm in}}{{\tilde r}}}\right)^{7/10}   \left(\frac{0.1}{\alpha }\right)^{4/5} \left(\frac{M}{10^6 M_\odot}\right)^{1/5}   {\rm kg~m^{-2}}     \,,\label{surfacedensity}
\end{equation}
we find that a combination of the two can. Indeed, we find that the superposition of a Kusmin-Toomre disk with 
$M_{\rm disk}\approx 63 M_\odot {f_{\rm Edd}}^{7/10} [M/(10^6 M_\odot)]^{11/5}(\alpha/0.1)^{-4/5}$ and $R\approx1016 GM/c^2$ with an exponential
disk with $M_{\rm disk}\approx 1.5 M_\odot {f_{\rm Edd}}^{7/10}\times$ $ [M/(10^6 M_\odot)]^{11/5}(\alpha/0.1)^{-4/5}$ and $R\approx74 G M/c^2$ approximates well (for our purposes) the surface density of a thin disk as given in Eqs.~\eqref{eq:rhoF} and \eqref{eq:H}. We will therefore
use this composite model to calculate the Newtonian potential of a thin accretion disk, and we will superimpose to it a spherically symmetric potential describing the BH.

Before venturing into such a detailed calculation, however, let us estimate the expected magnitude of this effect. For an EMRI with separation $r$
of a few gravitational radii (i.e. $\tilde{r} \sim{\cal O}(1)$), the time needed for the satellite to fall into the central massive BH is $\Delta t\sim (M/m_{\rm sat}) (GM/c^3)$, 
which is comparable to or smaller
than $\sim 1$ yr for $M\sim 10^6 M_\odot$ and $m_{\rm sat}= 1$ -- $10 M_\odot$. Therefore, during the lifetime of an eLISA-like mission the separation of a typical EMRI changes by $\Delta \tilde{r} \sim {\cal O}(1)$. Let us
then calculate the change $\Delta M$ in the mass of the accretion disk contained in a radius $r$ 
during the mission lifetime.
(At least at lowest order in a monopolar expansion, the matter 
at radii larger than the radius $r$ of the EMRI exerts no force on the satellite.) Therefore, using the densities given above, we find
\be
\frac{\Delta M}{M} \sim \frac{2 \pi \rho r H  \Delta{r}}{M} \sim
 5\times 10^{-9} \left(\frac{0.1}{\alpha}\right)^{4/5} \left(\frac{M}{10^6 M_\odot}\right)^{6/5} f_{\rm Edd}^{7/10} 
 \left(1-\sqrt{\frac{\tilde{r}_{\rm in}}{\tilde{r}}}\right)^{7/10} \tilde{r}^{1/4} {\Delta \tilde{r}}\,,
\ee
for a thin-disk model
and
\be\frac{\Delta M}{M} \sim \frac{2 \pi \rho r H \Delta{r}}{M} \sim 
3.5\times 10^{-14} \left(\frac{0.1}{\alpha}\right) \left(\frac{M}{10^6 M_\odot}\right) {f_{\rm Edd}}{\tilde{r}^{1/2}} \Delta \tilde{r}\,,
\ee
for a thick-disk model. Therefore, it is clear that this effect can be more important than the second-order self force for thin disks, but is negligible in the thick disk case even relative to second-order self-force corrections.
Note that these results, although derived in the extreme-mass ratio case, can be extrapolated (at least as orders of magnitude) to comparable-mass binaries by
taking $\tilde{r}\sim\Delta \tilde{r}$ to be the separation at which such systems enter the eLISA band, i.e. $\tilde{r} \sim \Delta \tilde{r}\sim 200$. 
It is thus clear that these gravitational effects
are completely negligible for comparable-mass binaries, since ${\Delta M}/{M}\ll {\cal O}(\nu)\sim 1/4$. 
\subsection{The effect of dynamical friction} \label{sec:DF}
The easiest way to compare the effect of dynamical friction to that of the self-force is through the changes induced 
on the energy of the satellite. The self-force is known to preserve energy balance and to decrease the satellite's energy exactly by the energy carried away 
by GWs~\cite{Gal'tsov:2010cz}. The GW energy flux is given, at Newtonian order, by the quadrupole formula~\cite{Peters:1963ux}
\begin{equation}
\dot{E}_{\rm GW}=\frac{32}{5} \frac{G^4}{c^5} \nu^2 \left(\frac{M}{r}\right)^5\,.\label{quadrupole0}
\end{equation}
%
Dynamical friction exerts a drag force in the direction of motion, and its magnitude is given, at Newtonian order\footnote{See Ref.~\cite{Barausse:2007ph} for the special-relativistic corrections to this formula.}, by
\begin{equation}
F_{\rm DF}= \frac{4\pi\rho (G m_{\rm sat})^2}{v^2} I
\end{equation}
where $v$ is the satellite's velocity relative to the gas (which we will take, as a rough approximation, to be given by the Keplerian velocity), and $I$ is given, for a satellite in circular motion with radius $r$, by~\cite{Kim:2007zb}
\begin{equation}\label{Idef}
  I = \left\{\begin{array}{l l@{\ }r@{\;}c@{\,}l}
     0.7706\ln\left(\frac{1+{\cal M}}{1.0004-0.9185{\cal M}}\right) -1.4703{\cal M},
     &\mbox{ for }&&{\cal M}&<1.0,\\
     \ln\left[330 (r/r_{\min}) ({\cal M}-0.71)^{5.72}{\cal M}^{-9.58} \right],
     &\mbox{ for }&1.0\leq&{\cal M}&<4.4, \\
     \ln\left[(r/r_{\min})/(0.11{\cal M}+1.65)\right],
     &\mbox{ for }&4.4\leq&{\cal M}, 
  \end{array}\right.
\end{equation}
Here, ${\cal M}\equiv v/v_s$ is the Mach number, and $r_{\min}$ is an unknown fitting parameter which we take to be the capture impact parameter of the satellite, e.g. for a 
BH, $r_{\min}\sim 2 G m_{\rm sat} [1+(v/c)^2]/v^2$. Our results show a very mild dependence on the exact value of $r_{\min}$.

The energy lost by the satellite because of dynamical friction is therefore
\begin{equation}
\dot{E}_{\rm DF}= F_{\rm DF} v_K \sim 4 \pi \rho \frac{(G m_{\rm sat})^2}{v_K} I  \bar{K} \,, \label{dotEDF}
\end{equation}
where $\bar{K}$ is the fraction of time the satellite spends inside the accretion disk (e.g. $\bar{K}=1$ for orbit on the equatorial plane, and
$\bar{K}\sim H/r$ for more generic orbits), while $I$ is given by Eq.~\eqref{Idef} with ${\cal M}\sim v_K/v_s \sim r/H \gg1 $ in the case of a thin disk, or
with   ${\cal M}\sim v_K/v_s \sim r/H\sim 1$ for thick disks. 
Using the density estimates above, we obtain
\begin{equation}\label{DFthin}
\frac{\dot{E}_{\rm DF}}{\dot{E}_{\rm GW}}\sim
5\times 10^{-7} f_{\rm Edd}^{11/20} \left(\frac{M}{10^6 M_\odot}\right)^{13/10} 
 \left(1-\sqrt{\frac{\tilde{r}_{\rm in}}{\tilde{r}}}\right)^{11/20} \left(\frac{0.1}{\alpha}\right)^{7/10}  \tilde{r}^{29/8} I  \bar{K}
\end{equation}
for thin disks, and
\begin{equation}\label{DFthick}
\frac{\dot{E}_{\rm DF}}{\dot{E}_{\rm GW}}\sim  10^{-14} f_{\rm Edd} \left(\frac{M}{10^6 M_\odot}\right) \left(\frac{0.1}{\alpha}\right) \tilde{r}^4 I \bar{K}
\end{equation}
for thick disks. This shows, in agreement with the results of Ref.~\cite{Barausse:2007dy}, that in thin disks dynamical friction may
be as important as GW emission (i.e. as important as the first-order gravitational self-force) in EMRIs at $\tilde{r}\gtrsim 40$,
but is subdominant compared to it at smaller separations, which are the most relevant for eLISA. Nevertheless, the consequent dephasing may
have an impact on GW detections, cf. Section~\ref{dephasing_DF_disks}.

If one considers the second-order self-force, dynamical friction clearly dominates over it for EMRIs in thin disks, essentially at all separations.
However, dynamical friction is always negligible relative to the second-order self force for thick disks at all separations
$\tilde{r}\lesssim 800$. 

We stress that our results can be extrapolated (at least as orders of magnitude) to comparable-mass binaries by taking $\nu\sim 1/4$ and replacing $M\to M_T$.
Therefore, for these systems the effect of dynamical friction is completely negligible compared to GW emission
in the case of thick disks, because at separations inside the eLISA
band (i.e. $\tilde{r}\lesssim 200$), ${\dot{E}_{\rm DF}}/{\dot{E}_{\rm GW}}\ll 1$.
In the case of thin disks, instead, dynamical friction has a magnitude comparable to GW emission in the early stages of the inspiral, but becomes smaller at separations $\tilde{r}\lesssim 60-70$ (but see Section~\ref{dephasing_DF_disks}). However, it is likely that any thin-disk structure would be destroyed by a massive BH binary. Also, in any case, the inspiral would probably not happen inside the thin disk, 
since the massive BH's size would be typically larger than the disk's height $H$, thus resulting in $\bar{K}\sim0$ in Eq.~\eqref{DFthin}.
Nevertheless, we note that the thin-disk estimates above can be useful to gauge the order of magnitude of the effect of the circumbinary disks that are believed to form 
around massive BH binaries in gas-rich galactic mergers~\cite{circumbinary_disks}.
Using the results above, in Secs.~\ref{dephasing_DF} and \ref{dephasing_DF_disks} we will 
estimate the GW dephasing introduced by dynamical friction more precisely, and compare it to that due to accretion and self-gravity effects.

\subsection{The effect of planetary migration}
\label{sec:migration}
The physical mechanism behind dynamical friction is the gravitational pull exerted on the satellite by the density perturbations (the ``wake'') excited in the gaseous medium
by the satellite itself. Dynamical friction, however, neglects the effect of the differential motion of the various annuli of the accretion disk. More specifically, 
it does not account for the fact that the part of the wake that lies at larger radii than the satellite lags behind it, 
while the part that lies at smaller radii trails the satellite. (This is because, at least for quasi-Keplerian disks, the rotational velocity is a decreasing function of the radius). 
As a result, the wake in the region ``exterior'' to the satellite's orbit will tend to reduce the orbit's angular momentum
--~thus causing the satellite to sink, while the ``inner'' part of the wake will tend to increase the orbit's angular momentum and stall the infall of the satellite. 
This effect is known as ``planetary migration'', and whether this will cause the satellite to migrate inwards or outwards depends on the detailed balance between the aforementioned effects,
and requires quite technical calculations (see e.g. Refs.~\cite{Ward:1997di,Armitage:2007gn} for a review). In particular, under appropriate conditions the satellite can create a gap in the disk. Namely, this happens if the  Hill sphere (or Roche radius) of the satellite becomes comparable to the height of the disk, i.e.~\cite{Armitage:2007gn}
\begin{equation}
\left(\frac{m_{\rm sat}}{3 M}\right)^{1/3} r\gtrsim H
\end{equation}
and if tidal torques remove gas from the gap faster than viscosity can refill it, i.e.~\cite{Armitage:2007gn}
\begin{equation}
\frac{m_{\rm sat}}{M} \gtrsim \alpha^{1/2} \left(\frac{v_s}{v_K}\right)^2\,.
\end{equation}
If such a gap forms,  the process
is dubbed ``type-II'' migration, i.e. the satellite moves inwards on the disk's viscous time scale, and loses angular momentum with rate~\cite{Syer:1995hk,Yunes:2011ws,Kocsis:2011dr}
\begin{equation}
\dot{L}_{\rm migr\,II}\approx m_{\rm sat} \left(\frac{4 \pi r^2 \Sigma}{m_{\rm sat}}\right)^{3/8} \frac{v_K v_r}{2}\,,
\end{equation}
where $\Sigma = H \rho$ is the surface density of the disk.
If instead a gap is not formed, 
this process goes under the name of ``type-I'' planetary migration, and the change in the angular momentum of the satellite is~\cite{2002ApJ...565.1257T}
\begin{equation}
\dot{L}_{\rm migr\,I}\approx\pm 0.65 \left(\frac{m_{\rm sat}}{M}\right)^2  \left(\frac{v_K}{v_s}\right)^2 \Sigma\, (v_K r)^2\,,
\end{equation}
where the $\pm{}$ sign reflects the stochastic nature of the process in a turbulent disk. 
Clearly, in the case of thick disks only type I migration can occur, while in thin disks both type-I and type-II migration can take place.\footnote{In the
context of planetary dynamics, a third type of migration occurs, but this type-III migration does not take place in the case of EMRIs~\cite{Yunes:2011ws,Kocsis:2011dr}.}
 Note that for AGN thin accretion disks, EMRIs are expected to open gaps at large separations and close them as the satellite gets closer to the massive BH. This is
the opposite behavior as in protoplanetary disks, and it is due to the fact that radiation pressure makes the height $H$ of AGN disks roughly constant~\cite{Kocsis:2011dr}. 

Using the density estimates for thin and thick disks given above, and the fact that the loss of angular momentum through GWs
is $\dot{L}_{\rm GW}= \dot{E}_{\rm GW}/\Omega_\phi$, we find
\begin{eqnarray}\label{eq:migr1}
\left(\frac{\dot{L}_{\rm migr\,I}}{\dot{L}_{\rm GW}}\right)_{\rm thin}&=& 10^{-5}
f_{\rm Edd}^{2/5} \left(\frac{M}{10^6 M_\odot}\right)^{7/5} \left(1-\sqrt{\frac{\tilde{r}_{\rm in}}{\tilde{r}}}\right)^{2/5}
\left(\frac{\alpha}{0.1}\right)^{-3/5} \tilde{r}^{7/2}\,,\\
\left(\frac{\dot{L}_{\rm migr\,II}}{\dot{L}_{\rm GW}}\right)_{\rm thin}&=&
2\times 10^{-4}
f_{\rm Edd}^{9/16} \left(\frac{M}{10^6 M_\odot}\right)^{1/4} \left(1-\sqrt{\frac{\tilde{r}_{\rm in}}{\tilde{r}}}\right)^{-7/16}
\left(\frac{\alpha}{0.1}\right)^{1/2} \left(\frac{\nu}{10^{-5}}\right)^{-11/8} \tilde{r}^{103/32}\,,\\
\label{typeI_thick}
\left(\frac{\dot{L}_{\rm migr\,I}}{\dot{L}_{\rm GW}}\right)_{\rm thick}&=&
6\times 10^{-16} f_{\rm Edd}  \left(\frac{M}{10^6 M_\odot}\right) \left(\frac{0.1}{\alpha}\right) \tilde{r}^4\,.
\end{eqnarray}
Therefore, for thin-disk accretion the effect of planetary migration can even be comparable to the first-order self-force in EMRIs (at least for satellites embedded in the disk).
More specifically, using reasonable values for EMRIs, e.g.
$M\sim10^6 M_{\odot}$, $\nu\sim 10^{-5}$, $f_{\rm Edd}\sim 0.1$, $\tilde{r}_{\rm in}\sim 6$, $\alpha\sim 0.1$, 
in the equations above, one finds that type-I and type-II migration dominates over the GW fluxes respectively
for $\tilde{r}\gtrsim 35$ and $\tilde{r}\gtrsim 18$, and they are larger than second-order self-force effects essentially at all separations.
This is in full agreement with the results of Refs.~\cite{Yunes:2011ws,Kocsis:2011dr}, where it was first noted that the effect of planetary migration 
in EMRIs embedded in thin disks can be comparable to that of the GW fluxes.

For thick disks, instead, Eq.~\eqref{typeI_thick} resembles very much Eq.~\eqref{DFthick} for dynamical friction, so as in that case
planetary migration is negligible up to separations of several thousand gravitational radii, even relative to the second-order self-force in EMRIs.
Again, this can be verified by using reasonable values $M\sim10^6 M_{\odot}$, $\nu\sim 10^{-5}$, $\alpha\sim 0.1$ and $f_{\rm Edd} \sim 10^{-4}$.

As in the dynamical-friction case, Eqs. \eqref{eq:migr1}--\eqref{typeI_thick} can be extrapolated (as orders of magnitude) to massive BH binaries by 
$M\to M_T$ and $\nu\sim1/4$, and considering separations $\tilde{r}\lesssim 200$ at which these systems emit in the eLISA band. 
As such, it is clear that planetary-migration effects are completely negligible relative to
GW emission in the thick disk case, while thin disks are probably destroyed by  massive BH binaries, for the same reasons that we mentioned in the dynamical friction case. However,
as already mentioned, one can interpret the thin disk as a model for the circumbinary disks that are expected to form around massive BH binaries in gas-rich environments~\cite{circumbinary_disks}.
In this latter case, the equations above show that the binary's inspiral may be driven by type-I planetary migration for $\tilde{r}\gtrsim 35$, while
type-II migration is important only before the binary enters the eLISA band.

\subsection{Gravitational interaction with stars}
The interaction of BH binaries (both comparable-mass and extreme mass ratio ones) with stars typically has negligible effects on the 
waveforms. This is because the binary's separation when it enters the eLISA band ($\sim 100-200$ gravitational radii for
a comparable mass binary, and even smaller values for EMRIs) is tiny compared to the average distance between stars (even in the high-density environment of
galactic centers). More precisely, the typical time scale of two-body interactions in galactic centers is $\sim 10^9$ yrs, hence a scattering event during the final year-long inspiral
is extremely unlikely. Equivalently, the typical time scale of two-body interactions becomes comparable to the GW-emission time scale only at separation of $\sim 0.01$ pc, much larger than the radius 
at which the binary enters the eLISA band. 
Still, Ref.~\cite{pau} showed that for a fraction (perhaps a few percent) of the EMRIs detectable with eLISA,
a star may be close enough to the binary system to significantly alter the binary's orbital evolution and GW emission. The fraction
of events for which such an effect might happen, however, depends strongly on the presence of a stellar cusp around the massive
BH, which is still debated even in the case of our own Galaxy~\cite{merritCore} (where for instance proper motion
data seem to point at a flat or declining stellar density toward SgrA$^*$~\cite{schoedel}).

As a side note, gravitational effects that depend on scales larger than those considered here may become sensitive to the galactic stellar content. This is the case for instance for GW detection with pulsar timing arrays, which probe smaller frequencies and therefore larger separations. A recent study does indeed suggest that stellar environments may affect detection in those setups~\cite{Ravi:2014aha}.

Finally, we also stress that interaction of an EMRI with a nearby massive BH (to within a few tenths of a pc) might also produce a detectable impact
on eLISA waveforms~\cite{nico}. The fraction of EMRIs subject to this effect is very uncertain, but may be on the order of a few percent (see the discussion in Ref.~\cite{nico}).

\subsection{Prospects for eLISA-like missions}\label{sec:prospects}
Future space-based detectors such as eLISA will observe EMRIs only at redshifts $z\lesssim0.7$~\cite{Gair:2008bx,Seoane:2013qna}. In the local universe, however, the fraction of massive
BHs accreting at high Eddington ratios is expected to be very small, as galactic nuclei are typically quiescent rather than active. More precisely, Ref.~\cite{Heckman:2004zf}
analyzes the distribution of Eddington ratios of BHs in local Sloan Digital Sky Survey galaxies, and finds that only a few percent of the BHs with masses $\sim 3\times 10^6 M_\odot$
are accreting with $f^L_{\rm Edd}\approx f_{\rm Edd}\gtrsim 10^{-2}$ (see Fig. 3 in Ref.~\cite{Heckman:2004zf}). Therefore, only a few percent of all EMRI events detectable by eLISA 
are expected to happen in
thin-disk environments. Qualitatively similar conclusions can be drawn from Ref.~\cite{Kelly:2010qc} (Fig. 7), which shows that at $z=1$ only a fraction $\lesssim 10^{-2}$ of BHs with mass $M\lesssim 10^9$
is in a quasar phase. 

Since eLISA is expected to detect between 5 and 50 EMRI events per year~\cite{Seoane:2013qna}, it is clear that detecting an EMRI event in a thin-disk environment will be 
rather unlikely. Therefore, for the astrophysical sources of interest for eLISA, matter effects should definitively be negligible compared to first-order self-force effects, and in most cases
also to second-order ones, with perhaps the exception of accretion, which as derived above can be comparable to  second-order self-force effects
if $f_{\rm Edd}\gtrsim 10^{-4}$.

Because the quasar luminosity increases from $z=0$ to its peak at $z\sim 2$ (see e.g. Ref.~\cite{Hopkins:2006fq}), one generally expects to have a larger fraction of BHs accreting at 
high Eddington ratios at higher redshifts, but making quantitative statements is more difficult as high-redshift samples will intrinsically be biased 
in favor of the most luminous objects. More importantly, what matters in order to attempt to estimate the effects of gas on eLISA sources at high redshift 
is not actually the Eddington ratio of \textit{isolated galactic nuclei}, but rather the accretion properties of massive BHs \textit{in binary systems}. In fact, eLISA will
be able to detect the inspiral, merger and ringdown signal of such systems up to high redshifts $z\sim 10$ and beyond, with expected event rates between 10 and 100 per 
year~\cite{Seoane:2013qna}. The accretion properties of binary systems
are even more difficult to observe as there are no direct electromagnetic observations of massive BH binaries (with the exception of dual AGNs, which are however
at kpc separations, much larger than the subparsec separations of eLISA sources; see e.g. Ref.~\cite{Comerford:2013fha} for a recent search of dual-AGN candidates, and references therein
for previously discovered candidates). Furthermore, the accretion properties of binaries are  
probably only weakly correlated to those  of isolated
galactic nuclei, because major galactic mergers (i.e. those between comparable-mass galaxies) are thought to trigger bursts of star formation and to feed gas
to the binary BH systems that form after the two galaxies merge. Some indications about accretion in binaries, however, come from the models of galaxy formation
that are used, for instance, to estimate massive BH merger event rates for eLISA. Ref.~\cite{2012MNRAS.423.2533B} presented a semianalytical galaxy formation model
tracking the evolution of the baryonic component of galaxies (stars and gas in spheroids and disks), as well as that of massive BHs, along dark-matter merger
trees. This model was validated by comparing its predictions to galactic and quasar observations in the local $z=0$ universe and at higher redshifts $z\lesssim 7$.
Because it allows one to track the coevolution between the massive BHs and their galactic hosts, this model was used to estimate how many BH mergers happen
in gas rich or in gas poor scenarios, finding that the fraction of gas-rich mergers generally tends to decrease with redshift\footnote{This can be understood by the fact that galactic nuclei get drier and drier as
gas is consumed by star formation, accretion and AGN feedback.} but remains generally non-negligible.  In particular, the predicted fraction of BH mergers 
happening in gas-rich environments is shown in Figs. 10 and 11 of  Ref.~\cite{2012MNRAS.423.2533B}.
Clearly, while the exact fraction depends on the model's details (e.g. on the choice of the seeds of the BH population at high redshifts),
it is safe to conclude that
the thin-disk estimates worked out in the previous sections may apply to a sizable fraction of the massive BH inspirals, mergers
and ringdown that will be detectable with eLISA. 

Thus, the results derived in the previous sections suggest that the effect of matter on binaries of
massive BHs can be safely ignored for a detection with eLISA. A possible exception is given by inspiralling binaries at separations $\gtrsim 60-70$ gravitational radii in a thin-disk (i.e. gas-rich) environment. 
In these systems the effect of dynamical friction and planetary migration may be comparable to or even larger than GW emission, while the GW signal may still be strong enough to be detected with eLISA.

\subsection{The effects of Dark Matter}\label{sec:DM}

Let us first consider the effect of DM for a binary of massive BHs with comparable masses (i.e. $\nu\sim1/4$) and total mass $M_T$ during the inspiral. We recall that these are the systems where we expect QNMs to be excited during mergers.
To obtain the order of magnitude of the purely gravitational (i.e. self-gravity) effect of  DM we can then assume a roughly constant DM density given by \eqref{rhoDM},
and calculate the change $\Delta M$ in the mass contained between the two BHs during eLISA's lifetime, i.e. from the separation $\tilde{r}\sim 200$
at which the binary enters the eLISA band until the merger-ringdown. One then obtains
\begin{equation}\label{dm1}
\frac{\Delta M}{ M_T} \sim 5\times 10^{-19} \left( \frac{M_T}{10^6 M_\odot}  \right)^2\left( \frac{\tilde{r}}{100}  \right)^3 
\left( \frac{\rho_{\rm DM}}{10^3 M_\odot {\rm pc}^{-3}}  \right)\,, 
\end{equation}
which is clearly negligible for all practical purposes. 

The nature of DM accretion onto BHs can be assessed as follows. If the horizon radius of the BH, $r_{+}\sim 2GM/c^2$, is much smaller than the mean free path $\ell$, 
then DM can be considered as effectively collisionless. On the other hand, if $r_{+}\gtrsim \ell$, cohesion forces and matter compressibility 
have to be taken into account~\cite{Shapiro:1983du} and accretion is described by the Bondi-Hoyle formula. 
Considering that $\ell=(n\sigma_{\rm DM})^{-1}$, where $n$ is the particle density and $\sigma_{\rm DM}$ is the scattering cross section, we find that the condition for collisionless behavior reads
\begin{equation}
 \frac{\sigma_{\rm DM}/(10^{-40}{\rm cm^2})}{m_{\rm DM}/{\rm GeV}}\ll 9\times 10^{14} \left(\frac{10^9 M_\odot{\rm pc}^{-3}}{\rho_{\rm DM}}\right) \left(\frac{10^6 M_\odot}{M}\right)\,, \label{BondiVScollisionless}
\end{equation}
where $m_{\rm DM}$ is the mass of the DM particle and we have normalized the result for typical values of the ratio ${\sigma_{\rm DM}}/{m_{\rm DM}}$. Note that we have considered a very large DM density, close to the annihilation limit discussed in Sec.~\ref{sec:dm_intro}. Even in the most favorable scenarios for Bondi-Hoyle accretion 
(large DM density and large accreting object), it is clear that DM can be treated as a collisionless fluid, 
unless either the DM cross section is ridiculously large, or the mass of DM particles is very small. An example of the latter situation are models of ultralight DM fields, like axions with masses of the order of eV or smaller~\cite{Arvanitaki:2009fg}. The effect of Bondi-Hoyle accretion of these particles onto a BH was studied in Ref.~\cite{Macedo:2013qea}.

Accretion of DM onto BHs is then regulated by the collisionless 
accretion rate for nonrelativistic particles, i.e. $\dot{M} = 16 \pi (G M)^2 \rho_{\rm DM}/(v_{\rm DM} c^2)$~\cite{Shapiro:1983du}. Here, 
$v_{\rm DM}$ is the \textit{typical} velocity of the  DM particles relative to the BH, i.e. $v_{\rm DM}=\max(v, \sigma_v)$,
where $\sigma_v=\sqrt{\langle v^2\rangle}$ is the  velocity dispersion of the DM particles, and $v$ is the 
BH velocity in the DM halo reference frame. 
The change $\Delta M$ in each BH's mass in a time $T$ is therefore
\begin{equation}\label{dm2}
\frac{\Delta M}{ M} \sim 4\times 10^{-16} \left( \frac{M}{10^6 M_\odot}  \right)\left( \frac{\rho_{\rm DM}}{10^3 M_\odot {\rm pc}^{-3}}  \right) 
\left(\frac{T}{1\, {\rm yr}}\right) \left( \frac{\tilde{r}}{100}  \right)^{1/2}\,,
\end{equation}
where we have used $v\sim c/ {\tilde{r}}^{1/2} \gg \sigma_v$. Thus, this effect is also negligible.

Let us now turn to dynamical friction. The latter produces a 
drag force $F_{\rm DF}\sim (4\pi\rho_{\rm DM} \nu^2 G^2 M_T^2/{v^2})\times \ln\ ({r}/{r_{\rm min}})$ ($r$ being the orbit's radius). Comparing 
the energy lost by the BH due to this drag, $\dot{E}_{\rm DF}\sim F_{\rm DF} v$, to the energy lost to GWs in the quadrupolar approximation, $\dot{E}_{\rm GW}\approx({32}/{5}) (c^5/G) (\nu^2/ \tilde{r}^{5})$, one finds
\begin{equation}\label{dm3}
\frac{\dot{E}_{\rm DF}}{ \dot{E}_{\rm GW}} \sim 2\times 10^{-14} \left( \frac{M_T}{10^6 M_\odot}  \right)^2 \left( \frac{\rho_{\rm DM}}{10^3 M_\odot {\rm pc}^{-3}}  \right) \left( \frac{\tilde{r}}{100}  \right)^{11/2} \ln\left (\frac{r}{r_{\rm min}}\right)\,,
\end{equation}
where we have used again $v\sim c/{\tilde{r}}^{1/2}$. As can be seen, dynamical friction also turns out to be negligible for comparable mass binaries.

The situation for EMRIs is slightly more complicated. For these systems, the estimates given by Eqs.~\eqref{dm1} -- \eqref{dm3} still describe the purely
gravitational effect of DM as well as the effect of accretion and dynamical friction on the satellite,
provided that $M$ is interpreted as the mass of the satellite in Eq.~\eqref{dm2}, while $M_T$ is still the total mass of the binary in the other equations.
As can be seen, these effects are negligible relative to both the first and second order self force.
The effect of accretion onto the more massive BH (which is almost at rest in the DM halo's frame) 
is also easily calculated using again the collisionless accretion rate for nonrelativistic particles: 
\begin{equation}\label{dm2bis}
\frac{\Delta M}{ M} \sim 5\times 10^{-14} \left( \frac{M}{10^6 M_\odot}  \right)\left( \frac{\rho_{\rm DM}}{10^3 M_\odot {\rm pc}^{-3}}  \right) 
\left(\frac{T}{1\, {\rm yr}}\right) \left( \frac{{\sigma_v}}{220\, {\rm km}/{\rm s}}  \right)^{-1}\,,
\end{equation}
where we have used the fact that $\sigma_v\gtrsim v$. 
Taking $\sigma_v$ to be of the order of the DM halo's virial velocity (e.g. $\sigma_v\sim 220\, {\rm km}/{\rm s}$ for the Milky Way),
one gets an effect which is completely negligible relative to both the first and second-order self force.

As mentioned in Sec.~\ref{sec:dm_intro}, EMRIs in satellite galaxies might possibly live in environments with much higher DM density, up to
a plateau density of $\rho\sim 10^9 -10^{12} M_\odot/{\rm pc}^3$. However, even inserting this (probably unrealistically large) DM density in
Eqs.~\eqref{dm1} -- \eqref{dm2bis}, one obtains that the DM effects on the waveforms
 may be \textit{at most} barely comparable to first-order self-force effects.

\subsection{Interference and resonant effects}\label{sec:resonances_interference}

We conclude this section by discussing two further effects that were not taken into account by our analysis.
 The first one is the interference in the GW radiation emitted by an extended distribution of matter around a BH. This effect was studied in detail by Saijo and Nakamura~\cite{Saijo:2000pe,Saijo:2000ee} and by Sotani and Saijo~\cite{Sotani:2005be} considering noninteracting (dust) thin disks, obtained as superpositions of noninteracting thin rings. They find peculiar ``resonances'' in the energy spectrum, which carry information about the disk's parameters, such as the location~\cite{Sotani:2005be}. Depending on the location of the disk, a low-frequency peak appears in the spectrum and the corresponding waveform can be detectable by a typical space-based detector with sensitivity higher than about $10^{-22}$ in the mHz band. It would be interesting to extend this kind of study to the case of a binary inspiral.

In addition to this effect, in an EMRI one might also expect to excite the characteristic frequencies of the matter distribution. Such an analysis was not included in Refs.~\cite{Saijo:2000pe,Saijo:2000ee,Sotani:2005be} because the disk was made of noninteracting point particles. However, in more realistic situations the polar modes of the disk will be coupled to the gravitational perturbations and can be excited during the inspiral. This situation is akin to the case of the fluid modes in relativistic neutron stars~\cite{Kokkotas:1999bd}, which can be excited by a point particle (see e.g.~\cite{Pons:2001xs}). Although such detailed analysis is beyond the scope of our work, we anticipate that fluid modes of disks, shells, and other localized distributions will be generically excited during the inspiral and would correspond to sharp resonances in the energy flux, in addition to the new class of modes of the composite BH-matter system that were discussed in Part~\ref{part:ringdown}. The characteristic frequencies of extended matter distributions are smaller than the gravitational BH modes, so that they can be excited also by particles in quasi-circular orbits outside the ISCO. A detailed analysis of this process is left for future work.
 
\section{Orbital changes}\label{sec:orbitalchange}
Given the estimates above, it is important to model more precisely the effects of matter around compact binaries. In this section we start this analysis in the limit of extreme mass ratios, by computing the periastron precession and the change in the ISCO frequencies introduced by environmental effect. The next section is instead devoted to the study of GW observables.

\subsection{Periastron shift}\label{sec:periastron}
Periastron precession is caused by a mismatch between radial $\Omega_r$ and azimuthal $\Omega_\phi$ oscillation frequencies and can be computed via
\be
\delta\phi=2\pi\left(\frac{\Omega_\phi}{\Omega_r}-1\right) \label{defshift}
\ee
For the generic spherically symmetric line element~\eqref{ansatz},
one can express the motion of timelike particles in the form
\be
\dot{r}^2=V_r=\frac{B(r)}{A(r)}\left[E^2 - A(r)\left(1 + \frac{L^2}{r^2}\right)\right]\,, \label{radialpotential}
\ee
where dots stand for derivatives with respect to proper time, and $E,L$ are (conserved) energy and angular momentum
parameters respectively. If we focus on nearly circular orbits, we obtain
\beq
\Omega_r&=&\frac{1}{\dot{t}}\sqrt{-V_r''/2}\,,\\
\Omega_\phi&=&\sqrt{A'/(2r)}\,,\label{Omegaphi}
\eeq
to be evaluated at the circular orbit radius $r_c$ and where $\dot t=E/A(r)$.

Although the solution~\eqref{eq:ernst} for a magnetized BH spacetime is not spherically symmetric, it is easy to show that geodesic motion is still planar at $\theta=\pi/2$ and that the radial potential reads:
\begin{equation}
\dot r^2=V_r=\frac{E^2}{K^4}-\left(1-\frac{2M}{r}\right)\left(\frac{1}{K^2}+\frac{L^2}{r^2}\right)\,, \label{radialpotentialB}
\end{equation}
where $K$ is to be evaluated on the orbital plane, $K=1+B^2r^2/4$.

Inserting the expressions above into Eq.~\eqref{defshift}, we obtain~\cite{Schmidt:2008qi}
\begin{equation}
 \delta\phi=2\pi\left[\frac{\sqrt{A/B}}{\sqrt{3A-2r A'+r A'' A/A'}}-1\right]\,. \label{shift_general}
\end{equation}
Substituting $A=B=1-2M/r$ in the formula above we get the leading-order GR prediction
\be
\delta\phi=\frac{6\pi M}{r_c}+{\cal O}\left(\frac{M}{r_c}\right)^2\,.
\ee

Let us now discuss the corrections to the periastron shift induced by BH dirtiness and by other effects. For simplicity we focus on the large $r_c$ limit, but the general formula~\eqref{shift_general} is valid also in the strong-curvature region. When $r_c\gg M$, we parametrize the deviation from the isolated Schwarzschild case by a parameter $\delta_{\rm per}$ defined by
\be
\delta\phi=\frac{6\pi M}{r_c}\left(1+\delta_{\rm per}\right)+{\cal O}\left(\frac{M}{r_c}\right)^2\,.
\ee

Substituting the parametrized metric~\eqref{PRA1}--\eqref{PRA2} into Eq.~\eqref{shift_general}, and using the conditions~\eqref{PPNconditions0} we get, at large distances,
\begin{equation}
 \delta\phi=\frac{2\pi M(2(1+\gamma)-\beta)}{r_c}\left[1+\frac{M \left(\beta  (8+\beta )-4 (5+4 \gamma )-\alpha _1 \left(12+6 \beta -4 \gamma +3 \alpha _1 \left(4+\alpha _1\right)\right)+3 \alpha _3-\beta _2\right)}{2 (\beta -2 (1+\gamma )) r_c}\right] +{\cal O}\left(\frac{M}{r_c}\right)^3\,.
\end{equation}
Assuming $\beta=1+\delta\beta$ and $\gamma=1+\delta\gamma$ then, to first order in the corrections
$\delta\beta$, $\delta\gamma$, $\alpha_1$, $\alpha_3$ and $\beta_2$, the change in the periastron shift reads
\begin{equation}
 \delta_{\rm per}=\frac{9M}{2 r_c}+\frac{2 \delta \gamma }{3}-\frac{\delta \beta }{3}+\frac{M}{6 r_c}\left(16 \delta \gamma-10 \delta \beta +14 \alpha _1-3 \alpha _3+\beta _2\right)+{\cal O}\left(\frac{M}{r_c}\right)^2\,, \label{periastron_shift}
\end{equation}
where the first term is the 2PN GR correction. Note that higher-order terms in the weak-field  expansion~\eqref{PRA1}--\eqref{PRA2} do not contribute at this order. The expression above is the most general correction for the periastron shift in a spherically symmetric, deformed Schwarzschild background.

In Table~\ref{tab:periastron} we present the corrections to the periastron shift in a variety of contexts. The first entries were computed as particular cases of the previous analysis. The estimate for the corrections due to a magnetic field is slightly more involved, because it was performed by using the Ernst solution~\eqref{eq:ernst}. That solution is not spherically symmetric,
but geodesic motion is planar and we obtain, for small magnetic fields and large orbital distance,
\begin{equation}
 \delta_{\rm per}=\frac{B^2 r_c^4}{4 M^2}\,.
\end{equation}

Finally, let us discuss the periastron shift induced by the presence of an accretion disk. To keep the analysis simple, we focus on Newtonian disks. We note that in Newtonian theory the periastron shift is expressed simply in terms of the Newtonian potential $\Phi(r)$ by
\be
\delta \phi=2\pi\left(\sqrt{\frac{\Phi'}{3\Phi'+r_c\Phi''}}-1\right) \,.\label{shiftN}
\ee

By applying the formalism above to the two Newtonian models described in Eqs.\eqref{disk1}-\eqref{disk2}, we obtain:
\begin{eqnarray}
  \delta_{\rm per}^{\rm Kusmin}&=& -\frac{M_{\rm disk} r_c^4 R^2}{2 M^2 \left(r_c^2+R^2\right)^{5/2}}   \,,\\
 \delta_{\rm per}^{\rm exp}&=& \frac{4 M_{\rm disk} R y_c^4}{3\pi  M^2 } \left\{I_1(y_c) [K_1(y_c)-2 y_c K_0(y_c)]+I_0(y_c) [2 y_c K_1(y_c)-3 K_0(y_c)]\right\}  \,.
\end{eqnarray}
where $y_c=r_c/(2R)$ and we recall that $I_n,\,K_n$ are modified Bessel functions of the first and second kind.\footnote{In the $R/r\ll1$ limit these equations reduce to $\delta_{\rm per}=\left(1,3\right)\frac{M_{\rm disk} R^2}{2 M^2 r_c}$,
for the two models, respectively.} However, as we previously discussed, to model the effects of a realistic thin disk, we use a superposition the two Newtonian models of Eqs.\eqref{disk1}-\eqref{disk2}, with parameters given below Eq.~\eqref{surfacedensity}. 
Plugging this superposition into Eq.~\eqref{shiftN} we obtain, in the $M_{\rm disk}\ll M$ limit,
\begin{equation}
 \delta_{\rm per}=6\times 10^{-9} {f_{\rm Edd}}^{7/10} \left(\frac{M}{10^6 M_\odot}\right)^{6/5}\left(\frac{\alpha}{0.1}\right)^{-4/5}\,, \label{per_disk_fit}
\end{equation}
where we assumed $r_c\approx 10 M$. This estimate is also presented in Table~\ref{tab:periastron}.
A more comprehensive list of Newtonian disk models can be found in Ref.~\cite{2000MNRAS.316..540C}, the periastron precession for some of them and their relativistic counterparts is presented in Ref.~\cite{Vogt:2008zs}, and they confirm our overall conclusions.

Finally, for the DM profile~\eqref{powerlawdensity}, we obtain
\begin{equation}
 \delta_{\rm per}= \frac{2 \pi}{3} \rho_0 M^2  \left(\frac{R}{M}\right)^{\hat\alpha } {\tilde{r}_c}^{4-\hat\alpha} \,.
\end{equation}
These results are summarized in Table~\ref{tab:periastron}. Notice that the equation above applies also to the case of thick disks when $\hat{\alpha}=3/2$, $R=M$ and $\rho_0=3.4\times 10^{-6} f_{\rm Edd} 
\left(\frac{0.1}{\alpha}\right) \left(\frac{10^6 M_\odot}{M}\right) \mbox{ kg}/\mbox{m}^3$, as can be checked by comparing the density profile of thick disks, Eq.~\eqref{rhoThick}, with the profile in Eq.~\eqref{powerlawdensity}.

Notice that in the absence of ``dirtiness'', the only extra contribution to the ``geodesic'' precession is due to self-force effects, which scale as
$\delta^{\rm SF}_{\rm per}\sim a_1 \nu +a_2 \nu^2$, with $\nu\sim 10^{-5}$ the mass ratio. The first term comes from first-order self-force calculations, the second would require second-order self-force calculations~\cite{Pound:2005fs,Barack:2010ny,LeTiec:2011bk}.
Most profiles in Table~\ref{tab:periastron} yield contributions larger than second-order self force, but they are typically smaller than first order self-force corrections. For example, even the most optimistic scenario which considers self-gravity of thin disks gives corrections which are three orders of magnitude smaller than first-order self-force effects.  As discussed in Sec.~\ref{sec:prospects}, EMRIs detectable by eLISA are expected
to be surrounded by thick disks, whose self-gravity is even smaller than the estimates above. On the other hand, as mentioned in Secs.~\ref{sec:DF} and \ref{sec:migration} and as discussed in detail in Sec.~\ref{sec:phase_dirty} below, the corrections to the GW phase due to dissipative effects such as dynamical friction and planetary migration can be larger than first-order self-force corrections in thin disks, and  should be taken into account to generate accurate templates.

\subsection{ISCO changes}
The presence of matter surrounding massive BHs would also affect other geodesic quantities, like the ISCO frequency and location. The deviations introduced by a ring distribution localized 
outside the ISCO  of a Schwarzschild BH were computed in Refs.~\cite{Will:1974zz,Will:1975zza}, and found to be 
\begin{equation}
\frac{\delta \Omega_{\rm ISCO}}{\Omega_{\rm ISCO}}\lesssim 2\times 10^{-4}\frac{\delta M}{10^{-3}M}\,.\label{ISCOdisk}
\end{equation}
Therefore, a thin ring of mass $\delta M\sim 10^{-3} M$ would affect the ISCO frequency to the level of $\sim 0.01\%$.

Furthermore, by using our previous analysis, it is straightforward to compute the corrections introduced by the presence of charges, magnetic fields and a cosmological constant. Imposing $V=V'=V''=0$ in Eqs.~\eqref{radialpotential} and \eqref{radialpotentialB}, we obtain 
\begin{equation}
 \frac{\delta \Omega_{\rm ISCO}}{\Omega_{\rm ISCO}}=\frac{7}{24}q^2+279 B^2 M^2-396 \Lambda M^2\,, \label{ISCO_qBLambda}
\end{equation}
to leading order in the corrections.
Likewise, the correction due to the parametrized metric~\eqref{PRA1}--\eqref{PRA2} for the first nine terms reads
\begin{equation}
 \frac{\delta \Omega_{\rm ISCO}}{\Omega_{\rm ISCO}}=\frac{5 \delta \beta }{36}-\frac{5 \delta \gamma }{36}-\frac{4 \alpha _1}{9}+\frac{19 \alpha _3}{432}+\frac{41 \alpha _4}{2592}+\frac{71 \alpha _5}{15552}+\frac{109 \alpha _6}{93312}+\frac{155 \alpha _7}{559872}+\frac{209 \alpha _8}{3359232}+...\,. \label{ISCO_parametrized}
\end{equation}
Note that this ISCO shift does not depend on the coefficients $\beta_i$ in Eq.~\eqref{PRA2} (because circular motion is not affected by the $g_{rr}$ component of the metric) and 
that the coefficients of the $\alpha_i$ terms decrease as $i$ increases, i.e. higher-order terms (denoted with $...$) are again suppressed. This hierarchy is likely due to the fact 
that in Schwarzschild coordinates $M/r_{\rm ISCO}=1/6<1$. Indeed, we expect all corrections to be comparable for a nearly extremal BH (for which $r_{\rm ISCO}\sim M$). 

Finally, let us compare these corrections with those introduced by self-force effects. Self-force introduces a correction to the ISCO frequency that, at leading order in the mass-ratio $\nu$, reads $\delta \Omega_{\rm ISCO}/\Omega_{\rm ISCO}\sim 0.487\nu$~\cite{Barack:2009ey,letiec}. 
By comparing to Eq.~\eqref{ISCOdisk}, a ring of mass
\begin{equation}
 \frac{\delta M}{M}\gtrsim 2\times 10^{-5}\frac{\nu}{10^{-5}}\,,
\end{equation}
would affect the ISCO frequency more than self-force effects of an object of mass $\nu M$.

Furthermore, by comparing to the first term in Eq.~\eqref{ISCO_qBLambda}, we obtain that
\begin{equation}
 q\gtrsim 4\times 10^{-3}\left(\frac{\nu}{10^{-5}}\right)^{1/2}\,,
\end{equation}
in order for electric-charge effects to be larger than those due to the self force. Similar lower bounds can be derived for the effects due to magnetic fields and a cosmological constant. Likewise, for the dominant corrections of the parametrized metric~\eqref{PRA1}--\eqref{PRA2}, we obtain
\begin{eqnarray}
 \delta\beta &\gtrsim& 4\times 10^{-5}\frac{\nu}{10^{-5}}\,,\\
 \delta\gamma &\gtrsim& 4\times 10^{-5}\frac{\nu}{10^{-5}}\,,\\
 \alpha_1 &\gtrsim& 10^{-5}\frac{\nu}{10^{-5}}\,.
\end{eqnarray}
This analysis shows that a precise measurement of the ISCO frequency might strongly constrain deviations from the Schwarzschild vacuum geometry. Such corrections can be competitive with self-force corrections if the parameters are larger than the estimates above. Extending our general treatment to the rotating case is therefore important to disentangle spin effects from environmental corrections.
\section{Gravitational waves from ``dirty'' EMRIs} \label{sec:phase_dirty}
In this section we study the effects of matter on EMRIs. To lowest order, we consider a point particle of mass $\nu M$ in quasicircular orbit around a supermassive object of mass $M$ and we compute the GW dephasing associated to environmental effects and the corresponding gravitational waveforms in the stationary phase approximation (see e.g. Chapter 4 in Ref.~\citep{Maggiore:1900zz}).

In the adiabatic approximation, the motion in the radial direction is governed by dissipative effects only. For a generic spherically-symmetric spacetime~\eqref{ansatz}, the orbital energy for quasicircular timelike geodesics reads
\begin{equation}
 E=\frac{\sqrt{2} \nu M A}{\sqrt{2 A-r A'}}\,, \label{Eorb}
\end{equation}
Also in the presence of matter the emission of GWs is governed by the GR quadrupole formula
\begin{equation}
 \dot E_{\rm GW}=\frac{32}{5}\nu^2 M^2 r^4\Omega_\phi^6\,, \label{quadrupole}
\end{equation}
which is equivalent to Eq.~\eqref{quadrupole0} where $\Omega_\phi$ is defined as in Eq.~\eqref{Omegaphi}. 
Differentiating Eq.~\eqref{Eorb} with respect to time and using the balance law $\dot E=-\dot E_{\rm GW}$, one gets the evolution equation for the orbital radius
\begin{equation}
 \dot r=-\frac{4 \sqrt{2} \nu M r A'^3 \left(2 A-r A'\right)^{3/2}}{5 \left[A \left(3 A'+r A''\right)\right]-2 r A'^2}\,. \label{rdot}
\end{equation}
Note that this equation depends only on the metric coefficient $A$ and not on $B$. In the Newtonian limit, $A=1+2\Phi\sim1$, we get
\begin{equation}
 \dot r= -\frac{64 \nu M r \Phi '^3}{15 \Phi '+5 r \Phi ''}\,.\label{rdotN}
\end{equation}
This equation can be solved for $r(t)$ and, using Eq.~\eqref{Omegaphi}, one also obtains the frequency $\Omega_\phi$ as a function of time.

Once the frequency evolution is known, the GW phase simply reads
\begin{equation}
 \varphi(t)=2\int^t\Omega_\phi(t')dt' \,.\label{GWphase}
\end{equation}


At Newtonian level, it is possible to obtain analytical templates of the waveforms in Fourier domain. Neglecting dissipation, $r$ and $\Omega_\phi$ are constant and the Newtonian waveforms simply read:
\begin{eqnarray}
 h_+(t)&=&\frac{r^2 \nu M  \omega_{\rm GW}^2 }{D}\left(\frac{1+\cos^2\iota}{2}\right)\cos( \omega_{\rm GW}t) \,, \label{hp}\\
 h_\times(t)&=&\frac{r^2 \nu M  \omega_{\rm GW}^2 }{D}\cos\iota \sin( \omega_{\rm GW}t)\,, \label{hm}
\end{eqnarray}
where $\omega_{\rm GW}=2\Omega_\phi$, $D$ is the distance to the source and $\iota$ is the viewing angle. Then, dissipative effects can be included by replacing the constant parameters $\omega_{\rm GW}$ and $r$ by $\omega_{\rm GW}(t)$ and $r(t)$, where the secular time evolution is governed by the GW emission~\citep{Maggiore:1900zz}.

\subsection{Warm-up: Vacuum}
In order to illustrate the computation, let us review the standard results to lowest order in the case of isolated binaries. If $\Phi(r)=-M/r$, then Eq.~\eqref{rdotN} can be solved for $r(t)$ as:
\begin{equation}
 r(t)=r_0(1-t/t_c)^{1/4}\,,
\end{equation}
where $t_c=5 r_0^4/(256 \nu M^3)$. Consequently, the orbital frequency reads
\begin{equation}
 \Omega_\phi=\frac{1}{ 4^{3/2}M}\left[\frac{\nu }{5M}(t_c-t)\right]^{-3/8}\,.
\end{equation}
Note that $\Omega_\phi(t_c)\to\infty$ and $t_c$ is commonly referred to as coalescence time.  
Equivalently, one can solve the exact equations~\eqref{rdot} and \eqref{Omegaphi} and expand them to lowest order. The GR phase reads
\begin{equation}
 \varphi_{\rm GR}=-\frac{2}{\nu }\left[\frac{\nu }{5M}(t_c-t)\right]^{5/8}\sim -2.3\times 10^6 {\rm rads}\left(\frac{10^{-6}}{\nu}\right)^{3/8}\left(\frac{10^6 M_\odot}{M}\right)^{5/8}\left(\frac{t_c-t}{{\rm yr}}\right)^{5/8}\,. \label{GRphase}
\end{equation}

It is probably more useful to express the phase in terms of the final orbital radius $r_f$ and the observation time $T$. In order to do so, we write Eq.~\eqref{GWphase} as
\begin{equation}
\Delta \varphi=2\int_{r_i}^{r_f} dr\, \frac{\Omega_\phi}{\dot{r}}=\frac{1}{16\nu}\left[\left(\frac{r_i}{M}\right)^{5/2}-\left(\frac{r_f}{M}\right)^{5/2}\right]=\frac{1}{16\nu}\left(\frac{r_f}{M}\right)^{5/2}\left(\chi^{5/8}-1\right)\,,
\end{equation}
where in the last step we have expressed $r_i$ in terms of $\chi$, which is defined as
\begin{equation}
 \chi=1+\frac{256 M^3 T\nu}{5 r_f^4}\sim 1+0.25\left(\frac{M}{r_f}\right)^4\frac{T}{{\rm yr}}\frac{\nu}{10^{-6}}\frac{10^6M_\odot}{M}\,.\label{chi}
\end{equation}
Note that, in a significant region of the parameter space (namely for large central mass, small mass ratio or large final radius $r_f$) the dimensionless quantity $\chi$ is close to 1. In this region, we may expand our final results in the limit $\chi\sim1$.

Finally, standard treatment to compute the gravitational waveforms within the stationary-phase approximation allows one to write Eqs.~\eqref{hp} and \eqref{hm} to lowest order as~\citep{Maggiore:1900zz} (see also Refs.~\cite{Eda:2013gg,Macedo:2013qea})
\begin{eqnarray}
 h_+(t)&=&A_+(t_{\rm ret})\cos\varphi(t_{\rm ret}) \,,\\
 h_\times(t)&=&A_\times(t_{\rm ret}) \sin\varphi(t_{\rm ret})\,,
\end{eqnarray}
where $t_{\rm ret}$ is the retarded time. The final result yields
\begin{equation}
 \tilde{h}_+= {\cal A}_+e^{i\Psi_+}\,,\qquad  \tilde{h}_\times={\cal A}_\times e^{i\Psi_\times}\,,
\end{equation}
where 
\begin{eqnarray}
  {\cal A}_+&=& \frac{4\sqrt{2/3} M^{7/3} \nu  t_c^{1/2}}{\pi^{2/3} r_0^2 D}f^{-7/6}\frac{1+\cos^2\iota}{2} \,,\\
 {\cal A}_\times&=& \frac{2\cos\iota}{1+\cos^2\iota}{\cal A}_+\,, \\
 \Psi_+&=& -\frac{16 t_c }{5 r_0^4} M^{4/3}(\pi f)^{-\frac{5}{3}}+2\pi f\left(t_c+\frac{r}{c}\right)-\varphi_0-\frac{\pi}{4}\,, \label{dephGR}\\
 \Psi_\times&=&\Psi_+ + \pi/2\,,
\end{eqnarray}
with $f=\Omega_\phi/\pi$ and $\varphi_0=\varphi(t_c)$.

In the next section we shall repeat this computation in various contexts. Although a relativistic computation using the full Eq.~\eqref{rdot} is in principle possible, to simplify the final expression we shall consider a small perturbation around the Newtonian solution above. This means that we shall superimpose small corrections such as charge, magnetic fields, and disks on the top of the unperturbed Newtonian potential $\Phi=-M/r$.

Finally, to quantify the effect of the environmental effects for the dephasing, we define $\delta_\varphi$ as
\begin{equation}
 \varphi=\varphi_{\rm GR}+\delta_\varphi\,,
\end{equation}
where $\varphi_{\rm GR}$ is the unperturbed GR value at lowest order, Eq.~\eqref{GRphase}. Note that we are considering here the absolute correction, not the relative one as previously done for other quantities. Indeed, we consider an eLISA-like mission with accuracy to phase measurements $|\delta_\varphi^{\rm exp}|>10/{\rm SNR}$, where ${\rm SNR}$ is the signal-to-noise ratio~\cite{AmaroSeoane:2012je,Seoane:2013qna}. We assume the EMRI is detected when ${\rm SNR}>10$, so that the detector is at least sensitive to (absolute) dephasings larger than 1 rad over the mission time. This is a conservative assumption because high-${\rm SNR}$ events may allow for higher accuracy. 

Furthermore, we define the relative correction of the full Fourier-phase in the stationary phase approximation as
\begin{equation}
 \Psi_+=\Psi_+^{\rm GR}(1+\delta_{\Psi_+})\,.
\end{equation}

\begin{figure}[thb]
\begin{center}
\begin{tabular}{cc}
\epsfig{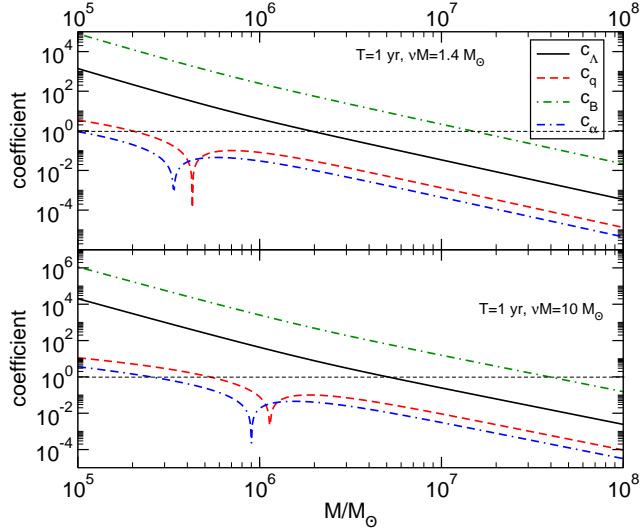}
\end{tabular}
\caption{Coefficients appearing in the formulas~\eqref{cLambda}, \eqref{cq}, \eqref{cB} and \eqref{calpha} as defined in Table~\ref{tab:dephasing} as functions of $M$. We have set the observation time $T=1{\rm yr}$, the final orbital radius $r_f=6M$, and the mass of the small particle reads $\nu M=1.4 M_\odot$ and $\nu M=10 M_\odot$ in the top and  bottom panels, respectively. The coefficient $c_{\hat \alpha}$ refers to ${\hat \alpha}=3.5$, other values give qualitatively similar results.
\label{fig:coefficients_dephasing}}
\end{center}
\end{figure}

The corrections in various contexts are presented in Tables~\ref{tab:dephasing} and~\ref{tab:dephasing_fourier} and computed below. In Table~\ref{tab:dephasing} we have introduced some coefficients $c_i=c_i(\chi)$ which are given in Eqs.~\eqref{cLambda}, \eqref{cq}, \eqref{cB} and \eqref{calpha} below and shown in Fig.~\ref{fig:coefficients_dephasing} as functions of $M$.

\begin{table}
\caption{Corrections $\delta_{\Psi_+}$ to the GW phase in the Fourier space computed within the stationary phase approximation for a quasicircular EMRI, due to
the same effects considered in Table \ref{tab:dephasing}, as well as dynamical friction from a power law density profile.}
\begin{tabular}{c|c}
 \hline\hline
Correction             & $\delta_{\Psi_+}$\\
\hline
accretion              & $\frac{4}{3}f_{\rm Edd}\dot{M}_{\rm Edd}\frac{t_c-t}{M}$	      \\
charge 		       & $-\frac{10}{3}q^2 (\pi f M)^{2/3}$	    \\
cosmological constant  & $-\frac{50}{99}\frac{\Lambda}{(\pi f)^2}$     \\
magnetic field         & $\frac{5}{22}\frac{B^2}{(\pi f)^2}$         \\
galactic halos	       & $-\frac{50}{792\pi}\frac{\rho}{(\pi f)^2}$	\\
thick accretion disks  & $-\frac{50}{792\pi}\frac{\rho}{(\pi f)^2}$, $\rho$ given by Eq.~\eqref{rhoThick}\\
DM distribution $\rho\sim r^{-{\hat \alpha}}$ &$\frac{40 (\pi f)^{\frac{2 {\hat \alpha} }{3}-2} M^{-{\hat \alpha} /3}  R^{{\hat \alpha} } (2 {\hat \alpha}-5 ) \rho_0}{3 \left(33-17 {\hat \alpha} +2 {\hat \alpha} ^2\right)}$	\\
dynamical friction for thin disks	& $0.1 \bar{K} f_{\rm Edd}^{11/20} \,  \left(\frac{0.1}{\alpha}\right)^{7/10} \left(\frac{10^6 M_\odot}{M}\right)^{1.12}\left(\log\left[\frac{v_s}{\pi f   r_{\rm min}}\right]-\frac{12}{49}\right)\left(\frac{\pi  f}{{\rm mHz}}\right)^{-29/12}$\\
dynamical friction for distribution $\rho\sim r^{-{\hat \alpha}}$     & $ -\frac{25 \pi \bar{K} R^{{\hat \alpha} } \rho_0 \left(3+2 ({\hat \alpha} -8) \log\left[\frac{v_s}{\pi f  r_{\rm min}}\right]\right)}{32 ({\hat \alpha} -8)^2M^{\frac{1}{3} ({\hat \alpha} +5)}}(\pi  f)^{\frac{2 {\hat \alpha} }{3}-\frac{11}{3}}$\\
\hline\hline
\end{tabular}
\label{tab:dephasing_fourier}
\end{table}

The explicit expressions shown in the main text are typically valid only in the small frequency, large-separation regime.
However, the estimates for the dephasing in Table~\ref{tab:dephasing_fourier} were obtained using relativistic calculations.

\subsection{Accretion of gas and DM onto the central BH}
In order to estimate the dephasing due to accretion onto the central BH during the inspiral, we can compute the total phase $\varphi$ in the absence of accretion (following the procedure explained above),
and evaluate the adiabatic correction $\delta_\varphi=(d\varphi/dM) \Delta M$, where $\Delta M$ is given in Eq.~\eqref{eq:accr} for gas accretion and in Eq.~\eqref{dm2bis} for DM accretion. The results are presented in Table~\ref{tab:dephasing}. We note that the effect of DM accretion onto the central object is usually negligible, unless extreme DM configurations with $\langle \rho_{\rm DM}\rangle\gtrsim 10^{12} M_\odot/{\rm pc}^3$ exist near the central BH, e.g. in dwarf satellite galaxies (cf. sec.~\ref{sec:dm_intro}). In that case, the total dephasing might have a marginal impact on detection 
and parameter estimation with an eLISA-like detector.

These estimates do {\it not} take possible dynamical effects into account. As we pointed out in Section~\ref{sec:ringdownaccretion}, a more realistic analysis of ringdown in time-varying geometries~\cite{Abdalla:2006vb,Abdalla:2007hg,Chirenti:2011rc,Shao:2004ws} shows that dynamical effects are negligible. It would be interesting to extend those calculations to BH inspirals.
\subsection{Cosmological Constant}
Let us start with the simplest case and consider $\Phi=-M/r+\Lambda r^2/6$. Solving the equations above in this particular case yields, to first order in $\Lambda$ and in the small frequency $f$ limit,
\begin{eqnarray}
 \delta_\varphi&=&-\frac{5}{7392}\left(\frac{r_f}{M}\right)^{11/2}\frac{\Lambda M^2}{\nu}\left[\frac{11-21 \chi ^{3/8}+10 \chi ^{7/4}}{\chi^{3/8}}\right]\sim -\frac{T}{3M}\left(\frac{r_f}{M}\right)^{3/2}\Lambda M^2\,,\label{cLambda}\\
 \delta_{\Psi_+}&=&-\frac{50}{99}\frac{\Lambda}{(\pi f)^2}\,.
\end{eqnarray}
where in the last step of the first equation we assumed $\chi\sim1$ and used the definition~\eqref{chi}. Our final result for the dephasing depends on final radius $r_f$ and on the observation time $T$. Equivalently, one might express the results in terms of the initial and final frequencies.
A correction similar to $\delta_{\Psi_+}$ can be straightforwardly computed also for the GW amplitude. 

\subsection{Charge}
Similarly to the previous case, we take into account the effects of an electric charge by considering the Newtonian potential $\Phi=-M/r+Q^2/(2r^2)$. We obtain:
\begin{eqnarray}
\delta_\varphi&=& \frac{5}{96 \nu } \left(\frac{r_f}{M}\right)^{3/2} q^2\left(5-3\chi^{-3/8}-2\chi^{3/8}\right) \sim\frac{T}{M}\left(\frac{M}{r_f}\right)^{3/2}q^2\,,\label{cq}\\
 \delta_{\Psi_+}&=&-\frac{10}{3}q^2 M^{2/3}(\pi f)^{2/3}\,.
\end{eqnarray}
\subsection{Magnetic field}
By requiring quasi-circular orbits and applying the balance law, the equation governing the radial motion due to GW dissipation reads, in the large distance and small magnetic field limit,
\begin{equation}
 \dot{r}=-\frac{64}{5}\frac{M^3\nu}{5 r^3}+\frac{32 }{5}M^2\nu B^2\,.
\end{equation}
This equation can be solved perturbatively in the small-$B$ limit. The standard treatment discussed above yields:
\begin{eqnarray}
 \delta_\varphi&=&\frac{5 B^2 M^2}{44352 \nu  }\left(\frac{r_f}{M}\right)^{9/2}\left[77\left(\chi ^{9/8}-1\right)-9 \frac{r_f}{M} \left(21-11\chi ^{-3/8}-10 \chi ^{11/8}\right)\right]\nn\\
 &\sim& \frac{T B^2M}{2}\left(\frac{r_f}{M}\right)^{1/2}\left(1+\frac{r_f}{M}\right)  \,,\label{cB}\\
 \delta_{\Psi_+}&=&\frac{5}{22}\frac{B^2}{(\pi f)^2}\,.
\end{eqnarray}
\subsection{Power-law density distributions}
Let us consider the case where we add to the central BH a Newtonian density distribution
\begin{equation}
 \rho_{\rm DM}\equiv\rho(r)=\rho_0(R/r)^{\hat \alpha}\, \qquad r_{\rm ISCO}<r<R\,.\label{powerlawdensity}
\end{equation}
This profile was studied in Ref.~\cite{Eda:2013gg} in the context of DM effects in intermediate-mass ratio binaries. In the following we shall use reference values for $\rho_0$ and $R$ comparable to those considered in Ref.~\cite{Eda:2013gg}, namely we use $\langle\rho_{\rm DM}\rangle\approx 10^3 M_\odot/{\rm pc^3}$ and $R\approx 0.33 {\rm pc}\sim 7\times 10^{12} M_\odot$.

The Newtonian potential which corresponds to Eq.~\eqref{powerlawdensity} reads
\begin{equation}
 \Phi=-\frac{M}{r}+\frac{4 \pi  \rho_0 R^2}{(2-{\hat \alpha} ) ({\hat \alpha}-3 )} \left(\frac{R}{r}\right)^{{\hat \alpha}-2}\,.
\end{equation}
Consistently with our perturbative approach, we assume that the corrections to the Newtonian potential are small. By using the potential above, Eq.~\eqref{rdotN} reduces to
\begin{equation}
 \dot{r}=\frac{64 M^3 \nu}{5 r^3}+\frac{256 M^2\nu \pi  ({\hat \alpha}-1 ) \rho_0 }{5 ({\hat \alpha}-3 )}\left(\frac{R}{r}\right)^{{\hat \alpha} }\,,
\end{equation}
which can be solved perturbatively in the small $\rho_0$ limit, although the solution is not illuminating. The dephasing in this case reads
\begin{eqnarray}
\delta_\varphi&=& \frac{5 \pi  \rho_0 M^2}{8 ({\hat \alpha} -3) \nu }\left(\frac{r_f}{M}\right)^{11/2-{\hat \alpha} }\left(\frac{R}{M}\right)^{{\hat \alpha} }\left(\frac{ (2 {\hat \alpha} -3) \chi ^{-{\hat \alpha} /4} \left(\chi ^{11/8}-\chi ^{{\hat \alpha} /4}\right)}{2 {\hat \alpha} -11}+\frac{\text{  }({\hat \alpha} -1) \chi ^{-\frac{3}{8}-\frac{{\hat \alpha} }{4}} \left(\chi ^{{\hat \alpha} /4}-\chi ^{7/4}\right)}{{\hat \alpha} -7}\right)\nn\\
&\sim&\frac{4 \pi  }{{\hat \alpha} -3}\left(\frac{R}{M}\right)^{{\hat \alpha} } \left(\frac{r_f}{M}\right)^{\frac{3}{2}-{\hat \alpha} } T \rho_0 M\,.\label{calpha}
\end{eqnarray}
Note that, by comparing the density profile of thick disks, Eq.~\eqref{rhoThick}, with the profile in Eq.~\eqref{powerlawdensity}, the dephasing due to the gravitational pull of thick disks can be expressed as a particular case of Eq.~\eqref{calpha} with the identifications $\hat{\alpha}=3/2$, $R=M$ and $\rho_0=3.4\times 10^{-6} f_{\rm Edd} 
\left(\frac{0.1}{\alpha}\right) \left(\frac{10^6 M_\odot}{M}\right) \mbox{ kg}/\mbox{m}^3$. The final result for the dephasing is given in Table~\ref{tab:dephasing}.


Finally, the relative correction to the Fourier-domain phase reads:
\begin{equation}
 \delta_{\Psi_+}=-\frac{40 (\pi f)^{\frac{2 {\hat \alpha} }{3}-2} M^{-{\hat \alpha} /3}  R^{{\hat \alpha} } (2 {\hat \alpha}-5 ) \rho_0}{3 ({\hat \alpha}-3)(2{\hat \alpha}-11)}\,,\label{dephalpha}
\end{equation}
which agrees with the results of Ref.~\cite{Eda:2013gg}. We normalize the DM density by computing the volume average $\langle\rho_{\rm DM}\rangle=\int_{r_i}^{r_f}  r^2 \rho(r)dr/\int_{r_i}^{r_f} r^2dr$, where $r_i$ and $r_f$ are the initial and final orbital radii, respectively.
In Table~\ref{tab:dephasing} we consider a typical DM density, $\langle\rho_{\rm DM}\rangle\sim 10^3 M_\odot/{\rm pc}^3$, and the radius $R\sim7\times 10^6M$~\cite{Eda:2013gg}. The latter are typical values to ensure that the corrections to the Newtonian potential introduced by the matter distribution are small. For example, when $\hat\alpha>3$ the Newtonian mass of the matter distribution localized outside the ISCO, $\delta M=\int_{6M}^\infty 4\pi r^2\rho \mbox{d}r$, converges. In this case, requiring $\delta M\ll M$ we get
\begin{equation}
 \frac{\delta M}{M}=\frac{2^{5-{\hat \alpha}}3^{3-{\hat \alpha}} \pi}{{\hat \alpha}-3}\left(\frac{R}{M}\right)^{\hat \alpha} \rho_0 M^2\sim 0.01\left(\frac{R}{7\times 10^6 M}\right)^{3.1}\left(\frac{M}{10^6M_\odot}\right)^2\frac{\rho_0}{10^3 M_\odot/{\rm pc}^3}\ll1\,,
\end{equation}
where in the second step we have assumed a representative value ${\hat \alpha}=3.1$. In the case $\hat\alpha<3$ the Newtonian mass does not converge, but the corrections to the Newtonian potential are still small compared to $M/r$ for these typical values of $\langle\rho_{\rm DM}\rangle$ and $R$, hence our perturbative approach is justified.
%
%

\subsection{Thin disks\label{sec:dephasing_selfgravity_thindisks}}
In order to estimate the dephasing introduced by the self-gravity of a thin disk we consider the most extreme 
case in which the entire inspiral occurs on the disk plane, which maximizes the dephasing.

\begin{figure}[thb]
\begin{center}
\begin{tabular}{cc}
\epsfig{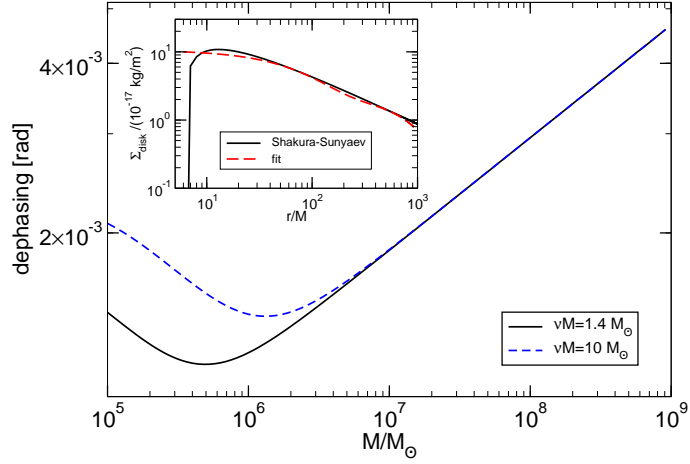}
\end{tabular}
\caption{Dephasing due to thin-disk self-gravity. The disk is modeled by superimposing the potentials in Eqs.~\eqref{disk1} and \eqref{disk2} and by fitting the model parameters to match the Shakura-Sunyaev surface density~\eqref{surfacedensity}. We have set the observation time $T=1{\rm yr}$, the final orbital radius $r_f=6M$, and the mass of the small particle reads $\nu M=1.4 M_\odot$ and $\nu M=10 M_\odot$ for the solid and the dashed line, respectively.
The dephasing is shown as a function of the central mass $M$ and for $f_{\rm Edd}=1$ and $\alpha=0.1$. The fit parameters are $M_{\rm disk}\approx 63 M_\odot {f_{\rm Edd}}^{7/10} [M/(10^6 M_\odot)]^{11/5}(\alpha/0.1)^{-4/5}$, $R\approx1016 GM/c^2$ for the Kusmin-Toomre model, and $M_{\rm disk}\approx 1.5 M_\odot {f_{\rm Edd}}^{7/10} [M/(10^6 M_\odot)]^{11/5}(\alpha/0.1)^{-4/5}$, $R\approx74 G M/c^2$ for the exponential model. The inset shows a comparison of the surface density~\eqref{surfacedensity} and of the fit. 
\label{fig:dephasing_disks}}
\end{center}
\end{figure}

A representative result is shown in Fig.~\ref{fig:dephasing_disks}. As discussed in Sec.~\ref{sec:disks}, to approximate a realistic Newtonian disk we have used a superposition of the potentials in Eqs.~\eqref{disk1} and \eqref{disk2} whose parameters $M_{\rm disk}$ and $R$ have been chosen to match the surface density~\eqref{surfacedensity}. As shown in the inset of Fig.~\ref{fig:dephasing_disks}, the fit well approximates the model~\eqref{surfacedensity}, at least in the intermediate region $6M<r<10^3M$ where the Newtonian model is accurate. 
The dephasing cannot be computed analytically in terms of simple functions, hence   we have resorted to a numerical integration of the equations presented above.
In Fig.~\ref{fig:dephasing_disks} we show the dephasing as a function of the central mass $M$ and for $f_{\rm Edd}=1$ and $\alpha=0.1$.
As we can see, even in this rather extreme case the dephasing is always quite small. This is consistent with our previous estimate. 
%
\subsection{Parametrized metric}
\begin{figure}[thb]
\begin{center}
\begin{tabular}{cc}
\epsfig{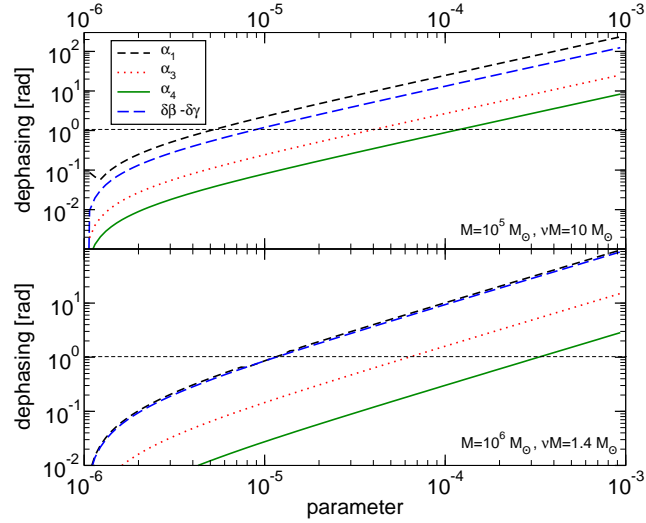}
\end{tabular}
\caption{Dephasing due to the parametrized metric element~\eqref{PRA1} and assuming GW dissipation through the lowest-order GR formula~\eqref{quadrupole}. Note that in this approximation the dephasing depends only on the metric coefficient $A$ and not on $B$. Higher order corrections in $1/r$ correspond to smaller dephasing. We consider two different choices of the masses of the binary components and assume 1-year observation time terminating at $r=6M$.
\label{fig:parametrized}}
\end{center}
\end{figure}

Modeling generic matter distributions outside BHs is a tremendously difficult problem. One ``quick and dirty'' way out is to
start with the modified background metric~\eqref{PRA1} and~\eqref{PRA2}, and {\it assume} geodesic motion in this modified metric. Such an inconsistent approach has its merits, as it can model
realistic situations with \textit{ad hoc} matter content in the case in which accretion and the hydrodynamic drag (i.e. dynamical friction and  planetary migration) are negligible.
This approach is also instructive to assess whether a hierarchy of corrections exists in the expansion~\eqref{PRA1} and~\eqref{PRA2}. At first sight, this is not obvious because the metric is expanded in the weak-field limit, whereas the dominant contribution to the GW phase comes from the latest stage of the inspiral, where the satellite probes the strong-field region of the central object. Figure~\ref{fig:parametrized} shows that, nonetheless, such a hierarchy of terms exists. That is, the lowest order corrections in the weak-field expansion are associated with the larger dephasing. In particular, this suggests that, in dealing with modified BH solutions, a large distance expansion of the background might be sufficient for capturing the dominant corrections. 

Another interesting question is the extent to which the metric corrections can be constrained by a putative detection of an inspiral waveform. A dephasing of 1 rad is associated with $\alpha_1\sim {\cal O}(10^{-5})$, depending on the other parameters, whereas the higher-order terms are constrained roughly at the level of $10^{-4}$. Note also that the dephasing does not depend on the $g_{rr}$ metric component so that, to this level of approximation, the $\beta_i$ coefficients in Eq.~\eqref{PRA2} are unconstrained. In Fig.~\ref{fig:parametrized} we have assumed that the observation stops when the orbital radius is $r=6M$ regardless of the value of $\alpha_i$.

Note that, because of the sole dependence on $A(r)$, the dephasing depends on the combination $\delta\beta-\delta\gamma$. However, it is interesting to notice that such a combination can be constrained at the level of $10^{-5}-10^{-4}$ depending on the parameters. This is one-two orders of magnitude more stringent than current PPN constraints on $\beta$ from Mercury's periastron shift~\cite{Will:2005va}.

\subsection{Dephasing due to DM accretion and dynamical friction} \label{dephasing_DF}
In addition to self-gravity effects, DM can affect the gravitational waveforms due to accretion and dynamical friction effects. As an example, we consider the density profile~\eqref{powerlawdensity}, whose self-gravity introduces a dephasing computed in Eq.~\eqref{dephalpha} (see also Ref.~\cite{Eda:2013gg}).

The dephasing introduced by accretion can be computed by extending the analysis done in Ref.~\cite{Macedo:2013qea}. To lowest order, accretion affects the radial motion by adding an extra term~\cite{Macedo:2013qea}
\begin{equation}
 \dot{r}_{\rm accretion}\sim -\frac{2}{1+{\hat \alpha}}\frac{{\dot m}_{\rm sat}}{m_{\rm sat}(t)}r\,. \label{rdotDF}
\end{equation}
Depending on the DM mass and cross section, accretion can be collisionless or of Bondi-Hoyle type~\cite{Shapiro:1983du,Macedo:2013qea}. However, as shown by Eq.~\eqref{BondiVScollisionless}, for typical values of DM density, cross section and DM particle mass, we expect accretion to be collisionless. For completeness, we consider both cases, where (see e.g.~\cite{Shapiro:1983du})
\begin{equation}
 \dot{m}_{\rm sat}=\left\{\begin{array}{l}
                   4\pi \frac{\rho m_{\rm sat}^2}{v} \quad {\rm collisionless}\\
                   4\pi \lambda \frac{\rho m_{\rm sat}^2}{v^3} \quad {\rm Bondi}
                  \end{array}\right.\,,
\end{equation}
where for the Bondi-Hoyle accretion $\lambda\sim 1$ and we assumed supersonic motion, $v\sim \sqrt{M/r}\gg v_s$. 
Using the equations above, the radial motion can be solved analytically in the small density regime. With the stationary phase technique discussed above, we obtain the perturbative corrections:
\begin{equation}
 \delta_{\Psi_+}^{\rm accretion}\sim \left\{\begin{array}{c}
                   \frac{25 \pi ({\hat \alpha} -4)  R^{{\hat \alpha} } \lambda  \rho_0}{8 ({\hat \alpha} -7) (2{\hat \alpha} -9)M^{1+\frac{{\hat \alpha} }{3}}}(\pi  f)^{\frac{2 {\hat \alpha} }{3}-3} \quad {\rm collisionless}\\
                   \frac{25\pi ({\hat \alpha} -5) R^{{\hat \alpha} }  \lambda \rho_0 }{8 ({\hat \alpha} -8) (2 {\hat \alpha} -11)M^{\frac{5}{3}+\frac{{\hat \alpha} }{3}}} (\pi  f)^{\frac{2 {\hat \alpha} }{3}-\frac{11}{3}} \quad {\rm Bondi}
                  \end{array}\right.\,,\label{deltaPhiaccretion}
\end{equation}
and it is clear that, in the small frequency limit -- i.e, small $Mf$ or large distances --, Bondi-Hoyle accretion gives stronger corrections than collisionless accretion. 

Let us compare the effect of accretion with the dephasing introduced by self-gravity, Eq.~\eqref{dephalpha}. In the small-frequency limit we obtain
\begin{equation}
 \frac{\delta_{\Psi_+}^{\rm Bondi}}{\delta_{\Psi_+}^{\rm self-gravity}}\sim -\frac{15 ({\hat \alpha}-5 ) ({\hat \alpha}-3 ) \lambda }{64  ({\hat \alpha} -8) (2 {\hat \alpha}-5 )}(\pi f M)^{-5/3}\,.
\end{equation}
Therefore, the effects of Bondi-Hoyle accretion generically dominate over self-gravity effects for any value of ${\hat \alpha}$. 

However, since DM likely behaves as a collisionless fluid (cf. Sec.~\ref{sec:DM}), a more realistic estimate is to compare self-gravity with collisionless accretion. In this case we get
\begin{equation}
 \frac{\delta_{\Psi_+}^{\rm collisionless}}{\delta_{\Psi_+}^{\rm self-gravity}}\sim \frac{64 (7-{\hat \alpha} ) (2 {\hat \alpha} -9) (2 {\hat \alpha} -5)}{15 ({\hat \alpha} -4) ({\hat \alpha} -3) (2 {\hat \alpha} -11) }\pi M f\,.
\end{equation}
so that accretion is typically negligible with respect to self-gravity in the small-frequency regime.

In general, the effect of accretion is generically small. For example, for ${\hat \alpha}=2$, Eq.~\eqref{deltaPhiaccretion} in the more favorable Bondi-Hoyle case can be written as
\begin{equation}
 \delta_{\Psi_+}^{\rm Bondi}\sim-6\times 10^{-8}\lambda\left(\frac{R}{7\times 10^6 M}\right)^{2} \left(\frac{\rho_0}{10^3 M_\odot{\rm pc}^{-3}}\right) \left(\frac{f}{{\rm mHz}}\right)^{-7/3}\left(\frac{M}{10^6 M_\odot}\right)^{-7/3}\,,
\end{equation}
and therefore, even though such a correction enters at lower PN order than self-gravity effects, its actual normalization in the mHz band is tiny, as mentioned in Sec.~\ref{sec:DM}. Other values of ${\hat \alpha}$ give the same order of magnitude for the relative correction. 


The dephasing introduced by dynamical friction can be also estimated by using this framework and the results of Sec.~\ref{sec:DF}. In this case, the time dependence of the orbital radius is regulated by the standard Newtonian equation of motion with an extra term $\dot r_{\rm DF}\sim-\frac{dr}{dE} \dot{E}_{\rm DF}$, where ${\dot E}_{\rm DF}$ is given in Eq.~\eqref{dotEDF}. This equation can be solved analytically in the limit ${\cal M}\sim v_K/v_s\gg1$ and for equatorial motion. The final dephasing in the small frequency limit reads
\begin{equation}
 \delta_{\Psi_+}^{\rm DF}\sim -\frac{25 \pi \bar{K} R^{{\hat \alpha} } \rho_0 \left(3+2 ({\hat \alpha} -8) \log\left[\frac{v_s}{\pi f  r_{\rm min}}\right]\right)}{32 ({\hat \alpha} -8)^2M^{\frac{1}{3} ({\hat \alpha} +5)}}(\pi  f)^{\frac{2 {\hat \alpha} }{3}-\frac{11}{3}}\,. \label{deltaDFDM}
\end{equation}
where $\bar{K}\sim1$. This shows that, besides the logarithmic term, the dephasing introduced by dynamical friction has the same frequency dependence as that introduced by Bondi-Hoyle accretion, i.e.
\begin{equation}
 \frac{\delta_{\Psi_+}^{\rm DF}}{\delta_{\Psi_+}^{\rm Bondi}}\sim \frac{\bar{K} (11-2 {\hat \alpha} )}{4 ({\hat \alpha} -8) ({\hat \alpha} -5) \lambda }\left(3+2 ({\hat \alpha} -8) \log\left[\frac{v_s}{ \pi f  r_{\rm min}}\right]\right)\,.
\end{equation}
At low frequencies the two effects are comparable but both much smaller than radiation-reaction effects. 
On the other hand, the results above show also that dynamical friction is dominant with respect to both collisionless accretion and self-gravity effects. Our analysis shows that the results of Ref.~\cite{Eda:2013gg} would be drastically modified, as they were obtained by neglecting both accretion and dynamical friction.

In order to estimate the total dephasing due to dynamical friction from the DM distribution during the inspiral, we have integrated the equations of motion numerically 
including the dissipative term~\eqref{rdotDF} and using $\dot r\sim-\frac{dr}{dE} \dot{E}$ to compute the energy flux. The GW phase is then given by Eq.~\eqref{GWphase}. We find that, in the small-mass ratio limit, the dephasing for an inspiral of one year terminating at $r_f=6M$ is
\be
\delta_\varphi^{\rm DM-DF} \sim 10^{-14}\left(\frac{\rho_{\rm DM}}{10^3 M_\odot/{\rm pc}^3}\right)\left(\frac{\nu}{10^{-5}}\right)^{0.65} \left(\frac{M}{10^6 M_{\odot}}\right)^{0.17}\,,
\ee
which is the value reported in Table~\ref{tab:dephasing}. This estimate is consistent with Eq.~\eqref{dm3} considering 
that the inspiral occurs at $\tilde{r}\sim 6$ for an EMRI. 

\subsection{Dephasing due to dynamical friction in thin and thick disks} \label{dephasing_DF_disks}
Our results for the dephasing due to dynamical friction from the DM distribution can be directly used to estimate the dephasing introduced by dynamical friction in thin and thick disks. For thin disks, assuming the inspiral occurs on the equatorial plane where the disk is located, and using the approximation $1-\sqrt{{\tilde{r}_{\rm in}}/{\tilde{r}}}\approx 1$ in Eq.~\eqref{eq:rhoF}, we can map the profile~\eqref{powerlawdensity} to the disk density \eqref{eq:rhoF} with the identification $\hat\alpha=15/8$, $R\to M$, $\rho_0\to 169 f_{\rm Edd}^{11/20} \,  \left(\frac{0.1}{\alpha}\right)^{7/10} \left(\frac{10^6 M_\odot}{M}\right)^{7/10} \mbox{ kg}/\mbox{m}^3$. Hence, by using Eq.~\eqref{deltaDFDM}, the dephasing associated to dynamical friction of a Shakura-Sunyaev thin disk in the stationary phase approximation reads
\begin{equation}
 \delta_{\Psi_+}^{\rm DF, disk}\sim 0.1 \bar{K} f_{\rm Edd}^{11/20} \,  \left(\frac{0.1}{\alpha}\right)^{7/10} \left(\frac{10^6 M_\odot}{M}\right)^{1.12}\left(\log\left[\frac{v_s}{\pi f   r_{\rm min}}\right]-\frac{12}{49}\right)\left(\frac{\pi  f}{{\rm mHz}}\right)^{-29/12}\,,
\end{equation}
which is the result reported in Table~\ref{tab:dephasing_fourier}.

To compute the total dephasing during the inspiral between the orbital distance $r_i$ and the orbital distance $r_f$ it is more convenient to integrate the equations of motion numerically. To isolate the dissipative effect of dynamical friction we neglect possible self-gravity effects of the matter. As we previously discussed, the latter are smaller than the hydrodynamic drag, so our assumption is well motivated. Thus, we assume that the satellite follows a quasicircular geodesic of the central object and that the orbital radius changes adiabatically due to the energy flux
\begin{equation}
 \dot E_{\rm T}\equiv \dot E_{\rm GW}+\dot E_{\rm DF}=\frac{32}{5} \frac{m_{\rm sat}^2}{M^2}\left(\frac{M}{r}\right)^5+4 \pi \rho \frac{m_{\rm sat}^2}{v_K} I  \bar{K}
\end{equation}
where we used Eqs.~\eqref{quadrupole} and \eqref{dotEDF}. For thin disks we consider the most extreme case in which the inspiral occurs entirely within the disk on the orbital plane, so $\bar K\sim 1$ and $\rho$ is given by Eq.~\eqref{eq:rhoF}, whereas for thick disks $\rho$ is given by Eq.~\eqref{rhoThick}. We also take $v_K\sim v\sim \sqrt{M/r}$ in both cases. As we previously discussed, differentiating Eq.~\eqref{Eorb} with respect to time and using the balance law $\dot E=-\dot E_{\rm T}$, one gets the evolution equation for the orbital radius
\begin{equation}
 \dot r(t)=-\dot E_{\rm T}\left[\frac{dE}{dr}\right]^{-1}\,.
\end{equation}
This equation can be straightforwardly integrated with initial condition $r_i=r(0)$. Finally, the GW phase can be computed as Eq.~\eqref{GWphase} with $\Omega_\phi$ given in terms of $r(t)$ by Eq.~\eqref{Omegaphi}.

In order to compare to the results shown in Table~\ref{tab:dephasing}, we fix $r_f=6M$ and choose $r_i$ such that the satellite reaches the ISCO after one year. For a thin disk with we obtain, in the small-mass ratio limit, the dephasing 
\be
\delta_\varphi^{\rm DF, thin} \sim 285 f_{\rm Edd}\left(\frac{0.1}{\alpha}\right)\left(\frac{\nu}{10^{-5}}\right)^{1/2} \left(\frac{M}{10^6 M_{\odot}}\right)^{-0.3}\,,
\ee
with respect to the case in which dynamical friction is neglected. This confirms that dynamical friction in thin disks 
might produce an observable dephasing in the gravitational waveforms. 
On the other hand, dynamical friction in thick-disk environments (such as those more relevant for eLISA-like missions) gives
\be
\delta_\varphi^{\rm DF, thick} \sim 3\times 10^{-9} \frac{f_{\rm Edd}}{10^{-4}}\left(\frac{0.1}{\alpha}\right) \left(\frac{\nu}{10^{-5}}\right)^{0.48} \left(\frac{M}{10^6 M_{\odot}}\right)^{-0.58}\,.
\ee
where again the result is valid when the mass ratio is small.
Although the dephasing is larger than that introduced by the gravitational pull of thick disks, the net effect is many orders of magnitude smaller than in the case of thin disks and can be always neglected (cf. Sec.~\ref{sec:DF}).

\clearpage
\newpage

\part{Blurred tests of General Relativity induced by astrophysical environments}\label{part:testsGR}
There are excellent monographs dealing with tests of GR in various regimes~\cite{Will,Will:2005va,Gair:2012nm,Yunes:2013dva,Clifton:2011jh}, and 
we will not attempt any comprehensive overview here. Rather than presenting a detailed analysis for each specific modified theory of gravity, we wish instead to provide a catalog of order-of-magnitude estimates of various effects that can be used to constrain alternative theories in a model-independent fashion. Although admittedly approximate, this approach is useful to understand which observations impose the most stringent constraints on a particular theory, thus paving the way for more rigorous treatments. To the best of our knowledge the impact of environment effects for strong-field tests of gravity has never been discussed in the literature, but it is of utmost importance to assess the potential of GW astronomy.

We consider an alternative theory whose deviation from GR is encoded in a new field $\bm{\Psi}$ of generic spin which, at the level of the action, is associated with two sets of coupling constants, $a_i$ and $b_i$. The first set corresponds to nonminimal interactions between the field and the metric, whereas the second set corresponds to a nonminimal coupling to the matter sector. We assume that GR is recovered when those couplings are vanishing so that a small-coupling expansion is well defined. In this limit, the most generic action can be schematically written as,
\begin{eqnarray}
 S&=&\frac{1}{16\pi{\cal G}}\int dx^4\sqrt{-g}\left[R+\partial^2\bm{\Psi}+\sum_ia_i U_i(\bm{\Psi},\bm{g},\partial\bm{\Psi},\partial\bm{g},...)\right]\nn\\
 &+&S_m^{(0)}[\bm{\Psi}_m,g_{\mu\nu}]+\sum_i b_i S_{m,i}^{(1)}[\bm{\Psi}_m,\bm{\Psi},\bm{g},\partial\bm{\Psi}_m,\partial\bm{\Psi},\partial\bm{g},...]\,,\label{modifiedGR}
+{\cal O}(a_i^2, b_i^2)
\end{eqnarray}
where ${\cal G}$ is an effective gravitational coupling, not necessarily equal to $G$, $\bm{\Psi}_m$ schematically represents any matter field, $U_i$ are some nonminimal interaction terms, and we have linearized the matter action $S_m$. Without loss of generality, we have rescaled the extra field $\bm{\Psi}$ in order to be canonically normalized. Finally, $U_i$ and $S_m^{(1,i)}$ are all vanishing when $\bm{\Psi}\equiv0$, thus recovering GR.

In the action above, the couplings in $S_{m,i}^{(1)}$ are such that an effective metric can be defined in terms of $\bm{g}$ and $\bm{\Psi}$ which is minimally coupled to the matter fields. This guarantees that the weak equivalence principle is generically satisfied, whereas the strong equivalence principle will in general be violated. 

Furthermore, the theory~\eqref{modifiedGR} is manifestly Lorentz invariant. Possible Lorentz violations in the gravitational sector can be accounted for by adding suitable Lagrangian multipliers which enforce a preferred time direction at each spacetime point. This is the case of \AE-theory~\cite{Jacobson:2000xp,AE} or Ho\v rava gravity~\cite{HL,Blas:2009qj}, which modify GR by adding a timelike vector field.
This produces BH solutions that differ from GR~\cite{Eling:2006ec,Barausse:2011pu,Blas:2011ni,Barausse:2013nwa,Barausse:2012ny,Barausse:2012qh}, 
thus giving a way of testing these theories, at least in principle, with GW observations.

The couplings $a_i$ and $b_i$ can have different physical dimensions, depending on the theory. We assume $[a_i]={\rm length}^{x_i}$ and $[b_i]={\rm length}^{y_i}$. Under these assumptions, all corrections to GR will be proportional to some power of $a_i$ and $b_i$. For example, spherically symmetric geometries are modified in these theories, and the coefficient $\alpha_i$ and $\beta_i$ in Eqs.~\eqref{PRA1} and \eqref{PRA2} would be proportional to some power of the fundamental couplings $a_i$ and $b_i$. Finally, the role of ${\cal G}$ is to rescale the gravitational constant. We set ${\cal G}=1$ in the following, but it can be easily reinserted by dimensional analysis.

Various extensions of GR are in fact low-energy expansions of some more fundamental (and often unknown) theory and, as such, they are rigorously valid only in an effective field theory sense (cf. e.g. Ref.~\cite{Yunes:2013dva}). In these theories, a perturbative approach in the coupling parameters is not only natural, but required for internal consistency. 
As we shall discuss, our approach does reduce to some of the most studied alternative theories in some particular cases.
However, we stress that the estimates on the couplings that we derive in this section are valid under the rather strong assumption that the theory behaves perturbatively in the regime where GW emission takes place. While that is true in most cases, in some scalar-tensor theories that assumption breaks down in certain binary systems~\cite{ST1,Palenzuela:2013hsa,Shibata:2013pra}. On the other hand, such possible strong-field effects --~neglected by our approach~-- have to be investigated on a case-by-case analysis. Rather than following this often-beaten path, here we attempt to describe beyond-GR effects in a novel (yet approximate) model-independent fashion.

The rest of this part is divided in two sections. In the next section we discuss tests of GR that involve ``conservative'' effects. We define the latter, with some abuse of terminology, as those corrections related to a deformation of the background geometries with respect to the GR solution. These tests include the classical ones, some of which we have previously discussed. Finally, in Sec.~\ref{sec:dissipative} we discuss tests that involve ``dissipative'' effects, i.e. those related with modified GW emission.

\section{Executive Summary}
%
\begin{table}[hbt]
\small
\centering \caption{Catalog of modified theories of gravity that we consider in this work and their relation with the action~\eqref{modifiedGR}. We consider the small-coupling limit away from GR, whose deviations are encoded in a single extra field $\bm{\Psi}$. The notation is taken from the corresponding references. In the entries below n.m. and m.d. stand for nonminimal and model dependent, respectively.} 
\vskip 12pt
\begin{tabular}{c|ccccc|c}
\hline \hline
Theory     & $\bm{\Psi}$ & $a_i$ & $x_i$ & $b_i$ & $y_i$ &Effect\\
\hline 
Brans-Dicke (BD) [Einstein frame]~\cite{Will:2005va}	& scalar 	& none 	& none 	& $\sim1/\omega_{\rm BD}$ 	& $0$		& \begin{tabular}{c} spacetime-dependent $G$  \end{tabular} \\
Einstein-Dilaton-Gauss-Bonnet (EDGB)~\cite{Yagi:2011xp,Pani:2009wy}	& scalar 	& $\xi_i$ 	& $2$ 	& none 	& none		& \begin{tabular}{c} quadratic-in-curvature corrections \\ topological invariant\end{tabular}	\\
Dynamical Chern-Simons (DCS)~\cite{Yagi:2011xp,Yunes:2009hc}		& pseudoscalar 	& $\xi_4$ 	& $2$ 	& none 	& none		& \begin{tabular}{c} quadratic-in-curvature corrections \\ modify reflection-invariant solutions \end{tabular}	\\
\AE-theory~\cite{Jacobson:2000xp,AE}, Ho\v rava gravity~\cite{HL,Blas:2009qj}			& vector 	& $c_i$ 	& $0$ 	& none 	& none	& \begin{tabular}{c} Lorentz violation in gravity \end{tabular}	\\
massive gravity~\cite{Brito:2013wya,Hassan:2011zd}			& tensor 	& $\mu_{\rm graviton}^2$ 	& $-1$ 	& n.m. 	& m.d.		& \begin{tabular}{c} large-distance corrections \\ break diffeomorphism invariance \end{tabular}	\\
\hline \hline
\end{tabular}
\label{tab:theories}
\end{table}
While our analysis is largely theory-independent, in the remaining we will discuss specific examples, by considering some of the most studied alternative theories of gravity. Table~\ref{tab:theories} presents a summary of 
some of these, and their relation with the action~\eqref{modifiedGR}.

We have considered a variety of effects that are affected by strong-field corrections to GR. They are divided in ``conservative'' and ``dissipative'' effects and discussed in Secs.~\ref{sec:conservative} and \ref{sec:dissipative}, respectively. Our main finding is that such effects might be degenerate with those induced by matter distributions. In order to be detectable, beyond-GR corrections should be larger than environmental effects. This poses intrinsic \emph{lower} bounds on the couplings $a_i$ and $b_i$, which are shown in Tables~\ref{tab:modifedVSdirty} and \ref{tab:dotPmodified_dirtiness} and discussed in what follows.

\begin{table}[hbt]
\centering \caption{Intrinsic lower constraints on generic deviations from GR due to environmental effects taking into account different processes. The limit was obtained by comparing the beyond-GR corrections parametrized by the metric~\eqref{PRA1}-\eqref{PRA2} and those induced to matter distributions within GR. The exponent $n$ depends on the theory: $n=2$ for all theories listed in Table~\ref{tab:theories} except for \AE-theory and Ho\v rava gravity, for which $n=1$. For the periastron advance, we consider the correction introduced by a thin disk as shown in Table~\ref{tab:periastron}. For the Shapiro delay, $D_e$ and $D_p$ are the positions of the emitter and the receiver, respectively, whereas $D$ is the impact parameter. Note that, even by assuming $D\sim M$ (i.e. that the classical tests are performed around a very compact object), tests based on ringdown or EMRIs provide the smaller lower bounds, showing that tests of GR are less affected by matter distributions in those cases. } 
\begin{tabular}{c|l}
\hline \hline
Effect     		            & Intrinsic lower bound 	\\
\hline 
Periastron advance     		& $\left(\frac{a_i}{M^{x_i}}\right)^n \gtrsim 10^{-7} {f_{\rm Edd}}^{7/10} \left[\frac{M}{10^6 M_\odot}\right]^{6/5}\left(\frac{\alpha}{0.1}\right)^{-4/5}$	\\
Shapiro delay     		& $\left(\frac{a_i}{M^{x_i}}\right)^n \gtrsim 10^{-4} \left[\frac{D}{M}\frac{\delta M}{10^{-6}M}\frac{\log\left({4 D_e D_p}/{D^2}\right)}{40} \right]^{}$ 	\\
Deflection of light     	& $\left(\frac{a_i}{M^{x_i}}\right)^n \gtrsim 10^{-6} \left[\frac{D}{M}\frac{\delta M}{10^{-6}M}\right]^{}$ 	\\
Ringdown     			& $\left(\frac{a_i}{M^{x_i}}\right)^n \gtrsim 10^{-6}\left(\frac{\delta M}{10^{-6}M}\right)^{}$ 	\\
EMRIs     			& $\left(\frac{a_i}{M^{x_i}}\right)^n \gtrsim 10^{-6}\left(\frac{\delta M}{10^{-6}M}\right)^{}$ 	\\
\hline \hline
\end{tabular}
\label{tab:modifedVSdirty}
\end{table}

\begin{table}[hbt]
\small
\centering \caption{Intrinsic lower constraints on various modified theories of gravity due to environmental effects in the orbital decay rate of a binary inspiral. The limit was obtained by comparing the results shown in Table~\ref{tab:dotPmodified} with Eq.~\eqref{deltaPoP_dirtiness} in various cases. Deviations from GR must be large enough in order for their effect to be larger than the environmental ones. For each environmental effect and for each theory, we have normalized the result by a parameter ${\cal P}$ and a parameter ${\cal T}$, respectively. These parameters are defined in the last row and last column, respectively, where we also defined $v_3=v/10^{-3}$, $\rho_3^{\rm DM}=\rho_0/(10^3 M_\odot/{\rm pc}^3)$, $\rho_2^{\rm disk}=\rho_0/(10^2 {\rm kg}/{\rm m}^3)$, $M_{10}=M_T/(10 M_\odot)$, $B_{8}=B/(10^{8}{\rm Gauss})$, $q_{3}=q/10^{-3}$, $\gamma_{\hat \alpha}=[{\hat \alpha}({\hat \alpha}-9)+12]/({\hat \alpha}-3)^2$. The limits for magnetic field and for electric charges are obtained using Eqs.~\eqref{dotPB} and \eqref{dotPq}, respectively. We considered Eq.~\eqref{powerlawdensity} as an example of spherically-symmetric DM density and baryonic-mass density profiles. For \AE-theory and Ho\v rava gravity, the function ${\cal F}$ depends on the coupling constants $c_i$ of the theory and on the sensitivities $s_i$. In the small coupling limit ${\cal F}$ is linear in the couplings, cf. Refs.~\cite{Yagi:2013qpa,Yagi:2013ava} for details.} 
\vskip 12pt
\begin{tabular}{c|cccc|c}

\hline \hline
&\multicolumn{4}{c}{Intrinsic lower bound}  \\ 
\hline
Theory     				& magnetic field & DM profile, ${\hat \alpha}=3/2$	&  baryonic matter 	& charge & coefficient ${{\cal T}}$		 \\
\hline 
BD		& $\omega_{\rm BD}^{-1}\gtrsim 10^{-6} {\cal P} {{\cal T}}    $ 	& $\omega_{\rm BD}^{-1}\gtrsim 10^{-19}{\cal P} {{\cal T}}$ & $\omega_{\rm BD}^{-1}\gtrsim 10^{-1-5{\hat \alpha}} {\cal P} {{\cal T}}$ & $\omega_{\rm BD}^{-1}\gtrsim 10^{-15} {\cal P} {{\cal T}} $ & $\left[\frac{0.1}{S}\right]^{2}$ \\
EDGB 		& $\zeta_3\gtrsim 10^{-12} {\cal P} {{\cal T}}$ & $\zeta_3\gtrsim 10^{-25} {\cal P} {{\cal T}}$ & $\zeta_3\gtrsim 10^{-7-5{\hat \alpha}} {\cal P} {{\cal T}}$ & $\zeta_3\gtrsim 10^{-21} {\cal P} {{\cal T}} $ & $\left[\frac{\nu}{0.1}\right]^{4}  \left[\frac{1}{\delta_m}\right]^{2}$\\
DCS		& $\zeta_4\gtrsim 10^{6} {\cal P} {{\cal T}}$ & $\zeta_4\gtrsim 10^{-7} {\cal P} {{\cal T}}$ & $\zeta_4\gtrsim 10^{-7-5{\hat \alpha}} {\cal P} {{\cal T}}$ & $\zeta_4\gtrsim 0.003 {\cal P} {{\cal T}}$ & $\left[\frac{\nu}{0.1}\right]^{2} v_3^{-6} \left[\frac{1}{\beta_{\rm dCS}}\right]$ \\
\AE/Ho\v rava 	& ${\cal F}\gtrsim 10^{-9} {\cal P} {{\cal T}}$	& ${\cal F}\gtrsim 10^{-22}{\cal P} {{\cal T}}$	& ${\cal F}\gtrsim 10^{-4-5{\hat \alpha}} {\cal P} {{\cal T}}$	& ${\cal F}\gtrsim 10^{-18}{\cal P} {{\cal T}}$ & 1\\
\hline
coefficient ${\cal P}$		& $B_{8}^{2} M_{10}^{2} v_3^{-4}$ & $\rho_3^{\rm DM} M_{10}^{2} v_3^{-1} \left[\frac{R}{7\times 10^6 M}\right]^{3/2}  $ & $\gamma_{\hat \alpha} \rho_2^{\rm disk}M_{10}^{2} v_3^{2{\hat \alpha}-4} \left[\frac{R}{10 M}\right]^{{\hat \alpha}} $ & $q_3^2 v_3^{4}$ \\
\hline \hline
\end{tabular}
\label{tab:dotPmodified_dirtiness}
\end{table}
%

\section{Classical tests based on ``conservative'' effects}\label{sec:conservative}

\subsection{Advance rate of the periastron}
In Eq.~\eqref{periastron_shift} we have derived the correction to the periastron shift in the generic spherically symmetric geometry~\eqref{ansatz} defined by Eqs.~\eqref{PRA1}-\eqref{PRA2}. In the theory~\eqref{modifiedGR}, corrections to a BH geometry would be proportional to some power of $a_i$ only, whereas corrections to neutron-star geometries would be proportional to some power of $a_i$ and of $b_i$. Since $\alpha_i$ and $\beta_i$ in Eq.~\eqref{periastron_shift} are dimensionless, they will be proportional to $a_i^n$ (or to $b_i^n$) normalized by some appropriate power of $M$. Thus we obtain the order of magnitude
\begin{equation}
 \delta_{\rm per}\sim {\cal O}(1)\frac{M}{r_c} \left(\frac{a_i}{M^{x_i}}\right)^n \,,
\end{equation}
for the BH case, and a similar expression involving $b_i$ for the neutron-star case. In the equation above we have considered $\delta\gamma=0=\delta\beta$. We recall that, although for various alternative theories solar-system tests require $\delta\gamma, \delta\beta\ll1$, these bounds can be evaded by possible screening in some particular theory, e.g. the Vainshtein mechanism in massive gravity and Galileon theories.

In theories containing a single scalar field nonminimally coupled to gravity and preserving Lorentz invariance, the small-field deviations from GR are typically proportional to the square of the fundamental couplings. This can be understood by noting that the extra field itself is proportional to the coupling $a_i$ and that the stress-energy tensor is quadratic in the field, so that the corrections to the metric are proportional to $a_i^2$, i.e. $n=2$ in the equation above. On the other hand, theories like \AE-theory~\cite{Jacobson:2000xp,AE} or Ho\v rava gravity~\cite{HL,Blas:2009qj} introduce fields which are of zeroth order in the couplings. In these theories the corrections to GR are linear in the couplings, i.e. $n=1$ in the equation above. (See e.g. Refs.~\cite{Eling:2006ec,Barausse:2011pu,Blas:2011ni,Barausse:2013nwa,Barausse:2012ny,Barausse:2012qh} for BH solutions in these theories, where these 
corrections do indeed appear at linear order in the couplings.)

Our results in Table~\ref{tab:periastron} can be used to compare the shift induced by environmental effects to that related to beyond-GR effects. Comparing with thin disks we obtain an intrinsic limit on the couplings $a_i$ for beyond-GR effects to {\it stand out} against environmental effects:
\begin{equation}
 \left(\frac{a_i}{M^{x_i}}\right)^n\gtrsim 6\times 10^{-8} {f_{\rm Edd}}^{7/10} \left[\frac{M}{10^6 M_\odot}\right]^{6/5}\left(\frac{\alpha}{0.1}\right)^{-4/5}\,,
\end{equation}
We recall that, in order to compare with the corrections introduced by thin disks, we have used a fit that approximates a Shakura-Sunyaev surface density~\eqref{surfacedensity}. This result shows that the beyond-GR couplings must be sufficiently large for their corrections to be more significant than those produced by environmental effects.

\subsection{Shapiro time delay}
Another conservative effect that has been widely used as a classical test of GR is the gravitational time delay of light. The idea is to measure the difference of the time travel of a light signal in a gravitational field with respect to the flat case.

Let us consider the radial equation of a null-like particle for the generic metric~\eqref{ansatz}:
\begin{equation}
\dot{r}^2=\frac{B(r)}{A(r)}\left[E^2 - A(r) L^2/r^2\right]\,, \label{radialpotentialnull}
\end{equation}
where the ``dot'' denotes derivatives with respect to the proper time.
Because $\dot\phi=L/r^2$, we obtain:
\begin{equation}
\left(\frac{ d\phi}{dr}\right)^2=\frac{D^2}{r^2 B(r^2/A-D^2)}\,,
\end{equation}
where $D=L/E$ is the impact parameter (which reduces to the distance of minimum approach in the weak-field regime). Using $ds^2=0$ for light rays, Eq.~\eqref{ansatz} and the equation above we get:
\begin{equation}
 \frac{dt}{dr}=\sqrt{\frac{1}{AB}+\frac{1}{B}\frac{D^2}{r^2-AD^2}}\,. \label{dtShapiro}
\end{equation}
This result holds for a generic spherically symmetric spacetime. Once the trajectory of the light ray is known, the time delay can be computed as $T=\int \frac{dt}{dr} dr$. Let us consider a light signal sent from an emitter to a receiver and reflected back. The total time delay in GR reads~\cite{Will}:
\begin{equation}
 T_{\rm GR}\sim 2(D_e+D_p)+4M\log\left(\frac{4 D_e D_p}{D^2}\right)+M^2\left(-\frac{8 (D_e+D_p)}{D_e D_p}+\frac{8 \pi }{D}\right)\,, \label{ShapiroGR}
\end{equation}
where $D$ is again the distance of minimum approach of the signal from the massive object, $D_e$ and $D_p$ are the positions of the emitter and the receiver, respectively, and we have considered a large distance expansion and also the most favorable situation when the receiver is on the far side of the central object from the emitter, $D\ll D_e, D_p$ (superior conjunction). The most precise measurement of $\delta_T$ was performed by the Cassini spacecraft~\cite{Cassini} that has measured the fractional Doppler frequency
\begin{equation}
 y_{\rm GR}\equiv\frac{dT_{\rm GR}}{dt}=-\frac{8M}{D}\frac{dD}{dt}\left(1+\frac{\pi M}{D}\right)\,.
\end{equation}
On the other hand, using the deformed metric~\eqref{PRA1}--\eqref{PRA2}, we obtain:
\begin{eqnarray}
 T-T_{\rm GR}&\sim& 2M\delta\gamma\log\left(\frac{4 D_e D_p}{D^2}\right)\nn\\
 &+&M^2 \left(\frac{(D_e+D_p) \left(2 \delta \beta -6 \delta \gamma -2 \alpha _1-\beta _2\right)}{D_e D_p}+\frac{\pi  \left(6 \delta \gamma-2 \delta \beta +2 \alpha _1+\beta _2\right)}{D}\right)\,.\label{ShapiroT}
\end{eqnarray}
which corresponds to a deviation in the fractional frequency
\begin{equation}
 \delta_y\equiv \frac{y}{y_{\rm GR}}-1\sim \frac{2 D+3 M \pi }{4 D+4 M \pi }\delta \gamma +\frac{M \pi }{4 (D+M \pi )}\left(\alpha _1+\frac{\beta _2}{2}-\delta \beta \right)\,.
\end{equation}
For the Cassini mission $M\sim M_\odot$, $D\sim R_\odot$ and $\delta\gamma$ has been constrained roughly to the level of $2\times 10^{-5}$. Using the result above, we can estimate the constraints on the other parameters in the case $\delta\gamma\equiv0$. Note that there is a threefold degeneracy in the second-order correction. Assuming only a single parameter is turned on in the metric, we obtain:
\begin{equation}
 |\delta\beta|\lesssim6\,,\quad |\alpha_1|\lesssim 6\,,\quad \beta_2\lesssim 12\,.
\end{equation}
Given our small-coupling assumption, these bounds are not interesting. Higher order parameters can be included at next order in the large-distance expansion, but the corresponding limits would be even less stringent. Hence, as expected, this test of GR in the solar-system regime is not useful to constrain the higher order terms in our parametrized metric. However, if the central object is much more compact than the Sun, then $D\sim M$ and the corrections proportional to $\delta\gamma$ would be comparable to those proportional to $\alpha_1$, $\beta_2$ and $\delta\beta$.

Finally, let us consider the degeneracy with environmental effects. To give a simple characterization of some matter distribution with total mass $\delta M$, let us differentiate Eq.~\eqref{ShapiroGR} with respect to $M$. To first order we obtain
\begin{equation}
 T_{\rm matter}-T_{\rm GR}\sim 4\delta M\log\left(\frac{4 D_e D_p}{D^2}\right)\,.
\end{equation}
We compare this correction to Eq.~\eqref{ShapiroT}. Assuming the small corrections $\delta\gamma$, $\delta\beta$, $\alpha_i$ and $\beta_i$ to be proportional to $a^n/M^{nx_i}$, we obtain the intrinsic lower bound
\begin{eqnarray}
 \left(\frac{a_i}{M^{x_i}}\right)^n&\gtrsim&  10^{-6} \left(\frac{\delta M}{10^{-6}M}\right)^{} 		\qquad \hspace{3.55cm}\delta\gamma\neq0\,, \\
 \left(\frac{a_i}{M^{x_i}}\right)^n&\gtrsim& 5\times 10^{-5} \left[\frac{D}{M}\frac{\delta M}{10^{-6}M}\frac{\log\left({4 D_e D_p}/{D^2}\right)}{40}\right]^{} \qquad \delta\gamma=0 \,.
\end{eqnarray}
where the last bound holds if at least one of $\alpha_1$, $\beta_2$ or $\delta\beta$ is nonvanishing. Since the PPN parameter $\gamma$ is already highly constrained, in Table~\ref{tab:modifedVSdirty} we have reported the second constraint. Note that the values above are normalized by a very compact object with $D\sim M$. Solar-system experiments would give values larger by roughly three orders of magnitude.

\subsection{Deflection of light}
Another classical effect that is modified by the deformed geometry~\eqref{PRA1}--\eqref{PRA2} is the deflection of a light signal, i.e. the bending of a null geodesic due to a gravitational field. By using Eqs.~\eqref{radialpotentialnull} and $\dot\phi=L/r^2$ (where again the dot denotes derivatives with respect to the proper time), it is straightforward to show that the light trajectory is
\begin{equation}
 \frac{d^2 u}{d\phi^2}+u=M\left[3(1+\delta\gamma)u^2-\frac{\delta\gamma}{D^2}\right]-M^2\left[\frac{u \left(2 \delta \beta -2 \delta \gamma +8 D^2 u^2 \delta \gamma +\left(2-4 D^2 u^2\right) \alpha _1+\beta _2-2 D^2 u^2 \beta _2\right)}{D^2}\right]\,,
\end{equation}
where $u\equiv1/r$ and we have expanded to second order in $M/r$ and to first order in the deformation parameters. The perturbative solution of this equation reads:
\begin{eqnarray}
 u(\phi)&=&\frac{\sin\phi}{D}+M\frac{1+(1+\delta \gamma  ) \cos^2\phi}{D^2}+M^2\left[-\frac{3 (20 \phi  \cos\phi-9 \sin\phi+\sin(3\phi))}{16 D^3}\right.\nn\\
 &+&\left. \frac{8 (2 \delta \beta -5 \delta \gamma ) \phi  \cos\phi-4 (2 \delta \beta +5 \delta \gamma  \cos(2\phi)) \sin\phi+(-4 \phi  \cos\phi+3 \sin\phi+\sin(3\phi)) \left(2 \alpha _1+\beta _2\right)}{16 D^3}\right]\,,\nn
\end{eqnarray}
where we have set the integration constants to zero since they are irrelevant for the rest. Finally, the deflection angle $\hat\eta$ can be computed as the sum of the two asymptotic angles $\hat\eta_1$ and $\hat\eta_2$ defined via $u(-\hat\eta_1)=0$ and $u(\pi+\hat\eta_2)=0$. In the small angle limit we obtain
\begin{equation}
 \hat\eta=\frac{2M(2+\delta\gamma)}{D}+\frac{M^2}{D^2}\frac{\pi}{4}\left[15-4 \delta \beta +10 \delta \gamma +2 \alpha _1+\beta _2\right]\,, \label{bending}
\end{equation}
so that the relative deviation with respect to GR reads
\begin{equation}\label{deltabending}
 \delta\hat\eta\equiv\frac{\hat\eta}{\hat\eta_{\rm GR}}-1=\frac{\delta\gamma}{2}+\frac{M}{D}\frac{\pi}{32}\left(5 \delta \gamma-8 \delta \beta  +4 \alpha _1+2 \beta _2\right)\,.
\end{equation}
Beside the standard correction proportional to $\delta\gamma$, all other corrections are suppressed by a factor $M/D$ which, for the Sun, is about $M_\odot/R_\odot\sim 2\times 10^{-6}$. Hence, by assuming $\delta\gamma=0$, a measurement of the bending of light will put constraints on $\alpha_1$, $\beta_2$ and $\delta\beta$ which are 6 orders of magnitude less stringent than those actually in place for $\gamma$. The best constraint to date reads $\delta\hat\eta\lesssim3\times 10^{-4}$~\cite{Fomalont:2009zg}. On the other hand, if the central object is as dense as a neutron star, $M/D\sim 0.1$ and the constraints on the higher-order coefficients would be only slightly less stringent than those on $\delta\gamma$. 

As done above, in order to estimate the corrections due to some matter distribution with total mass $\delta M$, we differentiate Eq.~\eqref{bending} with respect to $M$ and set all the small parameters to zero. To first order we obtain
\begin{equation}
 \hat\eta_{\rm matter}-\hat\eta_{\rm GR}\sim 4\frac{\delta M}{D}\,.
\end{equation}
We compare this correction to Eq.~\eqref{deltabending}. Again, assuming the small corrections $\delta\gamma$, $\delta\beta$, $\alpha_i$ and $\beta_i$ are proportional to $a^n$, we obtain the intrinsic lower bound
\begin{eqnarray}
 \left(\frac{a_i}{M^{x_i}}\right)^n &\gtrsim&  10^{-6} \left(\frac{\delta M}{10^{-6}M}\right)^{} 		\qquad \hspace{0.3cm}\delta\gamma\neq0\,, \\
 \left(\frac{a_i}{M^{x_i}}\right)^n &\gtrsim& 10^{-6} \left[\frac{D}{M}\frac{\delta M}{10^{-6}M}\right]^{} \qquad \delta\gamma=0 \,,
\end{eqnarray}
and we have reported only the latter in Table~\ref{tab:modifedVSdirty}.

\section{Gravitational-wave tests based on ``dissipative'' effects}\label{sec:dissipative}
\subsection{Ringdown}
Detecting several modes of the gravitational waveforms can help disentangle the effects due to a modified theory from those due to extended matter distributions around massive BHs. However, this would require a detailed model of environmental effects, including knowledge of the density profile of the matter distribution around massive BHs, which is currently unknown.
%
Spin measurements might also be useful to break such degeneracy, as discussed in Sec.~\ref{sec:conclusions}.

A robust prediction of a non-GR effect is the isospectrality breaking for nonspinning and slowly-spinning BHs. As discussed above, polar and axial modes of the Schwarzschild geometry in GR are remarkably isospectral but this property is very fragile~\cite{Pani:2013ija}. Thus, a smoking gun of deviations from GR would be the existence in the gravitational signal of two very close modes with opposite parity which branch off from the degenerate mode in the GR limit. However, in practice axial modes are difficult to excite, so that the signal would be dominated by the first two polar modes (the fundamental mode and the first overtone).\footnote{Furthermore, as previously discussed, also matter configurations around compact objects would generically break mode isospectrality. In this case, a more detailed computation would be required to disentangle environmental effects from beyond-GR corrections.}

In the setup~\eqref{modifiedGR}, let us now estimate the corrections to the ringdown frequencies introduced by the new terms proportional to $a_i$ and $b_i$. If the object is a BH, the only corrections would be proportional to $a_i$. Furthermore, the only scale in the problem is the light crossing time. It follows that all quantities are normalized by $M$ and, in the small-coupling limit, the correction with respect to GR reads
\begin{equation}
 \frac{\delta\omega}{\omega}\sim {\cal O}(1)\left(\frac{a_i}{M^{x_i}}\right)^n\,,
\end{equation}
where the prefactor and the exponent $n$ are theory dependent and they must be computed in a case-by-case analysis. A constraint similar to the one above is in place for $b_i$ and it comes from the ringdown of compact stars. In this case the prefactor depends also on the stellar compactness, on the equation of state and on the mode family considered.

Assuming an accuracy of roughly $1\%$ \footnote{The accuracy depends on the SNR of the event, and the scaling is given in Section~\ref{sec:detection} (see also Refs.~\cite{Berti:2005ys,Berti:2007zu,Berti:2009kk}).} in detecting the BH ringdown modes~\cite{Berti:2009kk}, the relation above sets the order of magnitude of the constraints:
\begin{eqnarray}
 a_i&\lesssim&1.5^{x_i}\times 10^{-2/n}\left(\frac{M}{M_\odot}\right)^{x_i}{\rm km}^{x_i}\,, \label{airingdown}\\
 b_i&\lesssim&1.5^{y_i}\times 10^{-2/n}\left(\frac{M}{M_\odot}\right)^{y_i}{\rm km}^{y_i}\,, 
\end{eqnarray}

Let us discuss some examples, recalling that all theories listed in Table~\ref{tab:theories} have $n=2$ except for \AE-theory and Ho\v rava gravity, in which $n=1$. In Brans-Dicke theory in the Einstein frame $a_i=0$ and there is a single nonvanishing $b_i$, say $b\sim 1/\omega_{\rm BD}$. Since the latter is dimensionless it follows that $1/\omega_{\rm BD}<0.1$, i.e., a measurement of ringdown waves compatible with GR at the $1\%$ level can at most constrain $\omega_{\rm BD}>10$, independently of the BH mass. This result is much less competitive than current bounds which set $\omega_{\rm BD}>4\times 10^4$~\cite{Will:2005va}. The latter comes from the Shapiro delay measurement of the Cassini spacecraft by noting that in Brans-Dicke theory the time delay would acquire the same first order correction as in Eq.~\eqref{ShapiroT} with the substitution $\delta\gamma\to 1-{\cal G}/G\sim1/\omega_{\rm BD}$ once physical units are reinserted in the Jordan frame~\cite{Will}. On the other hand, quadratic theories of gravities have $b_i=0$, and, in the notation of Ref.~\cite{Pani:2011xm}, they have $a_i=f_i={\rm const}$ for $i=1,2,3,4$ and $a_i=0$ for $i>4$. Because $n=2$ and $x_i=2$, the ringdown constraint on the nonvanishing $a_i$ reads
\begin{equation}
f_i\lesssim 0.2\left(\frac{M}{M_\odot}\right)^{2}{\rm km}^{2}\,. \label{ringdownBD}
\end{equation}

Another popular class of theories are massive gravities. In their bimetric formulation~\cite{Hassan:2011zd}, they have ~--at linear level~\cite{Brito:2013wya}~-- a single nonvanishing coupling, $a=\mu_{\rm graviton}^2$, where $\mu_{\rm graviton}$ is the graviton mass in natural units. Because in this case $x=-1$, we get:
\begin{equation}
 \mu_{\rm graviton}\lesssim 0.2\left(\frac{M}{M_\odot}\right)^{-1}{\rm km}^{-1}\,. 
\end{equation}
Interestingly in this case $x<0$, and supermassive objects provide better constraints.

Finally, \AE-theory and Ho\v rava gravity are also mapped in our parametrization. In the notation of Ref.~\cite{Barausse:2013nwa} these theories are defined in terms of a set of coupling constants $c_i$. Since these couplings are all dimensionless, we obtain 
\begin{equation}
 c_i\lesssim 0.01\,, \label{ringdownAE}
\end{equation}
independently from the BH mass. In this case the constraint is more stringent than in the case of Brans-Dicke theory, because corrections to GR coming from \AE-theory and Ho\v rava gravity are linear in the couplings $c_i$, i.e. $n=1$ in Eq.~\eqref{airingdown}.

The estimates above agree well with more detailed computations when the latter are available and, in general, they will provide the correct order of magnitude. This analysis does not include \emph{new} sets of modes which generically appear in some particular modified theory. This is, for example, the case of massive perturbations which allow for quasibound states~\cite{Dolan:2007mj,Pani:2012vp,Brito:2013wya}. However, although long lived, such modes are exponentially suppressed at large distances, so that the gravitational waveform will be dominated by the deformed fundamental GR modes.

\subsubsection{Intrinsic lower limits due to environmental effects}	
Finally, let us compare these beyond-GR effects with those due to realistic astrophysical environments. From our results summarized in Table~\ref{bstable}, a small matter distribution with mass $\delta M$ would affect the ringdown frequency roughly at the level of $0.05\% [\delta M/(10^{-3}M)]$. This places an \emph{intrinsic} lower bound on possible ringdown constraints on modified theories. A simple comparison yields the lower bound:
\begin{equation}
 \left(\frac{a_i}{M^{x_i}}\right)^n\gtrsim 5\times 10^{-7}\left(\frac{\delta M}{10^{-6}M}\right)^{}\,, \label{intrinsicbounds}
\end{equation}
where $\delta M$ represents the mass of a quite generic matter distribution (as shown in Table~\ref{bstable}, matter-bumpy BHs, rings and short-hair BHs give comparable corrections).
Alternative theories whose couplings are much smaller than the bound above would produce smaller deviations to the ringdown frequencies than matter configurations surrounding massive BHs.

\subsubsection{Example: Chern-Simons gravity}	
As a check of our generic procedure, let us consider a particular theory, dynamical Chern-Simons (DCS) gravity~\cite{Alexander:2009tp}. This theory is equivalent to GR in the spherically symmetric case so that conservative effects due to a single nonspinning object are absent.
The ringdown modes of Schwarzschild BHs in this theory were computed in Ref.~\cite{Molina:2010fb}. In the small-coupling limit (which is consistent with Chern-Simon gravity being an effective theory), the corrections to the fundamental $l=2$ gravitational mode read
\begin{equation}
 (\delta_R,\delta_I)=(-2.1,3.7)\frac{\alpha_{\rm DCS}^2}{M^4}\,.
\end{equation}
From our results summarized in Table~\ref{bstable}, this places an intrinsic lower bound on possible ringdown constraint on $\alpha_{\rm DCS}$:
\begin{equation}
 \alpha_{\rm DCS}\gtrsim 3.4\times{\rm km}^2\left(\frac{\delta M}{10^{-3}M}\right)^{1/2}\left(\frac{M}{10 M_\odot}\right)^2\,.
\end{equation}
Considering that $[\alpha_{\rm DCS}]={\rm length}^2$, this number differs from the estimate~\eqref{intrinsicbounds} only by $30\%$.
Projected bounds on the Chern-Simons constant implies $\alpha_{\rm DCS}\sim {\cal O}(0.01-1){\rm km}^2$~\cite{Yagi:2012vf}. Our analysis of extended matter distributions summarized in Table~\ref{bstable} shows that such level of accuracy cannot be reached with ringdown tests whenever $\delta M\gtrsim 10^{-3} M$.

\subsection{Orbital decay rate}
Some of the most stringent constraints on modified gravity come from the astonishingly precise measurement of the orbital decay rate of binary pulsars~\cite{Will}, in particular of the period derivative $\dot P$. Computing this quantity requires knowledge of both conservative and dissipative effects. The former modify the Hamiltonian of the binary system and, in turn, the relation between the period and the total energy. The latter modify the GW emission and, in turn, the secular changes of the orbital parameters. A precise computation of these corrections necessarily requires a case-by-case analysis. This program started with Brans-Dicke theory and has been recently extended to other cases~\cite{Alsing:2011er,Yagi:2011xp,Yagi:2013qpa}. We attempt here an approximate parametrization that should nonetheless accommodate most deformations\footnote{Some of our discussion has some overlap with the recent analysis of Ref.~\cite{Stein:2013wza}, where a precise characterization of theories with an extra scalar field has been developed. Our discussion in this section will be more generic and necessarily approximate.}. We follow the ppE formalism~\cite{Yunes:2009ke} and parametrize the orbital distance, the energy and the energy flux as~\cite{Chatziioannou:2012rf}, respectively,
\begin{eqnarray}
 r&=&\frac{M_T^{1/3}}{\Omega_\phi^{2/3}}\left[1+\frac{{\cal A}}{6}{\tilde{p}}(M_T\Omega_\phi)^{2{\tilde{p}}/3}\right]\,, \label{ppE1}\\
 E&=&\frac{\mu M_T}{2r}\left[1+{\cal A}\left(\frac{M_T}{r}\right)^{\tilde{p}}\right]\,, \label{ppE2}\\
 \dot E&=&\dot E_{\rm GR}\left[1+{\cal B}\left(\frac{M_T}{r}\right)^{\tilde{q}}\right] \label{ppE3}\,,
\end{eqnarray}
where $\mu\equiv\nu M_T$, $\nu$ is the symmetric mass ratio, $M_T$ is the total mass of the binary, ${\cal A}$ and ${\cal B}$ depend on the fundamental couplings of the theory and the GR terms were written to lowest PN order. The terms ${\tilde{p}}$ and $\tilde{q}$ are dimensionless quantities that depend on the specific theory. For simplicity we assume circular orbits whose orbital velocity is $v\equiv \Omega_\phi r$. Inverting this relation we get, to lowest order,
\begin{equation}
 \Omega_\phi=\frac{v^3}{M_T}\left[1-\frac{{\cal A}}{2}{\tilde{p}} v^{2{\tilde{p}}}\right]\,.
\end{equation}
It is now straightforward to compute the period derivative by using the chain rule:
\begin{equation}
 \dot P\equiv \frac{\partial P}{\partial\Omega_\phi}\frac{\partial\Omega_\phi}{\partial E}\dot E=-\frac{P}{\Omega_\phi}\left[\frac{\partial E}{\partial \Omega_\phi}\right]^{-1}\dot E\,. \label{dotPppE}
\end{equation}
Using the parametrization~\eqref{ppE1}--\eqref{ppE3} we obtain
\begin{equation}
 \frac{\dot P}{P}\sim \left(\frac{\dot P}{P}\right)_{\rm GR}\left[1+\frac{{\cal A}}{6}\left({\tilde{p}}({\tilde{p}}-3)-6\right)v^{2{\tilde{p}}}+{\cal B} v^{2{\tilde{q}}}\right]\,.\label{dotPppE2}
\end{equation}
In the simplest ppE model, the waveform in Fourier space is parametrized as:
\begin{equation}
 \tilde{h}=\tilde{h}_{\rm GR}\left(1+\alpha_{\rm ppE}u^{a_{\rm ppE}}\right)e^{i\beta_{\rm ppE}u^{b_{\rm ppE}}}\,,
\end{equation}
where $u=(M_c\Omega_\phi)^{1/3}$, $M_c$ is the chirp mass and the ppE parameters $\alpha_{\rm ppE}$, $\beta_{\rm ppE}$, $a_{\rm ppE}$ and $b_{\rm ppE}$ are related to ${\cal A}$, ${\cal B}$, ${\tilde{p}}$ and $\tilde{q}$. Using the explicit mapping provided in Ref.~\cite{Chatziioannou:2012rf} we can write the period derivative in terms  $\alpha_{\rm ppE}$, $\beta_{\rm ppE}$, $a_{\rm ppE}$ and $b_{\rm ppE}$. We obtain:
\begin{equation}
 \delta_{\dot P/P}\equiv\frac{\dot P/P}{(\dot P/P)_{\rm GR}}-1=C \nu ^{1+\frac{b_{\rm ppE}}{5}} v^{5+b_{\rm ppE}}\,, \label{dotPoP}
\end{equation}
with
\begin{equation}
 C=\left\{\begin{array}{c}
           -\left[\frac{43}{4}+\frac{9 b_{\rm ppE}}{4}-\frac{6}{5+b_{\rm ppE}}\right]\alpha_{\rm ppE}+\left[\frac{2 (b_{\rm ppE}-3) b_{\rm ppE} (151+b_{\rm ppE} (80+9 b_{\rm ppE}))}{15 (5+b_{\rm ppE})}\right]\beta_{\rm ppE}\qquad {\tilde{p}}=\tilde{q}\\
           \frac{8 (3-b_{\rm ppE}) b_{\rm ppE} [b_{\rm ppE} (4+b_{\rm ppE})-29] }{15 [81+b_{\rm ppE} (46+5 b_{\rm ppE})]} \beta_{\rm ppE} \qquad \hspace{6.2cm} {\tilde{p}}>\tilde{q}\\
           \frac{16}{15} (3-b_{\rm ppE}) b_{\rm ppE} \beta_{\rm ppE} \qquad \hspace{7.95cm} \tilde{q}>{\tilde{p}}  \\
          \end{array}\right.\,.
\end{equation}
We note that not all ppE parameters are independent in the case at hand. Indeed, $b_{\rm ppE}=a_{\rm ppE}-5$ for any value of ${\tilde{p}}$ and ${\tilde{q}}$~\cite{Chatziioannou:2012rf}. Furthermore, if ${\tilde{p}}\neq {\tilde{q}}$, $\beta_{\rm ppE}$ and $\alpha_{\rm ppE}$ are not independent:
\begin{equation}
 \alpha_{\rm ppE}/\beta_{\rm ppE}=\left\{\begin{array}{c}
           \frac{16 ({\tilde{p}}-4) (2 {\tilde{p}}-5) [{\tilde{p}} (5 {\tilde{p}}-4)-6]}{15 [{\tilde{p}} (5 {\tilde{p}}-2)-6]} \qquad {\tilde{p}}>{\tilde{q}}\\
           \frac{16}{15} ({\tilde{q}}-4) (2 {\tilde{q}}-5) \qquad \hspace{0.8cm} {\tilde{q}}>{\tilde{p}}  \\
          \end{array}\right.\,.
\end{equation}
Therefore, when $\tilde{p}\neq \tilde{q}$, $\alpha_{\rm ppE}\equiv0$ implies $\beta_{\rm ppE}\equiv0$ unless $p$ or ${\tilde{q}}$ acquire some specific values. 

Because we are interested in binary systems where $v\ll1$, Eq.~\eqref{dotPoP} implies that such systems can provide large corrections to $\dot P$ if the underlying theory of gravity corresponds to $b_{\rm ppE}<-5$. This is the case for theories that allow for monopole or dipole emission, like Brans-Dicke theory, Einstein-dilaton-Gauss-Bonnet (EDGB) theory, \AE-theory and Ho\v rava gravity. In Table~\ref{tab:dotPmodified} we present the correction $\delta_{\dot P/P}$ for some specific theory and we derive typical order-of-magnitude constraints on the coupling constant of the theories assuming a measurement of $\delta_{\dot P/P}$ at the level of $5\%$. These constraints are admittedly simplistic as a precise computation may have to account for eccentricity, higher post-Newtonian orders, and spin effects. Nonetheless, Eq.~\eqref{dotPoP} serves as an estimate of the potential of orbital decay measurements to test GR.

Finally, it is important to point that the parametrization~\eqref{dotPoP} does not include all possible deformations from GR. For example it does not include corrections due to massive scalar fields~\cite{Alsing:2011er}. These and other possible corrections can nonetheless be included by modifying the parametrization~\eqref{ppE1}--\eqref{ppE3}, although this is beyond our scope.

\begin{table}[hbt]
\centering \caption{Dominant corrections to the orbital decay rate for circular inspiral in some modified theories of gravity. The ppE parameters and some definitions are taken from Table~3 in Ref.~\cite{Yunes:2013dva} except for the definition of $\beta_{\rm dCS}$ which differs from the original by a factor $\zeta_4/\nu^{14/5}$. For \AE-theory and Ho\v rava gravity, the function ${\cal F}(c_i,s_i)$ depends on the coupling constants $c_i$ of the theory and on the sensitivities $s_i$. In the small coupling limit ${\cal F}$ is linear in the couplings, cf. Refs.~\cite{Yagi:2013ava,Yagi:2013qpa} for details. } 
\vskip 12pt
\begin{tabular}{c|cc}
\hline \hline
Theory     &$\delta_{\dot P/P}$ & Constraint \\
\hline 
BD			&$\frac{5 S^2}{48 v^2 \omega_{\rm BD}}$	& $\omega_{\rm BD}\gtrsim 3\times 10^3\left[\frac{S}{0.1}\right]^2\left[\frac{10^{-3}}{v}\right]^2\frac{0.3}{\delta_{\dot P/P}}$	\\
EDGB	&$\frac{5 \delta_m^2 \zeta_3}{48 v^2 \nu ^4}$	& $\zeta_3\lesssim 3\times 10^{-10}\left[\frac{v}{10^{-3}}\right]^{2}\left[\frac{\nu^4}{0.1}\right]^4\left[\frac{1}{\delta_m}\right]^{2}\left[\frac{\delta_{\dot P/P}}{0.3}\right]$	\\
DCS		&$\frac{32 v^4 \beta_{\rm dCS} \zeta_4}{3 \nu ^2}$& $\zeta_4\lesssim3\times 10^8 \left[\frac{10^{-3}}{v}\right]^{4} \left[\frac{\nu}{0.1}\right]^{2} \left[\frac{1}{\beta_{\rm dCS}}\right] \left[\frac{\delta_{\dot P/P}}{0.3}\right]$	\\
\AE-theory/Ho\v rava gravity	&$\frac{{\cal F}(c_i,s_i)}{v^2}$	& ${\cal F}(c_i,s_i)\lesssim 4\times 10^{-7}\left[\frac{10^{-3}}{v}\right]^2\frac{0.3}{\delta_{\dot P/P}}$	\\
\hline \hline
\end{tabular}
\label{tab:dotPmodified}
\end{table}
%

\subsubsection{Degeneracy with environmental effects}
The presence of matter distributions mostly introduces conservative corrections to the waveform. However, dynamical friction introduces \emph{dissipative} effects that modify the flux formula, cf. Eq.~\eqref{dotEDF}. Such corrections require an extension of the standard ppE formalism to account for logarithmic terms in Eq.~\eqref{ppE3}. To include dynamical friction effects, we generalize Eq.~\eqref{ppE3} as
\begin{equation}
 \dot E=\dot E_{\rm GR}\left\{1+{\cal B}\left(\frac{M_T}{r}\right)^{\tilde{q}}+{\cal C}\left(\frac{M_T}{r}\right)^{\tilde{s}}\log\left[\tilde{\gamma}\left(\frac{r}{M_T}\right)^{\tilde{t}}\right]\right\}\,, \label{dEDF}
\end{equation}
where $\tilde{s}$, $\tilde{t}$, $\tilde{\gamma}$ and ${\cal C}$ are extra dimensionless parameters. It is easy to show that the relative correction to the orbital decay rate~\eqref{dotPppE2} is generalized as
\begin{equation}
 \frac{\dot P}{P}\sim \left(\frac{\dot P}{P}\right)_{\rm GR}\left\{1+\frac{{\cal A}}{6}\left({\tilde{p}}({\tilde{p}}-3)-6\right)v^{2{\tilde{p}}}+{\cal B} v^{2{\tilde{q}}}+{\cal C} v^{2 \tilde{s}} \log\left[\frac{\tilde{\gamma}}{v^{2 \tilde{t}}}\right]\right\}\,. \label{dotPppE3}
\end{equation}
This expression contains both conservative corrections proportional to ${\cal A}$ and dissipative corrections proportional to ${\cal B}$ and ${\cal C}$. The results above can be reduced to the case of dynamical friction in two special cases: 
\begin{enumerate}
 \item For nonrelativistic supersonic motion, ${\cal M}\gg4$ and $v\ll1$, Eq.~\eqref{dotEDF} corresponds to ${\cal B}=0$, $\tilde{s}=11/8$, $\tilde t=1/2$, $\tilde{\gamma}=v_s M_T/(0.22 m_{\rm sat})$ and
 \begin{equation}
  {\cal C}\sim 3.4\times 10^{-18} f_{\rm Edd}^{11/10} (m_{\rm sat}/M_\odot)^2 (0.1/\alpha)^{7/10}(10^6 M_\odot/M_T)^{7/10} \,, \label{calC}
 \end{equation}

 \item In the case of subsonic motion, ${\cal M}\ll1$, Eq.~\eqref{dotEDF} corresponds to ${\cal C}=0$, $\tilde{q}=15/8$ and
 \begin{equation}
  {\cal B}\sim-4.8\times 10^{-18} f_{\rm Edd}^{11/10} (m_{\rm sat}/M_\odot)^2 v_s^{-1} (0.1/\alpha)^{7/10}(10^6 M_\odot/M_T)^{7/10}\,.  \label{calB}
 \end{equation}

\end{enumerate}
The intermediate case in which ${\cal M}\sim{\cal O}(1)$ is more involved because it cannot be recast in the form~\eqref{dEDF}.

Let us now focus on conservative corrections, setting ${\cal B}={\cal C}=0.$\footnote{The analysis of the dissipative corrections in Eq.~\eqref{dEDF} for the GW signal of an EMRI is performed in the next section. We remark that conservative corrections are important only in combination with dissipation, as provided by Eq.~\eqref{dEDF}. The latter coincides with GR when ${\cal B}={\cal C}=0$.} As a representative example, we consider the power-law density distribution~\eqref{powerlawdensity}. For that density profile and to lowest order, we get an equation similar to Eq.~\eqref{ppE1} if we identify
\begin{equation}
{\tilde{p}}={\hat \alpha}-3\,, \qquad {\cal A}=\frac{8\pi M_T^{2-{\hat \alpha}} R^{\hat \alpha} \rho_0}{({\hat \alpha}-3)^2}\,.
\end{equation}
We therefore obtain the changes in the orbital decay rate due to this matter configuration
\begin{equation}
 \delta_{\dot P/P}\sim \frac{4\pi M_T^{2-{\hat \alpha}}   R^{{\hat \alpha} }  (12+({\hat \alpha}-9 ) {\hat \alpha} ) \rho_0}{3 ({\hat \alpha} -3)^2}v^{2 ({\hat \alpha}-3)}\,.\label{deltaPoP_dirtiness}
\end{equation}
Interestingly, when ${\hat \alpha}>3$ the correction to the orbital decay rate enters at higher PN order than the leading order GR effect. This condition corresponds to a finite total mass for the density configuration. On the other hand, DM profiles are compatible with Eq.~\eqref{powerlawdensity} with ${\hat \alpha}<3$~\cite{Eda:2013gg}. In addition, the effects of a constant magnetic field $B$, of a cosmological constant $\Lambda$ and of an electric field with charge $Q\equiv q M$ are all described by Eq.~\eqref{deltaPoP_dirtiness} with the identifications:
\begin{eqnarray}
 {\hat \alpha}&=&0\,,\qquad \rho_0=-\frac{3B^2}{8 \pi }\sim -1.3\times 10^5 {\rm kg}/{\rm m}^3\left(\frac{B}{10^{12}{\rm Gauss}}\right)^2 \,, \label{rho0toB}\\ 
 {\hat \alpha}&=&0\,,\qquad \rho_0=\frac{\Lambda }{4\pi } \sim 10^{-26} {\rm kg}/{\rm m}^3 \frac{\Lambda}{10^{-52}{\rm m}^{-2}}\,,\\
 {\hat \alpha}&=&4\,,\qquad \rho_0=\frac{q^2}{4\pi }\left(\frac{M_T}{R}\right)^4\frac{1}{M_T^2}\sim 8\times 10^8 {\rm kg}/{\rm m}^3 \left[\frac{q}{10^{-3}}\right]^{2} \left[\frac{5M_T}{R}\right]^{4} \left[\frac{10 M_\odot}{M_T}\right]^{2}\,, \label{rho0toQ}
\end{eqnarray}
which correspond to
\begin{eqnarray}
 \delta_{\dot P/P}&\sim& -\frac{2 B^2 M_T^2}{3 v^6} \,, \label{dotPB} \\
 \delta_{\dot P/P}&\sim& \frac{4 M_T^2 \Lambda }{9 v^6} \,, \label{dotPLambda}\\
 \delta_{\dot P/P}&\sim& -\frac{8}{3} q^2 v^2 \,, \label{dotPq}
\end{eqnarray}
respectively.

In principle, this approach can be used to put \emph{intrinsic constraints} on the ppE parameters, i.e. constraints that are theory-independent. Such constraints can then be translated to lower bounds on the couplings of \emph{any} specific theory for which a ppE parametrization is available. For concreteness, in Table~\ref{tab:dotPmodified_dirtiness} we present the intrinsic limits on some modified theories of gravity due to the presence
of various forms of matter,
namely a magnetic field defined via Eq.~\eqref{rho0toB}, a DM profile given by Eq.~\eqref{powerlawdensity} with $\rho_0=10^3 M_\odot/{\rm pc^{3}}$, $R\sim 7\times10^6M$ and ${\hat \alpha}=3/2$,
a generic distribution~\eqref{powerlawdensity} normalized by a typical disk density $\sim 10^2{\rm kg}/{\rm m}^3$ [cf. Eq.~\eqref{eq:rhoF}]~\footnote{Because for simplicity we consider only spherically-symmetric distributions, the mass of our configuration is larger than that of a thin disk with profile given Eqs.~\eqref{eq:rhoF} and \eqref{eq:H}. Therefore, our estimates should be considered as an upper limit.},
and  an electric charge defined via Eq.~\eqref{rho0toQ}.

As shown in the previous sections (cf. Tables~\ref{tab:periastron} and \ref{tab:dephasing}), tests of GR will be very difficult to perform with EMRIs in thin-disk environments, where the effects of planetary migration, dynamical friction and accretion can be comparable to or even larger than GW emission itself. Therefore, here and in the next section we focus on environmental effects that are \textit{smaller} than GW emission. Table~\ref{tab:dotPmodified_dirtiness} shows that the deviations from GR have to be sufficiently large if their effect is to dominate over these environmental corrections.
The specific limits depends strongly on ${\hat \alpha}$ and on the matter content. Nonetheless, Tables~\ref{tab:dotPmodified} and \ref{tab:dotPmodified_dirtiness} show that, in the region which is phenomenologically allowed to date,
the effects due to modifications of GR can still be larger than the matter effects.
On the one hand, our analysis shows that a magnetic field $B\gtrsim 5\times 10^8 {\rm Gauss}$ would make it impossible to improve the current best bound, $\omega_{\rm BD}\gtrsim 4\times 10^4$, even assuming infinite precision in the measurement of $\dot P$.
On the other hand, except for such extreme situations, tests of GR using orbital decay rates are robust even in the presence of matter distributions. In principle, by assuming GR is correct, such tests can be used to probe matter around binary systems. To the best of our knowledge this is the first time that a similar analysis has been performed.

\subsection{Parametrized EMRIs}
In the extreme-mass ratio limit the smaller body orbits the massive central object along geodesics of the background spacetime to lowest order in the mass ratio. The orbital parameters evolve secularly due to dissipative effects. By combining our previous results, we can parametrize the background metric using the expansion~\eqref{PRA1}--\eqref{PRA2}, whereas we use Eq.~\eqref{dEDF} to parametrize the energy dissipation. It worth stressing that, in general, the coefficients $\alpha_i$, $\beta_i$ and ${\cal B}$ are not independent but related to each other through the coupling parameters of the theory. 

We focus on circular orbits whose frequency reads $\Omega_\phi=\sqrt{M/r^3}+{\cal O}(\delta\beta,\delta\gamma,\alpha_i)$. A relevant quantity is the total number of cycles during the inspiral from $r_i$ to $r_f$. This reads
\begin{equation}
 {\cal N}=\int_{t_i}^{t_f}dt f =\int_{r_i}^{r_f}dr \frac{f}{\dot E_p}\frac{\partial E_p}{\partial r}\,,
\end{equation}
where $f=\Omega_\phi/\pi$ is the GW frequency and $E_p$ is the binding energy of the particle in circular motion. We can evaluate the integral above using the balance law $\dot E_p=-\dot E$, where $\dot E$ is defined in Eq.~\eqref{dEDF}, and assuming for simplicity that $\dot E_{\rm GR}$ is given by the quadrupole formula~\eqref{quadrupole}. 

In order to obtain a simple analytical result let us consider $r_f=r_{\rm ISCO}$ and $r_i=r_{\rm ISCO}(1+\epsilon)$ with $\epsilon\ll1$. A similar analysis can be performed using gauge-invariant quantities like the frequency, but we found it convenient to use the orbital radius instead. By expanding all quantities to first order in the deformations, we obtain
\begin{eqnarray}
 \delta_{\cal N}&=&-6^{-\tilde{q}}{\cal B}-6^{-\tilde{s}}{\cal C}\log(6^{\tilde{t}} \tilde{\gamma})+\frac{\delta \beta -\delta \gamma }{9}+\frac{11 \alpha _1}{18}-\frac{\alpha _3}{54}-\frac{13 \alpha _4}{648}-\frac{\alpha _5}{108}-\frac{25 \alpha _6}{7776}-\frac{67 \alpha _7}{69984}-\frac{217 \alpha _8}{839808}+{\cal O}(\epsilon)\,, \label{deltaN}
\end{eqnarray}
where $\delta_{\cal N} \equiv \frac{\delta {\cal N}}{\delta {\cal N}_{\rm GR}}-1$ and we have truncated the sum in Eqs.~\eqref{PRA1}--\eqref{PRA2} to $N_\alpha=N_\beta=8$. 
The result above is the most generic correction within the assumptions of spherical symmetry, circular orbits, small deformations around Schwarzschild and assuming a correction to the quadrupole formula as in Eq.~\eqref{dEDF}. This result depends only on the metric function $A(r)$ and not on $B(r)$. The dissipative corrections are proportional to ${\cal B}$ and ${\cal C}$ and the prefactors $6^{-\tilde{q}}$ and $6^{-\tilde{s}}$ make these terms decrease quickly as $\tilde{q}$ or $\tilde s$ increase. In particular, Eq.~\eqref{deltaN} also contains the corrections due to dynamical friction which, in the case of supersonic motion, correspond to $\tilde{s}=11/8$, $\tilde{t}=1/2$ and ${\cal C}$ given by Eq.~\eqref{calC}.

Finally, even though we have truncated the series to $N_\alpha=8$, it is clear that there is a hierarchy of terms. This gives further confirmation that the parametrization~\eqref{PRA1}--\eqref{PRA2} is efficient, i.e. that higher-order terms give subdominant contributions.

Equation~\eqref{deltaN} shows that the main contributions to ${\cal N}$ come from the corrections proportional to ${\cal B}$ and ${\cal C}$ if $\tilde{q}\leq1$ or $\tilde{s}\leq1$, whereas they come from those proportional to $\alpha_1$ is $\tilde q$ and $\tilde s$ are large. On the other hand, possible deviations from the GR result accumulate during the inspiral, so larger corrections are expected if $\epsilon\gg1$. Using only the dominant correction proportional to $\alpha_1$, we obtain, to any order in $\epsilon$,
\begin{equation}
 \delta_{\cal N}\sim \frac{5 \left(-6 \epsilon^2+57 \epsilon^3+19 \epsilon^4+15 \left(-1+\sqrt{1+2 \epsilon}\right)+15 \epsilon \left(-3+2 \sqrt{1+2 \epsilon}\right)\right)}{18 (1+2 \epsilon) \left(1+\epsilon+7 \epsilon^2+3 \epsilon^3-\sqrt{1+2 \epsilon}\right)}\alpha_1\,.
\end{equation}
As expected the quantity above is an increasing function of $\epsilon$, and $\delta_{\cal N}\sim0.9 \alpha_1$ as $\epsilon\to\infty$.

Finally, to estimate the conservative effects of the environment in this case we can set ${\cal B}={\cal C}=0$ and differentiate the GR result with respect to $M$. We obtain $\delta_{\cal N}^{\rm matter}\sim \delta M/M$. Comparing this quantity with the equation above and assuming that $\alpha_1\propto a_i^n$, we obtain the intrinsic lower limit presented in Table~\ref{tab:modifedVSdirty}. 
On the other hand, dissipative effects are typically small, because they are proportional to the small quantities ${\cal B}$ and ${\cal C}$, cf. Eqs.~\eqref{calC} and \eqref{calB}.

\subsection{Monopole radiation}
In the previous sections we have discussed how some of the most stringent constraints on modified gravity arise from considering dipolar emission, since the latter is forbidden in GR. Indeed, GR being quadrupolar in nature, any monopole and dipole emission would be a clear signature of deviations from Einstein's theory. Monopolar emission has been much less explored, with some exception in the study of spherically symmetric collapse in scalar-tensor theories~\cite{Shibata:1994qd,Novak:1997hw,Nakao:2000ug}.

Spherically symmetric motion emits, generically, monopolar radiation in any theory with some (effective or fundamental) scalar degree of freedom. In particular, spherical oscillations of compact stars or BHs and spherically symmetric collapse can be strong emitters of monopolar radiation in modified gravity, whereas they do not source GWs in GR. Supernova explosions are believed to be nearly spherical and, therefore, in modified gravity they would release a considerable amount of their enormous energy through spherical scalar waves. 

Because of the spherical symmetry, it is easy to estimate the signal associated to monopole emission on dimensional ground. The amplitude of the wave scales as
\begin{equation}
 h\sim \bar\gamma\frac{M}{R}\frac{M}{D_L}\,,
\end{equation}
where $D_L$ is the luminosity distance of the source from the detector, $M$ and $R$ are the typical mass and radius, respectively, of the remnant produced in the collapse, and $\bar\gamma$ is a dimensionless coupling that parametrizes deviations from GR. Typically, $\bar\gamma$ is proportional to some power of $a_i$ and $b_i$ as appearing in the action~\eqref{modifiedGR}.\footnote{However, in theories in which weak-field corrections are suppressed, the response of a detector to scalar waves in the far zone can vanish at 
linear order, i.e. $\bar\gamma=0$. An example of one such theory is scalar-tensor gravity with a conformal factor $\sim e^{\beta \varphi^2}$, in which case the linear coupling of the scalar field to matter in the Einstein frame is exactly zero (cf. e.g. Refs.~\cite{Damour:1996ke,Novak:1997hw,ST1}).} The quantity to be compared with detector sensitivity ${\cal S}$ is $\tilde{h}\sqrt{f_{\rm GW}}$, where $\tilde{h}\sim h T$ is the Fourier spectrum of the signal, $T\sim R$ is the typical time scale of the problem and it is related to the typical length of the final object, $R$, whereas
\begin{equation}
 f_{\rm GW}\sim 1/R\,,
\end{equation}
is the characteristic frequency of the radiation. Then, assuming no detection of GWs from a spherical source at a luminosity distance $D_L$, we obtain an order-of-magnitude estimate for the upper bounds on $\bar\gamma$:
\begin{equation}
 \bar\gamma\lesssim 4\times 10^{-6}\left(\frac{D_L}{1{\rm kpc}}\right)\left(\frac{{\cal S}}{10^{-23}{\rm Hz}^{-1/2}}\right)\left(\frac{R}{30{\rm km}}\right)^{1/2}\left(\frac{10M_\odot}{M}\right)^2\,, \label{boundmonopole}
\end{equation}
where we have normalized the result by a typical peak sensitivity of Advanced LIGO and for typical source parameters. For example, if a supernova explosion will occur within $10{\rm kpc}$ during Advanced LIGO operational time, the absence of detection of corresponding GWs would imply a bound on $\bar\gamma$ roughly of the order of $4\times 10^{-5}$. However, such events are extremely rare within $10{\rm kpc}$ from Earth. Our analysis reduces to the well-studied case of Brans-Dicke scalar-tensor gravity when $\bar{\gamma}\sim1/\omega_{\rm BD}$. Indeed, the order of magnitude of the estimate above agrees with the detailed analysis of Ref.~\cite{Hayama:2012au}, where scalar-wave detectability with KAGRA has been studied (see also Ref.~\cite{PhysRevD.50.7304} for earlier work).

Similarly, we might retroactively analyze recent supernova explosions that occurred during the activity of first-generation GW detectors. Two examples are SN2008D and SN2011fe, which were located at about $9\times 10^7$ light years and $2\times 10^7$ light years from Earth, respectively. Assuming ${\cal S}\sim 10^{-21}{\rm Hz}^{-1/2}$, $M\sim 10 M_\odot$ and $R\sim 2M$, we obtain that the absence of GW detection from these sources implies a mild upper bound on $\bar\gamma$ of the order of unity.
\part{Conclusions and Appendices}\label{part:conclusions}
\section{Conclusions and extensions}\label{sec:conclusions}
The advent of GW astronomy demands for a careful quantification of the impact of realistic astrophysical environments on the GW signal.
Our results strongly suggest that GW astronomy can become a {\it precision} discipline: given an appropriate and sensitive detector,
astrophysical environmental effects are small and do not prevent a precise mapping of the compact-object content of the visible universe.
Moreover, if adequately modeled, GWs might be used in the most optimistic scenarios to study matter configurations around compact objects
as is routinely done in the electromagnetic band. We have presented a survey of several environmental effects in two situations of great interest for GW astronomy: the ringdown emission of massive BHs and the two-body inspiral of compact objects. We have studied the GW signal associated to the presence of electric charges, magnetic fields, cosmological evolution, matter disks and halos and finally the effects of possible deviations from GR.

Our analysis revealed novel effects related to the ringdown modes in the presence of environmental effects. The QNM spectrum of nonisolated BHs can be drastically different from that of isolated BHs, yet the BH response to external perturbations is unchanged at the relevant time scales for ringdown. This result is interesting on its own and would deserve an independent study. In particular, it would be interesting to see if such resonances can be excited during an extreme-mass ratio inspiral around BH surrounded by matter, thus providing a clean GW signature of matter configurations around compact objects. Toy models suggest this will happen at low frequencies when the inspiralling object is far away from the BH. Head-on collisions could also excite these modes which would then presumably show up at very late-times competing with Price's power-law tails. 
More realistic models are clearly necessary to understand the full implications of these results. A curiosity with possible observational effects is that matter {\it close}
to the event horizon also gives rise to such modified modes and response. This might have important implications for the so-called ``firewall'' proposal~\cite{Almheiri:2012rt}: the analysis in Sec.~\ref{sec:firewall} shows that any localized field with mass $\delta M\gtrsim 10^{-4}M$ close to the horizon of a massive BH would affect the ringdown frequencies to detectable levels. Investigating the implications of this effect for the gravitational waveforms in realistic scenarios is an interesting extension of our analysis. 

Self-force in BH spacetimes has been shown to be directly connected to the QNMs structure~\cite{Casals:2009zh,Casals:2013mpa}.
In view of the sometimes dramatic change in the QNM spectrum of nonisolated BHs, an open question is whether self-force outside ``dirty'' BHs also shows the same exquisite dependence on the environment configuration and distribution.

We note that the ringdown spectrum in the presence of matter at large distance shares many features with the spectrum of light massive fields around BHs (cf. Ref.~\cite{Cardoso:2013krh} for a review). In this case, the field is localized at $\sim1/\mu$, where $\mu$ is the mass of the field. It is known that also in this case the late-time behavior is drastically modified and a novel family of modes emerges which does {\it not} reduce to GR when the mass goes to zero. 

Overall, our results confirm the generic claim that environmental effects can be safely neglected for detection. However, in some situations such effects become important for parameter estimation and for searches of new physics. Environmental effects can in some cases be comparable to first- and second-order self-force corrections and can put intrinsic limits on our ability to test modified theories of gravity. We have provided order-of-magnitude estimates of these limits that are largely model- and theory-independent. Clearly, a precise analysis might in principle disentangle environmental effects from self-force and beyond-GR corrections but, given our ignorance of matter configurations around compact objects, a precise modeling of the dirtiness signal would be mandatory to perform such an analysis. 

In general the corrections to the ringdown signal are smaller than those affecting the inspiral waveforms. This is expected because the latter can accumulate during the last stages of the inspiral.  In addition, our analysis selects some ``smoking guns'' of dirtiness in the ringdown signal: isospectrality breaking, monopole and dipole radiation, novel families of modes, resonant effects in the GW flux and changes of the late-time GW signal.

The main limitation of our study is the fact that we neglected spin effects. Compact objects are usually spinning and angular momentum plays a crucial role in the GW signal. Spin modifies the multipolar structure of the background geometries as well as the geodesic motion and the GW emission. We have considered only static configurations and our approach has the merit of being very general within this strong assumption. Including spin effects for compact objects in a model- and theory-independent framework is a remarkable open problem which goes beyond the scope of this work. Because spin effects are typically dominant over environmental effects, one does not expect
any degeneracy problem. Furthermore, most of the effects we discuss are not directly related to the spin, so that we expect our analysis can capture the correct order of magnitude also in the case of spinning objects. 

One important exception to the above are near-extremal BHs. In this case the light-ring and the ISCO are located close to the event horizon, thus probing regions of stronger gravitational field with respect to the static case. We have shown that a parametrization of deformed Schwarzschild geometries in powers of $M/r$ [cf. Eqs.~\eqref{PRA1}--\eqref{PRA2}] provides a very efficient expansion, because higher-order corrections are suppressed even in the strong-field limit. We have shown this for nonspinning objects and our conclusions would likely remain valid in the slow-rotation limit. However the same is not true in the case of near-extremal objects, for which both ringdown and inspiral can probe the strong-curvature region $M/r\sim1$, where an expansion similar to~\eqref{PRA1}--\eqref{PRA2} would likely not converge. This suggests that strong-gravity effects are amplified in the near-extremal case, making highly-spinning compact objects the most promising astrophysical probes of strong gravity.

Another possible extension of this work would be to consider the impact of environmental effects on the eccentricity of compact-object binaries (see Refs.~\cite{Barausse:2007dy,gair} for
some results for the eccentricity evolution under the effect of dynamical friction and accretion).

\begin{acknowledgments}
We acknowledge Pau Amaro-Seoane, Leor Barack, Emanuele Berti, Marc Casals, Luciano Rezzolla, Joe Silk and Nico Yunes for a reading of the manuscript and useful correspondence.
E.B. acknowledges support from
the European Union's Seventh Framework Programme (FP7/PEOPLE-2011-CIG)
through the  Marie Curie Career Integration Grant GALFORMBHS PCIG11-GA-2012-321608.
V.C. acknowledges partial financial
support provided under the European Union's FP7 ERC Starting Grant ``The dynamics of black holes:
testing the limits of Einstein's theory'' grant agreement no. DyBHo--256667.
This research was supported in part by Perimeter Institute for Theoretical Physics. 
Research at Perimeter Institute is supported by the Government of Canada through 
Industry Canada and by the Province of Ontario through the Ministry of Economic Development 
\& Innovation.
P.P. acknowledges financial support provided by the European Community 
through the Intra-European Marie Curie contract aStronGR-2011-298297.
This work was supported by the NRHEP 295189 FP7-PEOPLE-2011-IRSES Grant, and by FCT-Portugal through projects
PTDC/FIS/116625/2010, CERN/FP/116341/2010, CERN/FP/123593/2011 and IF/00293/2013.
Computations were performed on the ``Baltasar Sete-Sois'' cluster at IST,
XSEDE clusters SDSC Trestles and NICS Kraken
through NSF Grant~No.~PHY-090003, Finis Terrae through Grant
CESGA-ICTS-234.
\end{acknowledgments}

\appendix
\addcontentsline{toc}{chapter}{Appendices}
\addtocontents{toc}{\protect\setcounter{tocdepth}{1}}
\section{Extreme ``dirtiness'': small black holes in anti-de Sitter backgrounds}\label{app:AdS}
An extreme, and very instructive, example of a ``dirty'' BH consists in placing a Schwarzschild BH
in a curved nonasymptotically flat background.
An example of such a spacetime which has attracted considerable interest in the past decade in the context of high-energy physics
is the Schwarzschild-anti-de Sitter geometry, 
\be
ds^2=-\left(|\Lambda|\,r^2/3+1-2M/r\right)dt^2+\left(|\Lambda|\,r^2/3+1-2M/r\right)^{-1}dr^2+r^2d\theta^2+r^2\sin^2\theta\, d\phi^2\,.
\ee
When $|\Lambda|>0$, this metric describes a BH living in a negatively curved spacetime with a negative cosmological constant. The boundary of AdS is timelike and therefore time-evolutions require boundary conditions there. For very small BHs, $\sqrt{|\Lambda|} M\ll 1$, the period $T$ of Schwarzschild ringdown modes scales as $T_{\rm Sch}\sim M$ and is very small. On other hand, the modes of pure anti-de Sitter have a period which roughly scales like the time $T_{\rm AdS}$ to reach spatial infinity, $T_{\rm AdS}\sim 1/\sqrt{|\Lambda|}$. Accordingly, for very small masses $M$ and times $t\lesssim 1/\sqrt{|\Lambda|}$ the system could not have had time to thermalize in the AdS modes, and should be ringing down in its pure Schwarzschild modes. In fact, for very small $M$ one should even see a decaying power-law tail as an intermediate stage.

To test this argument, we evolved a Gaussian wavepacket of unit amplitude, width $\sigma=0.06$, located at $r_0=0.1$, around a BH with 
radius $r_+=0.025$ and $|\Lambda|=3$. We impose regularity conditions at the event horizon, $\Phi(t,r)\sim \psi(t+r_*)/r$ and at spatial infinity, where the field now decays as $\Phi(t,r)\sim \psi(t)/r^3Y_{lm}(\theta,\phi)$. 
\begin{figure}[thb]
\begin{center}
\begin{tabular}{cc}
\epsfig{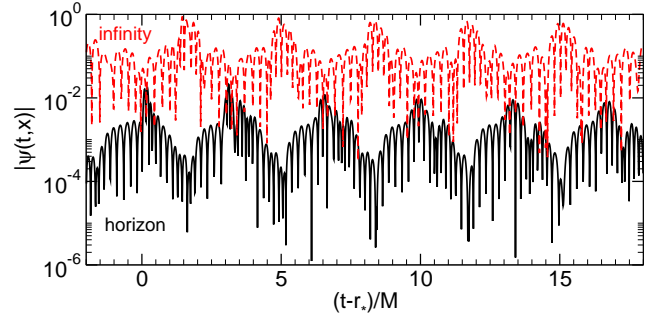}
\end{tabular}
\caption{Scattering of a scalar wavepacket in the Schwarzschild-AdS background. The wavepacket has a unit amplitude, width $\sigma=0.06$
and is centered at $r_0=0.1$, around a BH with radius $r_+=0.025$ (in units with $|\Lambda|=3$).
\label{fig:AdS}}
\end{center}
\end{figure}
The results are shown in Fig.~\ref{fig:AdS}, and show at early times the ringdown characteristic of {asymptotically flat} Schwarzschild BHs, modulated by the Schwarzschild-AdS frequency. Notice how each pulse is slowly absorbed on each interaction with the (small) BH. Eventually the first pulse gets distorted by multiple reflections off the boundary.

The main conclusion, trivial in retrospect, is that the modes of asymptotically flat Schwarzschild BHs do dominate the early time response.
We stress that this result also highlights the quasi-local nature of Schwarzschild ringdown, as the decay of light rays trapped at, and slowly leaking from, the null circular geodesic~\cite{Cardoso:2008bp}. It would be very interesting to see these intermediate states arising from a Green's function analysis in the same way that QNMs arise from poles of the corresponding Green's function~\cite{Leaver:1986gd,Berti:2009kk}.
\section{Gravitational perturbations of a Schwarzschild BH surrounded by a thin shell}\label{app:shell}
In this appendix we discuss how to compute the gravitational QNMs of a Schwarzschild BH surrounded by a thin shell. The relevant formalism has been developed in Ref.~\cite{Pani:2009ss}.

We start with the axial perturbations and discuss two (equivalent) methods to implement the junction conditions~\eqref{junctionaxial}.

\noindent {\bf Method A.} As discussed in Ref.~\cite{Pani:2009ss}, the junction conditions~\eqref{junctionaxial} assume that, at least along the worldline $r=r_0$, there exists a common time coordinate patch for the interior and the exterior metric. In practice, in a static, spherically symmetric spacetime this can be imposed by requiring continuity of $g_{tt}$ at the shell location. The metric~\eqref{metricshell} satisfies this criterion and we shall consider this background. 

It is easy to show that, in this background, the Regge-Wheeler function $\psi$ is related to the perturbation functions $h_0$ and $h_1$ by:
\begin{equation}
 \psi=\frac{A}{r}h_1\,,\qquad h_0=\frac{i}{\omega}B(r\psi)'\,. \label{defRW}
\end{equation}
Using the definitions above and Eqs.~\eqref{junctionaxial}, we obtain:
\begin{eqnarray}
 \psi_+&=&\frac{\psi_-}{\sqrt{{\bar\alpha}}}\,, \label{axial1}\\
 \psi_+'&=& \frac{\psi_-}{r_0}\left(\frac{1}{{\bar\alpha}}-\frac{1}{\sqrt{{\bar\alpha}}}\right)+\frac{\psi_-'}{{\bar\alpha}}\,. \label{axial2}
\end{eqnarray}
Thus, we can integrate the equations starting from the horizon outwards, extracting $\psi_-$ at the shell location, using the condition above to compute $\psi_+$ and $\psi_+'$, and finally continue the integration outwards to infinity, where the appropriate boundary condition is imposed.

\noindent {\bf Method B.} An equivalent method is based on recognizing that the metric in the interior is equivalent to the Schwarzschild metric after a time redefinition $t=\tilde t/\sqrt{{\bar\alpha}}$ (we use a tilde for any quantity in these coordinates). In these coordinates, $\tilde A=\tilde B$ and they are both equal to the standard Schwarzschild element with $M$ and $M_0$ in the interior and in the exterior, respectively. It is easy to show that Eq.~\eqref{defRW} becomes:
\begin{equation}
 \tilde\psi=\frac{B}{r}\tilde h_1\,,\qquad \tilde{h}_0=\frac{i}{\tilde \omega}B(r\tilde\psi)'\,, \label{defRW2}
\end{equation}
where the time dependence of each quantity has been factored out as $e^{-i\tilde\omega\tilde t}$ and we have used $\tilde{A}=\tilde{B}=B$. Now, in order to use the junction conditions~\eqref{junctionaxial}, we need to transform the Regge-Wheeler function to the coordinates where $g_{tt}$ is continuous. Under $\tilde{t}\to\sqrt{{\bar\alpha}}t$, we get:
\begin{eqnarray}
 \tilde{h}_0 e^{-i\tilde\omega\tilde t}d\tilde t &\to& h_0 e^{-i\omega t}dt\,,\\
 \tilde{h}_1 e^{-i\tilde\omega\tilde t} &\to& h_1 e^{-i\omega t}\,,
\end{eqnarray}
where $\omega=\tilde{\omega}\sqrt{{\bar\alpha}}$, $h_1=\tilde{h}_1$ and $h_0=\tilde{h}_0\sqrt{{\bar\alpha}}$. With this transformation, the perturbation functions are precisely in the form needed to apply Eqs.~\eqref{junctionaxial}. Substituting in Eqs.~\eqref{defRW2}, we get:
\begin{equation}
 \tilde{\psi}=\frac{B}{r}h_1\,,\qquad \frac{h_0}{\sqrt{{\bar\alpha}}}=\frac{i\sqrt{{\bar\alpha}}}{\omega}B(r\tilde \psi)'\,. \label{defRW2b}
\end{equation}
which are equivalent to Eqs.~\eqref{defRW} after the trivial rescaling $\tilde\psi=\psi/{\bar\alpha}$. Finally, we obtain again Eqs.~\eqref{axial1} and \eqref{axial2}, as we should.

Computing the junction conditions for polar perturbations is more involved. In the background~\eqref{metricshell}, using the linearized Einstein equations, we define the Zerilli function as:
\begin{eqnarray}
 Z&=&\frac{2 r^2 \omega  K-2 i (r-2 M_i) H_1}{\left(6 M_i+\lambda r\right) \sqrt{{\bar\alpha} } \omega }\,,\\
 Z'&=&\frac{i (r-2 M_i) \left(24 M_i^2+6 \lambda M_i r+(l-1) l (1+l) (2+l) r^2\right) H_1-2 r^2 \left(\lambda r^2-6 M_i^2-3 \lambda M_i r\right) \omega  K}{r (r-2 M_i) \left(6 M_i+\lambda r\right)^2 {\bar\alpha}  \omega }
\end{eqnarray}
where $\lambda=l(l+1)-2$ and $M_i$ takes the values $M$ or $M_0$ when the Zerilli functions are computed in the interior and in the exterior, respectively. Therefore, in order to integrate the Zerilli equation, we need only to compute the jump of the functions $K$ and $H_1$ across the shell. As for $K$, the first of the junction conditions~\eqref{junctionpolar} imposes continuity across the shell. In order to compute the jump of $H_1$, we use the equations of motion:
\begin{eqnarray}
 K'&=& \frac{2 (2 M_i-r) r \omega  H+i l (1+l) (2 M_i-r) H_1+2 r (r-3 M_i) \omega  K}{2 (2 M_i-r) r^2 \omega }    \,,\\
 H&=& \frac{i (2 M_i-r) \left(l (1+l) M_i {\bar\alpha} -2 r^3 \omega ^2\right) H_1+r \omega  \left(-6 M_i^2 {\bar\alpha} -2 \left(l(l+1)-3\right) M_i r {\bar\alpha} +\lambda r^2 {\bar\alpha} -2 r^4 \omega ^2\right) K}{r (r-2 M_i) \left(6 M_i+\lambda r\right) {\bar\alpha}  \omega }  \,.
\end{eqnarray}
The second of these equations is algebraic and it can be used to eliminate $H$ from the first equation. Finally, using the junction conditions~\eqref{junctionpolar} for $K$ and $K'$ we can convert the first equation above in a junction condition for $H_1$ only. The form of this equation is not very illuminating, but it is simply an algebraic equation in the form $H_{1+}=H_{1+}(Z_-,Z_-')$. Finally, these junction conditions allow us to write $Z_+$ and $Z_+'$ in terms of $Z_-$ and $Z_-'$ and to continue the integration to infinity.

An alternative --~and in fact more elegant~-- approach is the following. First, one could write the spacetime in Schwarzschild form by using the time coordinate $\tilde{t}$ defined in Method B above. Then, in this background one can use Chandrasekhar's transformation to relate the Zerilli and the Regge-Wheeler function. Thus, in the interior one can compute the Regge-Wheeler function outwards to the shell, transform to the Zerilli functions, apply the junction conditions across the shell in the way explained in Method B above, transform back to the Regge-Wheeler function and finally integrate the Regge-Wheeler equation to infinity. To test our method, we checked that these procedures are equivalent.

\bibliography{DirtyEffects_BHs}
\end{document}